\documentclass[fleqn,usenatbib,useAMS]{rasti}

\usepackage{newtxtext,newtxmath}
\usepackage[T1]{fontenc}
\usepackage{ae,aecompl}

\usepackage{graphicx}	% Including figure files
\usepackage{makecell}
\newcommand{\progeny}{{\sc ProGeny}}

\newcommand{\prospect}{{\sc ProSpect}}
\newcommand{\Rfits}{{\sc Rfits}}

\newcommand{\R}{{\sc R}}

\newcommand{\Cpp}{{\sc C++}}

\newcommand{\FITS}{{\sc FITS}}

\newcommand{\msol}{M$_{\odot}$}
\newcommand{\Zsol}{Z$_{\odot}$}

\title[ProGeny]{ProGeny I: a new simple stellar population spectra generator and impact of isochrones / stellar spectra / initial mass functions}

\author[A.~S.~G.~ Robotham et al.]{
A.~S.~G. Robotham$^{1}$\thanks{E-mail: aaron.robotham@uwa.edu.au} \&
S. Bellstedt$^{1}$
\\\\
$^{1}$ICRAR, M468, University of Western Australia, Crawley, WA 6009, Australia \\
}

\date{Last updated YYYY MM DD; in original form YYYY MM DD}

\pubyear{2024}

\begin{document}
\label{firstpage}
\pagerange{\pageref{firstpage}--\pageref{lastpage}}
\maketitle

\begin{abstract}
In this work we introduce \progeny, a new stellar population library (SPL) software package written in \R. This release encapsulates the core open source software (\url{github.com/asgr/ProGeny}) to generate simple/single stellar populations and their associated spectra (SSPs); the various data inputs required (in particular isochrones and stellar spectra); example scripts to generate the SSPs; and a number of pre-generated static SSPs available for immediate use. The most novel feature of \progeny{} is its ability to produce SSPs with evolving initial mass functions (IMFs), allowing functional dependencies on stellar age or metallicity. We perform both internal comparisons (within the \progeny{} SPL) and external comparisons (to SSPs associated with other public SPLs) and tests. The main conclusion is that the choice of isochrone has significantly more impact on the predicted SSP spectra than the choice of stellar spectra and/or IMF (comparing Chabrier and Kroupa variants). A number of limiting uncertainties and corrections for star formation rates and stellar masses are also presented.
\end{abstract}


%%%%%%%%%%%%%%%%%%%%%%%%%%%%%%%%%%%%%%%%%%%%%%%%%%

%%%%%%%%%%%%%%%%% BODY OF PAPER %%%%%%%%%%%%%%%%%%

% The MNRAS class isn't designed to include a table of contents, but for this document one is useful.
% I therefore have to do some kludging to make it work without masses of blank space.

\section{Introduction}

A core requirement when modelling the light output of a galaxy is an estimate of what a spectrum of a single (or simple) stellar population (SSP) looks like as a function of age, metallicity and initial mass function (IMF) of the stars \citep[see e.g.][]{1999ApJS..123....3L, 2003MNRAS.344.1000B, 2005MNRAS.362..799M, 2009ApJ...699..486C, 2016MNRAS.463.3409V, 2018MNRAS.479...75S, 2019MNRAS.490..978P}. Such an SSP is the base component of more complex modelling, which may include contributions or modifications from dust and active galactic nuclei (AGN). Various software packages are available that can combine SSPs with more complex models of star formation history in order to fit spectral energy distribution (SED, covering multi-band and/or spectra) observations of galaxies, e.g.: \prospect{} \citep{2020MNRAS.495..905R}; MAGPHYS \citep{2008MNRAS.388.1595D} and CIGALE \citep{2019A&A...622A.103B}; \citep{2021ApJS..254...22J}; Bagpipes \citep{2018MNRAS.480.4379C}; and Beagle \citep{2016MNRAS.462.1415C}. Such software often has different emphasis, and in detail no single software reproduces all the user functionality of another.

\prospect{} was written to work equally well in a pure generative mode \citep[for producing predicted spectral outputs for simulations, see][]{2018MNRAS.481.3573L, 2019MNRAS.489.4196L} and inference mode. The latter can also be thought of as `fitting' or `inversion' mode, and has been widely applied to observational data, with particular focus on the GAMA and DEVILS/COSMOS surveys (see Table \ref{tab:acronym} for survey descriptions). A number of works have tested, utilised and expanded on the functionality of \prospect, e.g.: \citet{2020MNRAS.498.5581B, 2021MNRAS.503.3309B, 2021MNRAS.505..540T, 2022MNRAS.509.4940T, 2022MNRAS.517.6035T, 2023MNRAS.522.6354T}. \prospect{} already comes with a broad range of popular SSPs from a few different stellar population libraries (SPLs). But while switching between SSPs is instructive re the robustness of the fitting, it is difficult to fully control the aspects that are varying because of how differently constructed the libraries are. It is also unfortunate that very few SPLs are accompanied by fully public and open source code \citep[an admirable exception being][]{2009ApJ...699..486C}, making design and code decisions almost impossible to assess beyond the cursory description proved in (often outdated) reference papers.

The experience of using \prospect{} for analysis has motivated a deeper investigation into the precise role SSPs play when analysing data. While \prospect{} does come with a number of available SSPs, their precise production varies in detail and it is not possible to fully control and understand the causes of any inference variations. To combat this, and in parallel provide more flexibility to the types of SSP available for use in \prospect{} and any similar SED generation or fitting software, we have produced the new open source (LGPL-3) \R{}\footnote{\citet{citeR}} package \progeny{} that we present in this work. It offers highly flexible production of SSPs via the variation of isochrones, stellar spectra\footnote{For clarity, this refers to the spectrum of a stellar photosphere, be it empirical or theoretical. In this paper when we refer to `stellar spectra' or similar, this is what is meant.} and IMFs. Perhaps most novel, compared to any current literature SSPs, \progeny{} provides the means to generate evolving IMF SSPs, either as a function of age or metallicity \citep[see][]{2012MNRAS.422.2246M, 2015ApJ...806L..31M}.

The paper provides a brief overview of various aspects of SSP spectra generation, but we note that more detailed references are readily available in the literature. We highlight the review of \citet{2013ARA&A..51..393C}, and the earlier work of \citet{1996ApJ...457..625C, 2005MNRAS.362..799M, 2007MNRAS.381.1329M, 2009ApJ...703.1123P, 2009ApJ...699..486C}. The literature consensus has long been that within the bounds of reasonable ranges, the choice of isochrones has the most impact on the computed spectral outputs of SSPs, outranking the impact of stellar spectra and IMF. The stability and sensitivity of these ingredients also has a knock-on in terms of the accuracy specific quantities can be measured to. Generally stellar masses can be measured with a good degree of confidence \citep[0.1--0.2 dex error being common, see][]{2011MNRAS.418.1587T, 2020MNRAS.495..905R}, but associated properties like age and metallicity are increasingly difficult to infer \citep{1996ApJ...457..625C}. In this work we discuss in detail the available isochrones, stellar spectra and IMFs provided with this initial release of \progeny.\footnote{v0.4.0 is the exact version used to create outputs for this paper and the associated Bellstedt \& Robotham in prep. companion paper.} A companion paper (\progeny{} II; Bellstedt \& Robotham in prep.) explores the impact of all these ingredients in depth, investigates how SSP choices affect the interpretation of observational data, and contains most of the \progeny{} relevant analysis regarding SSP modelling error etc.

While some fiducial SSPs outputs are generated in association with this paper (and are analysed and discussed in latter sections), detailed examples are also provided and documented so users can produce their own SSPs as required in under a minute of computational time. It is also relatively easy for advanced users to supplement the provided isochrones, stellar spectra and IMFs with their own, as long as they follow the required catalogue and FITS file formats. The package requires no compilation and has a small number of dependencies, so many common barriers to entry for experimentation have been removed. We note this paper is joined by a companion paper (Bellstedt \& Robotham in prep.) that investigates in depth the impact that switching SPLs and \progeny{} SSPs has when fitting galaxy SEDs.

In this work we discuss the main aspects of generative stellar population synthesis modelling for single stars (i.e.\ we neglect binary treatment in this initial release) (Section \ref{sec:genmodel}); the specific version of the \progeny{} SPL released in this work (Section \ref{sec:SPL}); a comparison between different ingredients in \progeny{} and other literature SSPs (Section \ref{sec:litcomp}); some overall conclusions of the work (Section \ref{sec:conclusions}); and a summary of the software, script and data released to coincide with this paper (`Data and Software Availability'). Because of the large number of software packages and astronomy acronyms used in this paper, a compact glossary is included in Appendix \ref{sec:glossary}. Where relevant we assume a H$_0$ = 67.8 (kms/s)/Mpc, $\Omega_M$ = 0.301 and $\Omega_\Lambda = 0.699$ cosmology \citep{2020A&A...641A...6P}. We note that when we refer to \Zsol{} we implicitly mean log$_{10}$(Fe/H) + 12 $\sim$ 7.46 and log$_{10}$(O/H) + 12 $\sim$ 8.69 \citep[as per][]{2009ARA&A..47..481A}. All isochrones and stellar spectra used are broadly consistent with this definition of \Zsol, and therefore implied metallicity (always in units of Z/\Zsol in \progeny) can be scaled as desired.

\section{Stellar Population Synthesis Methodology}
\label{sec:genmodel}

Stellar population synthesis (SPS) is generally used as the encompassing term for the steps required to make a synthetic spectrum of a distribution of stars \citep[e.g.][]{2013ARA&A..51..393C}. A specific approach to doing this is what we consider to be a stellar population library (SPL), and a specific combination of parameters from an SPL is what generates spectra for a SSP (in our nomenclature). For simplicity, when we refer to a SSP in this paper we are also referring to associated spectra generated \citep[this is consistent with the nomenclature of][]{2013ARA&A..51..393C}.

\progeny{} embodies two core concepts: a flexible software methodology for defining isochrones, stellar spectra, and interpolation preferences; and an SPL of semi-static SSPs\footnote{where in our usage an SPL is a family of related SSPs} which will be improved and expanded over time. We encourage inquisitive users to explore and modify default options to create customised SSPs suited to their needs, particularly the initial mass function (IMF). Users can also supplement the provided isochrones and stellar spectra with new or alternative versions. As \progeny{} evolves, we will expand base data options, creating new varieties while maintaining the original reference library for posterity.

In the following sub sections we explore the key components of \progeny, in particular the characteristics, units, and data formats required for seamless integration with the wider package ecosystem.

\subsection{Isochrone Format}

Isochrones are the standard means for conveying how stars evolve over time. Typically, they begin at the zero-age main-sequence (ZAM) point, marking the onset of star formation, and trace the evolution of a stellar population to a target age (and/or end of life). Some isochrones extend their coverage to pre-main-sequence (PMS) phases, modelling the evolution of low-mass objects that never reach hydrogen fusion (e.g.\ brown and red dwarfs). Others incorporate special treatment for more exotic periods of stellar evolution (e.g.\ horizontal branch and asymptotic giant branch stars), and/or the evolution of remnant material (e.g.\ black holes, white dwarfs and neutron stars).

The key information that isochrones must convey are the age, mass, luminosity (and potentially the effective physical size), temperature and surface gravity of the relevant stars. Ideally, this is also done for a range a metallicity (Z) and $\alpha$/Fe abundances, since both of these properties have a material impact on the evolution of stars. In a perfect world this information is provided in infinitesimally small increments of all the most interesting parameters for the purposes of constructing an SSP: age, mass and Z. However, pragmatically choices have to be made about the coarseness of the sampling in all dimensions either due to data and modelling limitations, or due to the lack of resulting spectral variability making higher resolution sampling unnecessary. \progeny{} only requires regular binning in terms of age and metallicity, i.e.\ the stellar masses for each of these subsets can vary. In this way, the target isochrone adjusts the discrete masses to focus on those undergoing the most dramatic phases of stellar evolution, such as the highly luminous horizontal branch or asymptotic giant branch phases. This is how all the included MIST, PARSEC, and BaSTI isochrones operate (discussed in detail in Section \ref{sec:FL_iso}).

\progeny{} requires the entire isochrone library to be stored in a single table, resulting in potentially large file sizes (typically several GB) to accommodate fine sampling in age, mass and Z. The subsetting of properties is carried out efficiently within \R{} by virtue of the {\tt data.table} package, which allows for highly efficient threaded table subsetting operations. This method proved to be orders of magnitude faster than splitting the isochrones across separate files of (e.g.) age and metallicity, as is more typical. An example isochrone table (using the MIST isochrone included with \progeny) is shown in Table \ref{tab:iso_ex}. This can be saved in any format readable by \R, but for high-speed loading the FST or Apache Parquet binary table formats are recommended.

\begin{table*}
\begin{center}
\caption{Example of the minimal isochrone tabular information required for \progeny. Columns are named as expected. `logZ' is the logarithmic metallicity relative to solar; `logAge' is the logarithmic age in years; `Mini' is the initial formation mass of the star; `Mass' is the current mass of the star (which will always be the same or lower than `Mini' due to various mass loss events); `Lum' is the luminosity in solar units; `Teff' is the effective temperature of the photosphere in Kelvin; `logG' is the logarithmic surface gravity in CGS units ($\text{cm}/\text{s}^2$); `label' is the stellar phase label (this is optional, but it can be used to match stellar spectra more accurately). This specific extract is from the 1,494,453 row MIST isochrone included with \progeny.}
\begin{tabular}{|r|r|r|r|r|r|r|r|}
 logZ & logAge & Mini & Mass & Lum & Teff & logG & label \\ 
  \hline
 -4.00 & 5.00 & 0.10 & 0.10 & 0.32 & 4140.11 & 3.35 & -1.00 \\ 
 -4.00 & 5.00 & 0.10 & 0.10 & 0.34 & 4149.91 & 3.35 & -1.00 \\ 
 -4.00 & 5.00 & 0.11 & 0.11 & 0.36 & 4164.44 & 3.34 & -1.00 \\ 
 -4.00 & 5.00 & 0.11 & 0.11 & 0.38 & 4178.90 & 3.34 & -1.00 \\ 
  -4.00 & 5.00 & 0.12 & 0.12 & 0.40 & 4193.29 & 3.34 & -1.00 \\ 
  ... & ... & ... & ... & ... & ... & ... & ... \\ 
\end{tabular}
\label{tab:iso_ex}
\end{center}
\end{table*}

\subsection{Stellar Photosphere Spectrum Format}

When making an SPL, the equally important complement to the choice of isochrones is the selection of stellar spectra, i.e.\ the predicted spectral outputs of stars for specific combinations of stellar evolutionary parameters. The main properties that determine the precise spectrum of a stellar photosphere are (in roughly descending order of importance) the: surface temperature, the surface gravity, the stellar metallicity, and the alpha abundance ($\alpha$/Fe). The ideal version of \progeny{} would have stellar spectra generated for the exact combination of these variables required by the isochrone at all ages, however this is never possible in practice because the isochrones track the evolution of stars (and therefore these key parameters) much more finely than could ever be generated by a stellar spectrum library. The reason for this depends on the source of the stellar spectra: for theoretical templates the limitation is really one of computation time and data volume \citep{1999ApJ...512..377H}, and for observed templates the limitation is the difficulty observing enough stars with a broad high resolution spectrograph \citep[e.g.\ X-Shooter on the VLT;][]{2011A&A...536A.105V}. The pragmatic solution is then to assemble stellar spectra as densely as is feasible, and at the desired resolution of the target SSP.

The fiducial set of stellar spectra that are provided with \progeny{} are discussed in more detail in Section \ref{sec:FL_atmos}. All stellar spectra share a common FITS file format that provides \progeny{} with the critical information needed for properly constructing an SSP. Should users wish to supplement \progeny{} with additional stellar spectra, or simply replace the fiducial libraries with different options, this is the multi-extensions FITS format that must be replicated:

\begin{itemize}
\item {\bf wave}: FITS vector of the wavelength grid (Angstroms). The vector elements must match 1--1 with the rows of the fourth extension {\bf spec}
\item {\bf info}: FITS table extension containing columns of Teff (Kelvin); logG (log$_{10}$ surface gravity, g/[cm/s$^2$]); logZ  (log$_{10}$ metallicity, Z) logA (log$_{10}$ alpha abundance, $\alpha$/Fe). The rows must match 1--1 with the rows of the fourth extension {\bf spec}.
\item {\bf rescale}: FITS vector of length three which specifies the rescaling to apply to the Teff, logG and logZ columns of {\bf info} when interpolating. The relevant columns are divided by the values in {\bf rescale} when interpolating, and in that sense they should correspond to the typical grid step-size in each dimension. The resulting search is then carried out in grid step-size units, rather than native absolute units which might differ hugely between parameters.
\item {\bf spec}: FITS matrix of the stellar spectra bolometrically normalised to 1. The rows should correspond to the properties of second extension table {\bf info}, and the columns are the relevant spectra that match 1--1 with the first extension vector {\bf wave}.
\end{itemize}

\subsection{Grid Interpolation Scheme}

The common problem all SPL software that generate SSPs have to solve is a mechanism to match the very fine resolution predictions of stellar properties generated by isochrones onto the necessarily coarser grids of available stellar spectra. High-quality isochrones capture small shifts in the temperate and the surface gravity, and ideally map out variations in metallicity and other properties such as $\alpha$/Fe abundances and stellar rotation.

Being pragmatic, the best solution is to map out a very broad range of these properties with a few different libraries (since different libraries tend to focus on different properties, and none cover all the parameter space generated by modern isochrones) and then use interpolation methods to approximate specific solutions that fall between available grid positions. Previous efforts have generally focussed on direct spectral interpolation \citep[where each wavelength element is treated as a multi-dimensional interpolation problem, e.g.][]{2022AA...661A..50V} or global interpolation \citep[where entire spectra are weighted depending on their `distance' to the objective stellar properties, e.g.][]{2003MNRAS.344.1000B}. Which method is used is normally informed by the spectral data in hand: empirically focussed libraries tend to be so sparse that direct spectral interpolation is critical, but theoretically focussed libraries tend to have much denser grids of spectra (especially in the crucial temperature and surface gravity dimensions) so global interpolation is preferable. \progeny{} has been predominantly developed using theoretical stellar atmospheric spectra as inputs, so we have focussed on global interpolation strategies.

Having chosen a global interpolation scheme, there are a number of additional subjective decisions to be made. How do we decide `distance' in such multi-dimensional space, and what type of penalty do we want to apply? It is clear that spectra which are natively closer to the target properties produced for a particular star in our isochrone would always be preferred, but preferring spectra that are, e.g.\, 100 K closer in Teff but 0.1 dex further away in surface gravity is not as obvious.

We carried out tests for how best we could replicate missing spectra. The optimal strategy in general is to define `distance' relative to the average gridding separation, in either linear of log units as appropriate. Distance in each parameter dimension ($d_i$) is then defined as relative to the mean grid distance ($\mu_{\Delta \nu_i}$) in this dimension $i$, where the upper Euclidian radius (or $L^2$ norm) we search over is 2 (by default) in this unit-less re-mapping of the parameter space. We then weight the spectra by the inverse of the re-scaled distance to the power 2 (by default), so if a library stellar spectrum is twice as close in this re-scaled parameter space it will contribute four times as much flux. For a particular spectrum and a particular parameter dimension with an absolute target to grid separation of $\delta \nu_i$, the weight will look like:

\begin{eqnarray}
d_i = \frac{\delta \nu_i}{\mu_{\Delta \nu_i}}, \\
w_i = \frac{1}{d_i^2}.
\end{eqnarray}

We can choose to limit the number of neighbouring spectra we allow to contribute (by default we limit this to 8, corresponding to cubic vertices in three-dimensional space) and enforce the sum of weights to equal 1 via a final renormalisation after all nearest matches have been combined. When determining the most appropriate spectrum to use \progeny{} can optionally use information from the isochrone regarding the stellar evolutionary phase. This means that only a star flagged as being in the AGB phase would be assigned an AGB spectrum etc. This option is recommended, on by default, and used for all work in this paper.

When producing a spectrum for a target age stellar population we can combine all the weights for common spectral interpolation matches. This means we can stack spectra much more efficiently than simply adding every occurrence with a small weight, instead adding a particular stellar spectrum once with all the weight terms summed accordingly. This interpolation method uses \citet{arya1998optimal} Approximate Nearest Neighbours (ANN) C++ library, so it is highly efficient for interpolating spectral grids even onto very large isochrone tables --- 1.5m MIST isochrone parameters takes about 2 seconds to map onto the 7.5k stellar spectrum grid values of the default \citet{2013AA...553A...6H} library discussed in detail below. The parameter interpolation only needs to be carried out once per generation of a new SSP, and in fact only needs to be redone if the search radius or weight power terms are being modified (i.e. a change of IMF would not require the regeneration of interpolation weights).

\subsection{Initial Mass Function}

The last critical part of generating an SSP (after choosing an isochrone, combining stellar spectra libraries, and creating a scheme to interpolate them) is to weight different stellar populations with a target initial mass function (IMF). This function simply encodes the original distribution of stellar masses that were generated by a single burst of star formation, and effectively this becomes an additional weighting term when we combine different stars in an isochrone \citep[for a detailed discussion on the physics and measurements of the IMF see][]{2024arXiv241007311K}.

IMFs in general share the characteristic that you expect to produce more (by number) stars of lower mass than higher mass. Many studies have established that there is some sort of natural lower bound to the star mass allowed, but debate continues over the exact shape of the IMF and the nature of the upper mass limit \citep[or if there is one at all, see][]{2004MNRAS.348..187W}. A further point of contention is how truly Universal and/or evolving the IMF might be. Is it self-similar over all time and space? Does it naturally evolve over time but not depend on the spatial environment? Does it only vary based on the spatial environment? All these questions are still regularly discussed in the modern astronomy literature \citep[see][]{2001MNRAS.322..231K, 2003ApJ...593..258B, 2008ApJ...674...29V, 2009ApJ...695..765M, 2010ARA&A..48..339B, 2023Natur.613..460L, 2024A&A...689A.221H}, and access to highly flexible SSPs is a vital component in better answering them.

A key reason \progeny{} was developed at all was our desire to have a simple interface to generate new SSPs with arbitrary IMFs. Generally speaking SPLs generate SSPs with a limited number of IMFs, which can be quite restrictive for analysis. We note Chabrier and Kroupa IMFs are produced most commonly, with some SPLs also generating Salpeter IMFs and various definitions of top and/or bottom-heavy IMF (these are less consistently defined, but reflect the IMF having a relative excess of high/low mass stars respectively). Even within a specific IMF you will get variants because a particular star formation event might have stars above or below a certain stellar mass limit entirely removed. Common limits used in published SSPs restrict the IMFs to only produce stars between 0.08--0.1 \msol{} (low limit) to 100--300 \msol{} (upper limit). With all these potential IMF shape and truncation variants you rapidly require dozens of IMFs to flexibly explore the impact of how robust any inference is to your IMF assumptions.

\progeny{} inherits the functional form of IMF used in \prospect{} already (general purpose variants of the Chabrier \citep{2003PASP..115..763C}, Kroupa \citep{2001MNRAS.322..231K} and Salpeter \citep{1955ApJ...121..161S} IMFs, with default parameters values that return the fiducial IMFs), and allows uses to modify the IMF to generate a new SSP very easily (and rapidly). As long as a similar functional format is used, users can also freely define their own entirely novel form of IMF. The only requirement to work in \progeny{} is that the input expected must be stellar mass in solar units, and the output must be the differential number counts defined such that the mass integral between extreme mass limits equals exactly 1 (reflecting a single solar mass of star formation, in principle). The included Chabrier, Kroupa, and Salpeter IMFs already obey this requirement even when the default parameters are modified, so users can inspect these functions to understand how similar functionality can be achieved for a completely new form of IMF.

\subsection{Output Stellar Populations Library Format}

Loading the \progeny{} package, loading all the relevant MIST isochrones and default moderate resolution stellar spectrum data, running the above steps via their respective \progeny{} functions, and generating a new in-memory SSP that can be directly loaded into \prospect{} takes roughly 3 minutes per metallicity output per core (this timing is based on using an Apple M3 Max processor). The creation of SSPs in \progeny{} is a natively parallel process, and by default uses 8 cores. For a particular isochrone if you have as many unique isochrone metallicity values as cores available, this should mean generating a new SSP should only ever take a couple of minutes from scratch, and much less time if you are only (e.g.) updating the IMF and not re-computing the interpolation grid weights etc.

The in-memory SSP can be saved directly to a single FITS file using the \Rfits{} package \citep[see][]{2024MNRAS.528.5046R}. The high-level \prospect{} package has a function available (speclib\_FITSload) to read the FITS SSP file back into memory with the correct structure. The details of the on-disk structure are described in the \prospect{} manual for interested readers, so we will not describe it here.

\section{The \progeny{} Stellar Population Library}
\label{sec:SPL}

Below we discuss the significant decisions made regarding which isochrones (\ref{sec:FL_iso}), stellar spectra (\ref{sec:FL_atmos}), and IMFs  (\ref{sec:FL_IMF}) are included with the initial public release of \progeny. We also outline the significant data formats for both input and output data, and how the input ingredients are computationally combined to create target SSPs.

\subsection{Isochrones}
\label{sec:FL_iso}

\begin{figure*}
\begin{center}
\includegraphics[width=5.8cm]{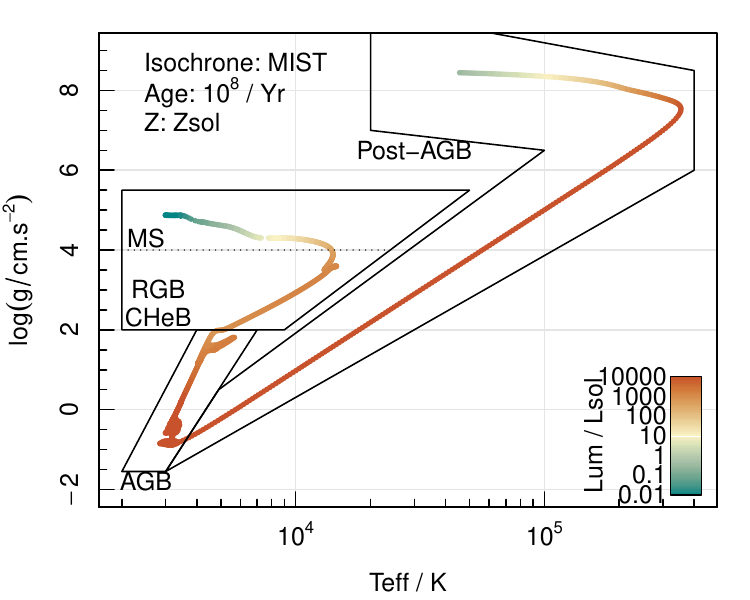}
\includegraphics[width=5.8cm]{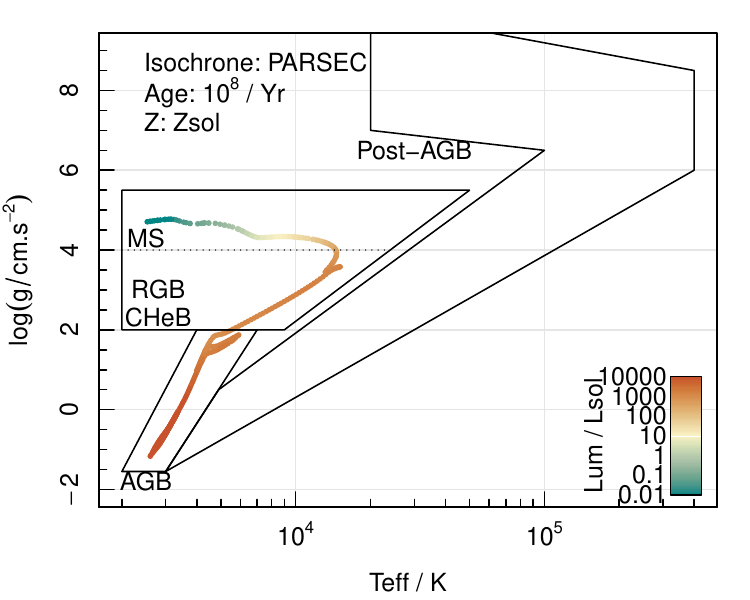}
\includegraphics[width=5.8cm]{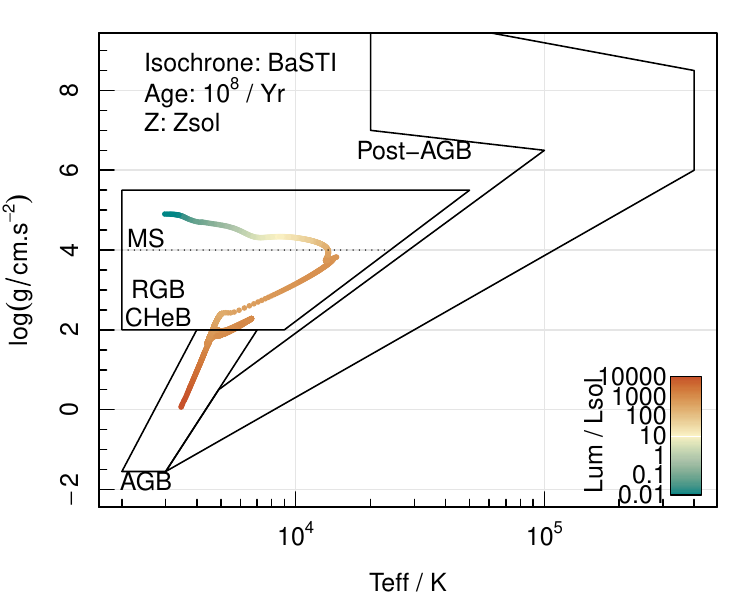}
\\
\includegraphics[width=5.8cm]{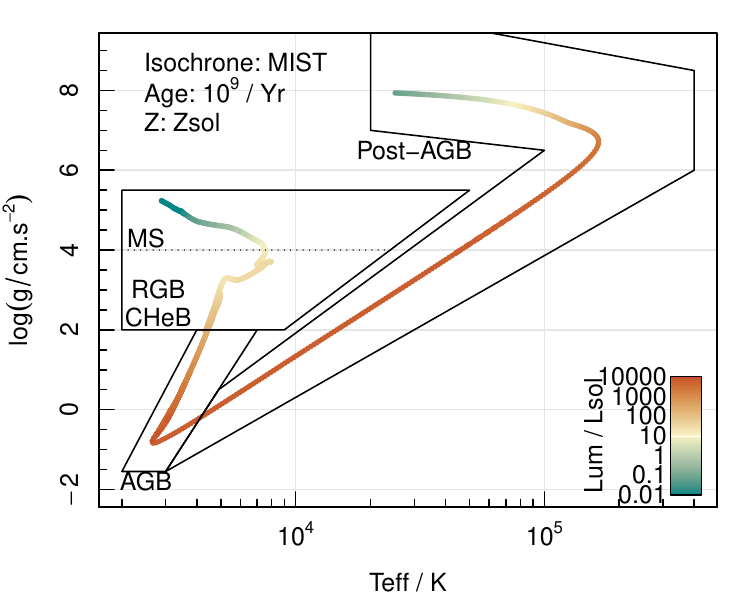}
\includegraphics[width=5.8cm]{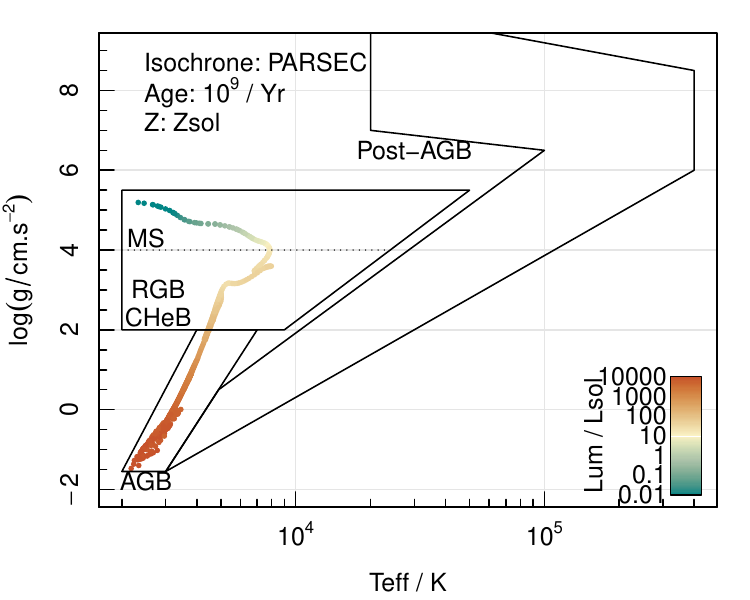}
\includegraphics[width=5.8cm]{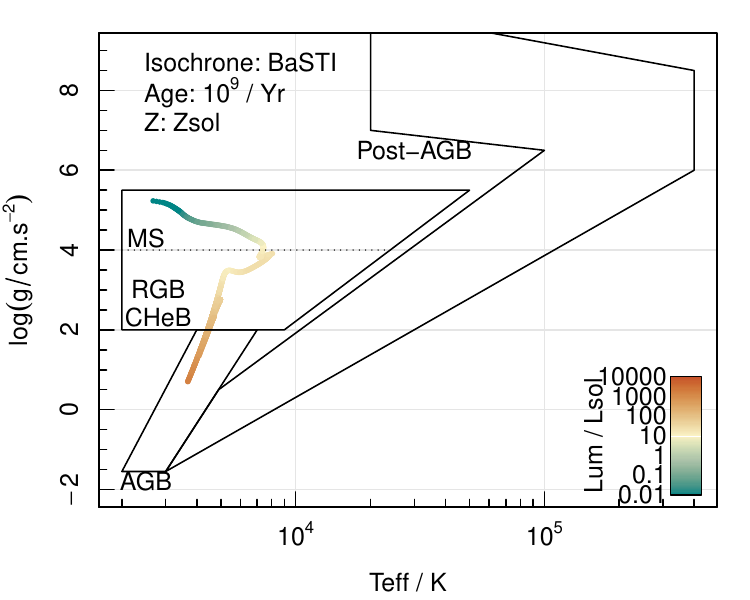}
\\
\includegraphics[width=5.8cm]{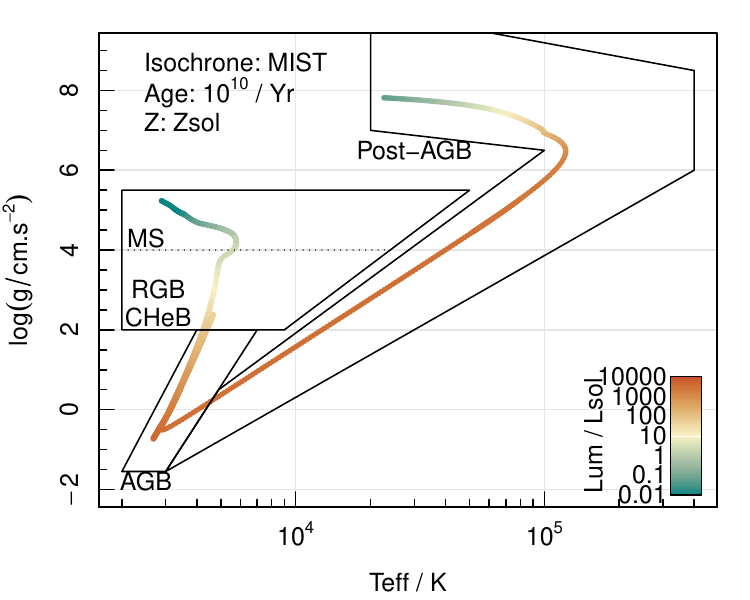}
\includegraphics[width=5.8cm]{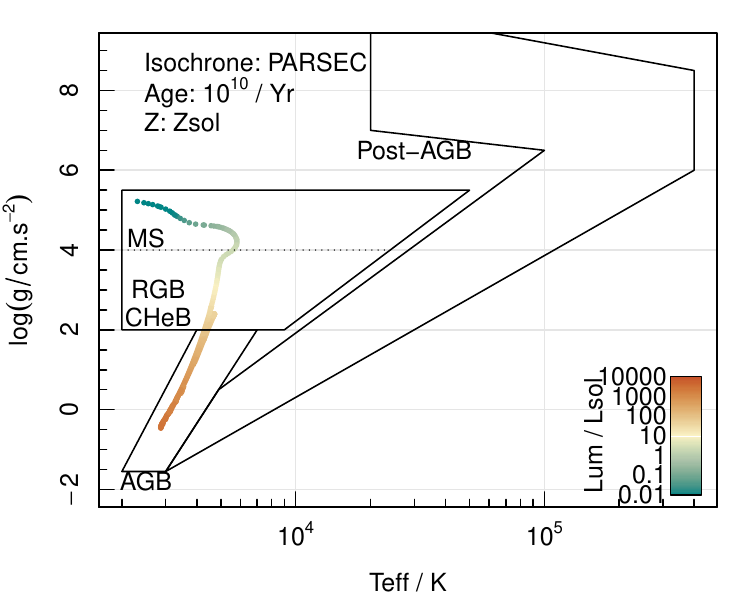}
\includegraphics[width=5.8cm]{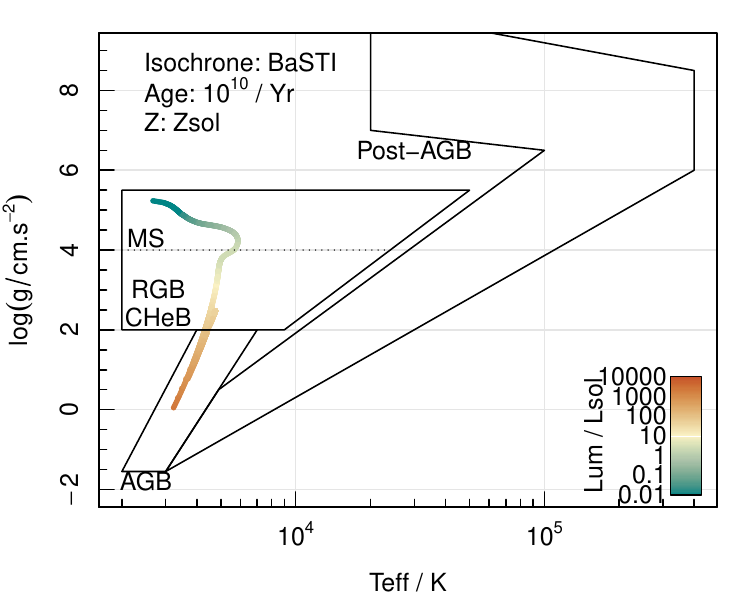}
\\
\caption{Comparison of isochrones. Fiducial MIST are the left panels; PARSEC are the middle panels; BaSTI are the right panels. The three rows represent (from top to bottom) 100 Myr, 1 Gyr and 10 Gyr populations. The labelled regions are `MS' for the stellar main sequence;  `RBG / CHeB' for the red giant branch and core helium burning phase; `AGB' for asymptotic giant branch (including early AGB evolution and thermally pulsating, aka TP-AGB); and `Post-AGB' for the post asymptotic giant branch (including stellar remnants, e.g.\ the stars with $\text{logG} \gtrapprox 7$ are hot, but cooling, white dwarfs).}
\label{fig:comp_iso}
\end{center}
\end{figure*}

\progeny{} initially comes with three sets of modern isochrones: the MIST isochrones produced with the MESA software \citep{2016ApJS..222....8D, 2011ApJS..192....3P}; the PARSEC isochrones that combine Padova and Trieste standard tracks with COLIBRI evolved tracks \citep{2012MNRAS.427..127B, 2017ApJ...835...77M}; and BaSTI \citep{2018ApJ...856..125H}. Table \ref{tab:iso} presents the main differences in terms of coverage and sampling between the isochrones included. We note that the youngest stellar population available in any current \progeny{} isochrone is 100,000 yr (MIST). This is appropriate for a large range of galaxy evolution use-cases, but might not be appropriate for very extreme bursts where even younger stellar populations can dominate the spectrum.

MIST has the largest range of ages and metallicities, but PARSEC offers the densest sampling in metallicity space. All three are discussed briefly below, but for reasons we will outline the MIST isochrones are generally preferred for creating new SSPs. 

\begin{table*}
\begin{center}
\caption{Top level comparison between the three initial isochrones available with \progeny. N (Age) is the total number of unique ages in each isochrone. N (Z) is the total number of unique metallicities in each isochrone. N (All) is the total number of isochrone entries (where this number is a function of N (Age), N (Z) and the stellar mass sampling resolution.}
\begin{tabular}{|c|c|c|c|c|c|}
Library & log(Age / Myr) & N (Age) & log(Z/Z$_\odot$) & N (Z) & N (All) \\
\hline
MIST & 5 -- 10.3 &107 & -4 -- 0.5 & 15 & 1,494,453 \\
PARSEC & 6.6 -- 10.1 & 36 & -2 -- 0.5 & 27 & 340,909  \\
BaSTI & 7.3 -- 10.3 & 30 & -3.2 -- 0.45 & 21 & 1,299,900
\end{tabular}
\end{center}
\label{tab:iso}
\end{table*}%

\subsubsection{MIST}

The MIST isochrones were obtained from the main web hosting server\footnote{\url{waps.cfa.harvard.edu/MIST/model_grids.html}}. We use the pre-packaged isochrone with v / vcrit = 0.4, and downloaded the basic version of the catalogue (25 columns). The MIST version number  was 1.2, and the MESA revision number was 7503.

The main advantage of the MIST isochrones compared to older varieties is that they cover all phases of stellar evolution including remnant stars, which can be important for older stellar populations since late phase evolution and remnant stars are what provide the UV upturn component of galaxy SEDs. Missing these entirely (as many older isochrones do) can cause fitting issues for older stellar populations in highly quenched systems, and previous efforts to generate SSPs had to develop their own bespoke methods to create realistic upturns \citep{2000ApJ...541..126M, 2003MNRAS.344.1000B}. Notably some SSPs do not produce an upturn for older stellar populations, e.g.\ \citet{2005MNRAS.362..799M} (as seen in Figure \ref{fig:comp_spec_Z0}, discussed in detail later in this work).

MIST comes pre-packaged with usefully high-resolution stellar evolution age cadences of 0.05 dex between $10^5$ -- $10^{10.3}$ years. This coupled with no worse than 0.5 dex metallicity gridding between $10^{-4}$ -- $10^{0.5}$ $\text{Z}_\odot$ makes MIST a very useful fiducial isochrone set for generating new \progeny{} SSPs.

\subsubsection{PARSEC}

PARSEC is a popular modern set of isochrones that really combine the efforts of two main groups: the Padova and Trieste isochrones that cover the majority of stellar evolution (e.g.\ the main sequence of stellar evolution) and later phases via the COLIBRI tracks \citep[see][for details]{2012MNRAS.427..127B, 2017ApJ...835...77M}. The latter covers the major asymptotic giant branch (AGB) phases that are extremely important for properly explaining the luminous output of infrared light in moderately old (few Gyr) old stars. Our compilation of the PARSEC isochrones was generated interactively via the main web server \footnote{\url{stev.oapd.inaf.it/cmd}}, and combines the v1.2S version of the Padova tracks with the S37 + S35 + PR16 combination of CALIBRI tracks (see the website for details on what these ingredients mean for the final isochrones). These isochrones cover ages in 0.1 dex bins between $10^{6.6}$ -- $10^{10.1}$ years, and metallicity in 0.1 dex bins between $10^{-2.2}$ -- $10^{0.5}$ $\text{Z}_\odot$. This means they generally cover a more limited range of ages and metallicity than MIST, and with coarser binning in age but finer in metallicity. In general for SED modelling, SSPs with finer and broader age sampling is advantageous.

The main limitation with PARSEC, and why they are not the recommended default set of isochrones, is that they lack proper remnant stellar evolution tracking. This is noted on the website, with a suggestion that this will be added longer term. As and when remnant stars are added to PARSEC, we will updated the standard isochrones available in \progeny. We note that it is possible for users to adapt the provided isochrones and bolt on their own prescriptions for remnant star evolution (e.g.\ the Geneva isochrones, or the appropriate part of MIST). Counter to other SPLs, \progeny{} avoids bespoke treatments for such evolutionary gaps. This means it relies more heavily on the evolutionary coverage of input isochrones, and lacks the explicit addition of features like planetary nebulae contributions that do exist in other libraries \citep{2003MNRAS.344.1000B}.

\subsubsection{BaSTI}

The Bag of Stellar Tracks and Isochrones project (BaSTI) is an alternative isochrone library that monitors the evolutionary properties of stars up until their main sequence turn off (approximately). We downloaded pre-computed isochrones from the main INAF database\footnote{\url{basti-iac.oa-abruzzo.inaf.it/isocs.html}}. The various files were converted into the appropriate format for \progeny, with a particular note that the surface gravity properties of the stars had to be derived from the Stefan-Boltzmann Law (which means they are assumed to be non-rotating). These isochrones cover ages in 0.1 dex bins between $10^{7.3}$ -- $10^{10.3}$ years, and metallicity in $\sim0.2$ dex bins between $10^{-3.2}$ -- $10^{0.45}$ $\text{Z}_\odot$. This means they generally cover a more limited range of ages and metallicity than MIST and PARSEC, and with coarser binning in age but finer in metallicity. The younger age limit is particular high (at roughly 20 Myr), making BaSTI a poor choice when computing properties of extreme star bursts (where sub 10 Myr sampling is critical).

The main limitation with BaSTI, and why they are not the recommended default set of isochrones, is that they almost entirely lack remnant stellar evolution tracking. In this respect they are more extreme than PARSEC since they cover a very short period of stellar evolution mass loss in general, and often none at all. This makes them both limited in terms of the spectral properties of remnant stars, but also lacking in data regarding the mass locked up in stellar remnants. Given the coverage and limitations of the main isochrones included with \progeny, we generally recommend users to prefer MIST, followed by PARSEC and then BaSTI (if selecting only one). But we suggest any thorough investigation considers the impact of changing isochrones also.

\subsubsection{Comparison}

It is instructive to investigate some of the major differences between the MIST, PARSEC, and BaSTI isochrones included in the first released version of \progeny. Figure \ref{fig:comp_iso} is a high-level comparison of the three families of isochrones at 100 Myr, 1 Gyr and 10 Gyr (the origin of the region polygons is made clear in Appendix \ref{sec:iso_regions} Figure \ref{fig:iso_regions}). It is immediately apparent that the PARSEC and BaSTI isochrones are missing post-AGB stellar evolution including remnants. The main sequence region (labelled `MS') is very similar between the three isochrone families, but the treatment of the RGB and TP-AGB region (bottom-left) does vary quite significantly, with PARSEC capturing the pulsations is relatively more detail. The more detailed tracking of the TP-AGB with PARSEC is not surprising, since it utilises the dedicated COLIBRI libraries for detailed modelling in this regime.

\begin{figure}
\includegraphics[width=\columnwidth]{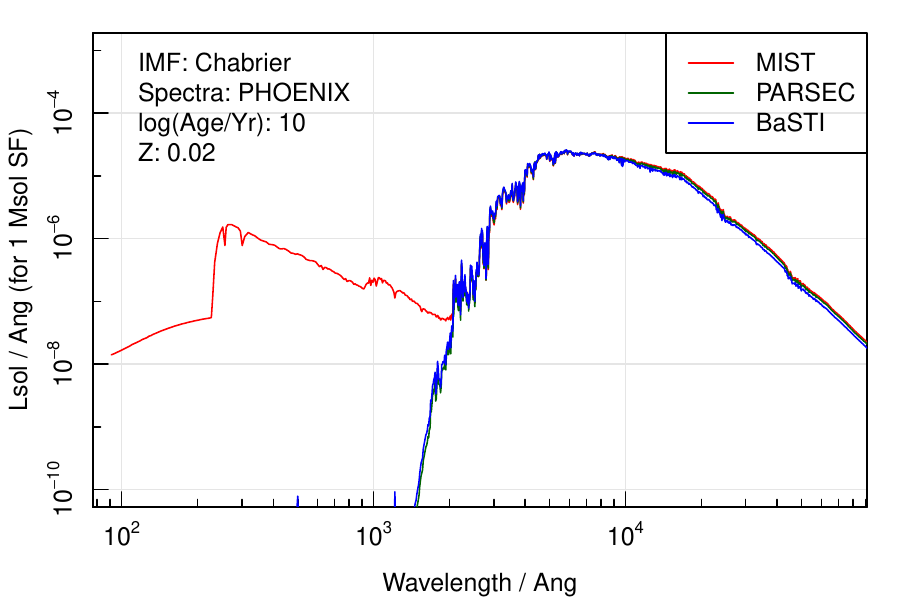}
\caption{Comparison of 10 Gyr ultraviolet upturn predictions for MIST, PARSEC, and BaSTI isochrones. The stellar spectra and interpolation used are otherwise identical, so the discrepancy in the ultraviolet is purely due to the inclusion (MIST) or lack of inclusion (PARSEC / BaSTI) of hot remnant stars. Note that the PARSEC and BaSTI spectra lie almost on top of each other in this plot.}
\label{fig:comp_UV}
\end{figure}

The impact of variation in the TP-AGB region seems to be very modest in terms of the eventual impact on the final SSPs, but the lack of remnant stars in PARSEC and BaSTI has a very notable impact on the ultraviolet emission of older (few Gyr) stellar populations. If we apply our default interpolation of stellar spectra onto the isochrones we can compare the spectral flux prediction for a target Chabrier IMF. The result of this is shown in Figure \ref{fig:comp_UV}, where there is a vastly different ultraviolet prediction due to the inclusion (MIST) or lack of inclusion (PARSEC, BaSTI) of hot remnant stars.

As discussed later in Section \ref{sec:litcomp}, there is still a huge deal of uncertainty in the details of the ultraviolet upturn for old stellar populations and exactly what processes produce it \citep[e.g.\ helium burning from Wolf-Rayet stars, post-AGB stars, horizontal branch stars, planetary nebulae, white dwarf remnants, and binary interactions; see][]{2016MNRAS.461..766L}. There is also a scarcity of empirical data informing our knowledge of ultraviolet stellar spectra. The main conclusion to make here is that merely the inclusion or not of hot remnant stars (e.g.\ white dwarfs) alone will have a dramatic effect in the ultraviolet, but less impact at all on the optical to infrared regimes (i.e.\ any data long-wards of 2000 \AA). Nebular emission can become significant in these wavelength regimes, but this will not be contributed to significantly by remnant stellar populations.

\subsection{Stellar Photosphere Spectra}
\label{sec:FL_atmos}

Equally as important as the isochrones, stellar spectra are the means by which we can convert our predictions of stellar evolution models to spectra that we can then use to model observed data, either directly as spectra, or passed through filters to create broadband photometry. Unlike isochrones, which are necessarily theoretical in nature (though informed by observations), stellar spectra can in principle be empirically derived using observations of many stars covering a large range of intrinsic properties, or generated theoretically. The main properties we usually wish to grid over are (in rough order of significance on the resulting spectrum): temperature; logarithmic surface gravity (referred to as `logG' in the \progeny{} package and this paper for compactness in Tables and Figures, where surface gravity has CGS units ($\text{cm}/\text{s}^2$) wherever mentioned); and metallicity (Z, always specified relative to solar in the \progeny package). More exotically, extensions are sometimes made to cover variations in stellar rotation and/or $\alpha$/Fe enhancement (or other abundance measures).

In the perfect library of stellar spectra, there would exist a predicted or measured spectrum for every star for every combination of properties predicted by our stellar evolution isochrones, but this is unfeasible in practice --- we would need millions of such spectra. The pragmatic solution is to map out a dense grid of these properties, with higher density where stars are most commonly found on the main sequence (since that will ultimately produce more accurate composite spectra). Rarer regions of temperate, surface gravity and metallicity can be covered more coarsely without biasing any resulting spectrum too much. Once we have established this dense grid, we can use the interpolation scheme (discussed above) to heuristically produce an approximate spectrum for any target combination of properties when moving along a given isochrone.

Stellar population libraries exist that choose to place more emphasis on either empirical or theoretical / synthetic stellar spectra, but the main tradeoffs are accuracy (assuming we can estimate stellar properties, observations should always be closer to `reality') and coverage. In the case of coverage, observed databases of stellar spectra tend to be much smaller (usually numbering hundreds), and only map out a limited range of parameter values. This makes interpolating or extrapolating these libraries onto target parameters quite complex, with spline interpolation and/or kriging schemes being necessary to achieve reasonable performance. In practice nearly all observationally focussed SPLs need to supplement the empirical stellar spectra with extreme/rare theoretical templates (e.g.\ for hot stars). The counter is also true, since some observed stellar spectra (e.g.\ TP\_AGB stars) cannot be accurately reproduced theoretically.

\begin{table*}
\begin{center}
\caption{Parameter ranges of the various stellar spectrum libraries initially includes with \progeny. N is the total number of unique spectra.}
\begin{tabular}{|c|c|c|c|c|c|c|c|}
Library & Type & Teff (K) & logG & logZ & $\lambda$ (\AA) & N & Reference \\
\hline
C3K (Conroy) & Base & 2,000 -- 50,000 & -1 -- 5.5 & -2.1 -- 0.5 & 100 -- $10^8$ & 8,602 & \citet{2018ApJ...854..139C} \\
PHOENIX (Husser)& Base & 2,300 -- 12,000 & 0 -- 6 & -4 -- 1 & 500 -- 55,000 & 7,559 & \citet{2013AA...553A...6H}  \\
\hline
PHOENIX (Allard)& Extend & 2,000 -- 70,000 & 0 -- 5.5 & -4 -- 0.5 & 12.6 -- 776,247 & 12,045 & \citet{2012RSPTA.370.2765A} \\
\hline
MILES (Vazdekis)& Alternative & 2,000 -- 50,000 & -1 -- 5.5 & -1.4 -- 0.2 & 3,525 -- 7,500 & 3,915 & \citet{2010MNRAS.404.1639V} \\
BaSeL (WLBC) & Alternative & 2,000 -- 50,000 & -1 -- 5.5 & -2 -- 0.5 & 91 -- $1.6 \times 10^6$ & 4,649 & \citet{2002AA...381..524W} \\
ATLAS9 (Castelli)& Alternative & 2,000 -- 50,000 & -1 -- 5.5 & -1 -- 0.5 & 91 -- $1.6 \times 10^6$ & 2,735 & \citet{2003IAUS..210P.A20C} \\
\hline
AGB (Lacon) & AGB & 2,000 -- 4,000 & NA & NA & 5,100 -- 24,900& 14 & \citet{2002AA...393..167L} \\
\hline
TMAP (Werner) & Hot & 20,000 -- 190,000 & 5 -- 9 & NA & 3 -- 400,000 & 124 & \citet{2003ASPC..288...31W} \\
\end{tabular}
\end{center}
\label{tab:atmos}
\end{table*}%

Below we discuss the main decisions made regarding the fiducial stellar spectra used within \progeny{} (see Table \ref{tab:atmos}). It is always possible for users to supplement these with their own preferences and use the \progeny{} software to generate new SSPs, but when we discuss the base \progeny{} SSPs they have been constructed with a mixture of the below libraries. All stellar spectra provided as part of the fiducial \progeny{} library encode the wavelength in angstroms and the luminosity in  $\delta \lambda$ units such that the integral over all wavelengths of $f(\lambda).\delta \lambda = 1$. This means all spectra can be scaled directly by the predicted stellar luminosity in an isochrone to achieve the correct output spectrum. Any additional stellar spectrum libraries added to \progeny{} by users should obey the same requirements. The spectra are also re-binned to match the exact wavelength sampling of CB19, since via inspection this appeared to offer a good compromise of moderately high resolution (especially around important features), coverage and data volume. We note that when loading stellar spectra it is possible to target any desired wavelength grid, with flux conserving spectral resampling handled within \progeny (written using efficient \Cpp code).

\subsubsection{Base and Extended Stellar Templates}

Within \progeny{} we include a dense grid of `base' templates that cover the main sequence of most stars (see Table \ref{tab:atmos}). These were noted to not extend high enough in temperature so smoothly connect with the `hot' stellar spectra used (discussed below). To improve coverage \progeny{} has the capability to supplement any `base' templates with an extension from a different library (`extend'). When the SSP is being generated the `base' templates are prioritised when matching parameters, with a user-controllable parameter specifying how much closer matched in temperature, surface gravity and metallicity space the `extend' stellar spectrum has to be before it is used instead.

\begin{figure*}
\includegraphics[width=8cm]{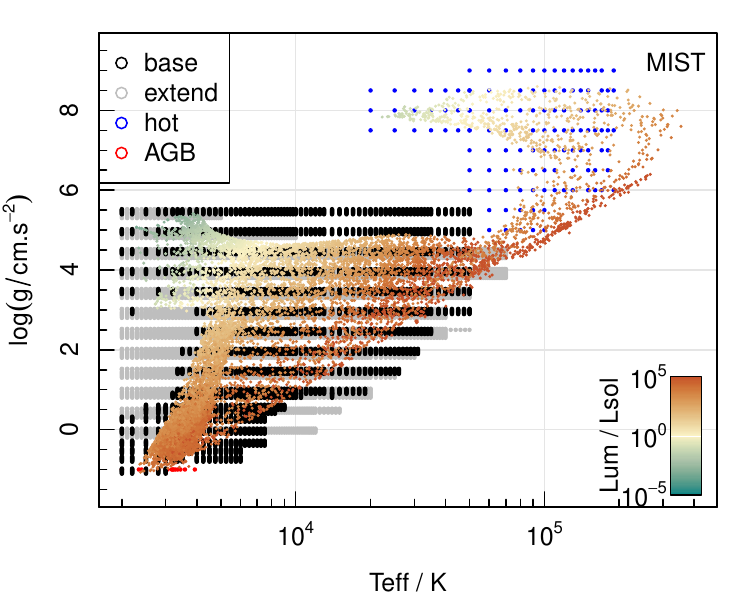}
\includegraphics[width=8cm]{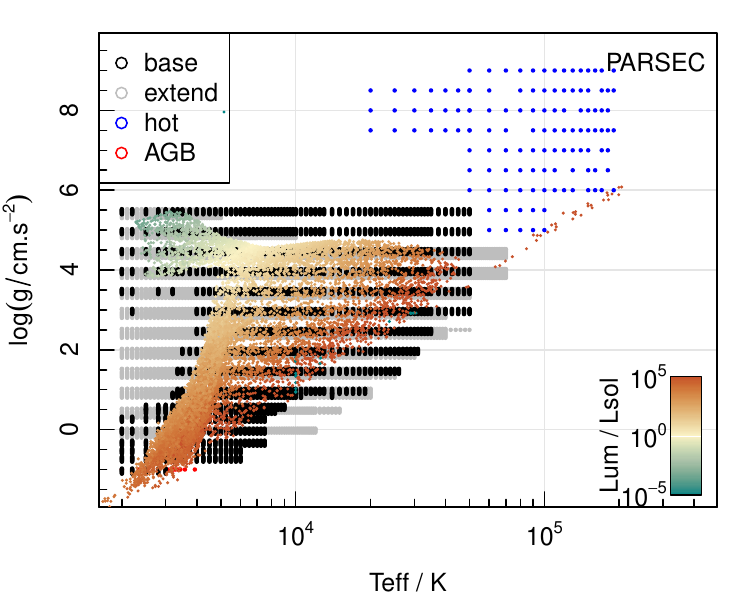}
\includegraphics[width=8cm]{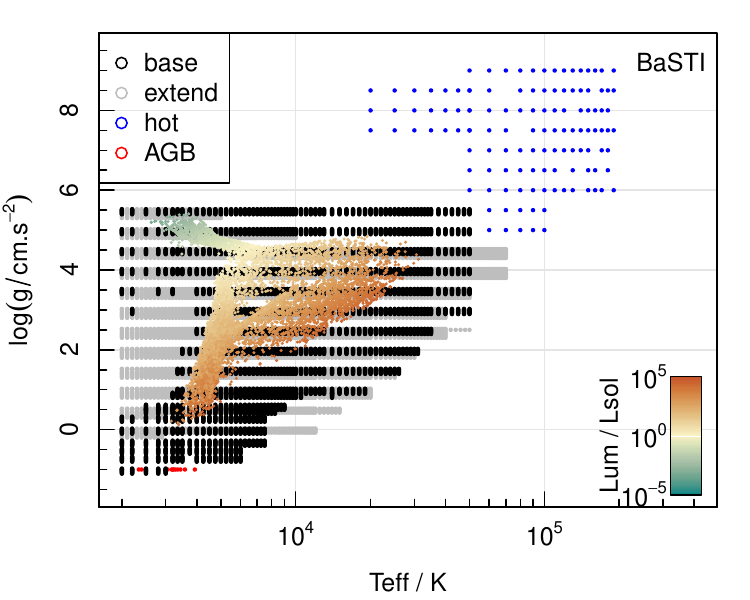}
\caption{Coverage of the fiducial stellar spectra used in \progeny{} with random samples from the MIST / PARSEC and BaSTI isochrones overlaid. The density of isochrone points is broadly indicative of which stellar spectra are the most important, and where most star light is coming from over the lifetime of stellar evolution. The small vertical dithering for `base' and `extend' templates are to reflect the presence of metallicity (Z) gridding (this is not available for `hot' or `AGB' stellar spectra).}
\label{fig:atmos_cover}
\end{figure*}

The result is that the `base' and `extend' templates generally cover the vast majority of the lifetime of stars for any typical IMF regardless of the isochrones used. This is clear when considering Figure \ref{fig:atmos_cover} where we display all stellar spectrum library grid points.

For good all-round coverage in the dimensions of temperature, surface gravity, and metallicity we use the C3K atmospheric spectra included with Flexible Stellar Population Synthesis as our fiducial `base' spectra \citep[FSPS;][]{2018ApJ...854..139C} . This covers temperatures from 2,000 to 50,000 K; surface gravity from -1 to 5.5; and metallicity from -2.1 to 0.5. Compared to alternatives, C3K has very broad coverage in temperature and surface gravity, but is a bit more limited in terms of the available metallicities. In general C3K offers good coverage for even our youngest OB stars, but we note that should isochrones be used in the future that have stellar evolution younger than 100,000 yr then additional spectral templates would be needed to cover more extreme massive/young OB star spectra.

The fiducial `extend' template spectra comes from the \citet{2012RSPTA.370.2765A} version of PHOENIX spectra\footnote{available at \url{archive.stsci.edu/hlsps/reference-atlases/cdbs/grid/phoenix/}}. For general details on the PHOENIX stellar spectrum generation software see \citet{1999ApJ...512..377H} and associated works. This library is often referred to as `BT-Settl', but we refer to it as `Allard' within \progeny{} and in this paper. The focus of this library was for very low mass stars (including brown dwarfs), but it covers more typical stellar populations also. Currently the Allard library covers 2,000 -- 70,000 K, which is reasonably hot (certainly hot enough for the youngest OB stars generated by any available \progeny{} isochrones), but means we do still need a dedicated `hot' stellar spectra library too to achieve the appropriate coverage of remnant stars (i.e.\ white dwarfs) when using the MIST isochrones. If using the MIST or PARSEC isochrones then the combination of `base' and `extend' libraries achieves good coverage of all but the AGB evolutionary phases. The Allard library covers surface gravity from 0 to 5.5; and metallicity from -4 to 0.5, giving it better metallicity coverage than C3K, but not extending as far into the low surface gravity AGB stellar spectra. Using C3K (as `base') and Allard (as `extend') gives excellent all-round parameter coverage for most phases of stellar evolution, but with additional treatment for AGB phases and ultra hot stars required (see below).

An alternative `base' template spectra comes from the \citet{2013AA...553A...6H} Gottingen version of the PHOENIX spectra\footnote{available at \url{ftp://phoenix.astro.physik.uni-goettingen.de/HiResFITS/}}. We used the log$_{10}$($\alpha$/Fe) = 0 (not enhanced) version of the library. The high resolution spectra (with better wavelength coverage than the mid resolution also available) was then re-binned to match the sampling resolution of other \progeny{} stellar spectra (which is the same as the CB19 high resolution library by default). Currently this library only extends to 12,000 K, where the website claims the plan is to eventually extend it to 25,000 K. Given the passage of time since the library's release (11 years) it appears this extension may never happen. This necessitates the use of a broader range `extend' library when using the Husser stellar spectra within \progeny, and is the reason we do not use the Husser library as our fiducial `base' template spectra.

One phase of stellar evolution that is not well captured with current default stellar templates in \progeny{} is Wolf-Rayet (WR) stars. This extreme phase of stellar evolution occurs between 1--10 Myr in the most massive (tens of solar masses) stars. Temperatures can reach millions of Kelvin, and in MIST the hottest WR stars evolve up to 320,000 K. With the default \progeny{} stellar spectral templates these hottest WR stars get represented with a blackbody spectrum, which is certainly not fully accurate. For better representing WR evolution additional spectral templates can be used \citep[e.g.\ CMFGEN, PoWR;][respectively]{1998ApJ...496..407H, 2015A&A...579A..75T}. Keen users can add their own spectral templates quite easily, but longterm we will aim to add appropriate WR templates.

\begin{figure}
\includegraphics[width=\columnwidth]{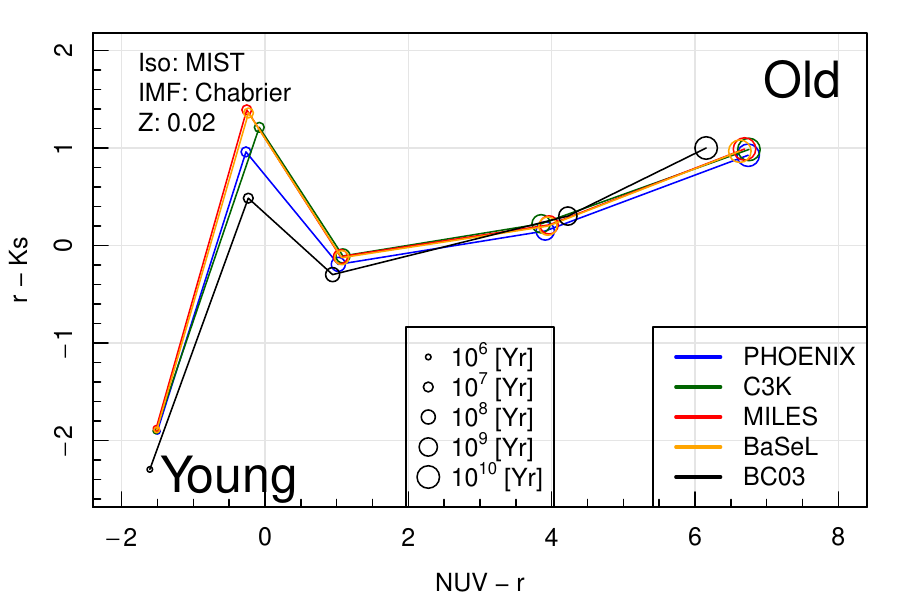}
\caption{Colour-colour comparison of different stellar spectra as a function of age for \Zsol{} using MIST isochrones. In all cases the `extend' library used is Allard PHOENIX (i.e.\ the blue PHOENIX label corresponds to the `base' Husser PHOENIX variant). Other versions of this Figure (for different metallicities and isochrones) can be viewed in Appendix \ref{sec:colour_extra}.}
\label{fig:atmos_colour}
\end{figure}

Figure \ref{fig:atmos_colour} presents the colour -- colour (NUV - r versus r - Ks) scatter between different combinations of stellar spectra of table \ref{tab:atmos} when using the same isochrone (MIST), IMF (Chabrier) and metallicity (solar). The variations peak for r - Ks colour at 10 Myr ($\sim0.5$ mag scatter). Beyond 100 Myr the colour scatter is $\sim0.1$ mag. Since only very young stellar populations have notable colour scatter due to the choice of stellar spectrum, we can generally assume the stellar mass is only marginally affected by this decision. Even the star formation rate (the integral of stellar mass formed up to 100 Myr in \prospect) will see little impact since it is the NUV - r axis (with very little scatter) the provides most of the constraining information when fitting a galaxy SED. We note that BC03 does show a bluer NUV - r colour for the oldest (10 Gyr) populations. This is reflective of it possessing a stronger UV-upturn for old stars, likely due it incorporating UV bright planetary nebulae spectra \citep[this is noted as a private communication in][]{2016MNRAS.463.3409V}.

\subsubsection{TP-AGB and AGB Stellar Templates}

A key extension to the `base' and the `extend' stellar spectra is the complex and highly variable asymptotic giant branch (AGB) phases of stellar evolution. These are both highly complex to model theoretically, and are generally not well understood in terms of the stellar spectra produced. In temporal order these phases cover the pre-AGB, early-AGB, AGB (including TP-AGB) and post-AGB periods of stellar evolution. The reason for the modelling difficulty is due to the rapid timescales of variability, and the complex interplay between the different layers of the stars (which both influences the behaviour of evolutionary tracks and the predicted spectra of the stellar spectra). A number of works have focussed on the complexity of this phase of stellar evolution, and there is a general consensus that for certain periods of stellar evolution (a few Gyr in age) the contribution can be highly significant due to the sheer size and luminosity of such stars, even though the amount of stellar mass involved might be small \citep[see][and discussions therein]{2005MNRAS.362..799M, 2010ApJ...708...58C}.

As noted above, different isochrone models have quite different predictions for the AGB phases (see Figure \ref{fig:comp_iso}). Common to other works, we choose an empirical library of AGB spectra since there is no theoretical template set that offers the AGB parameter coverage required for constructing an SSP. Specifically, we use the small template library released in \citet{2002AA...393..167L} as first used for SSP generation in \citet{2005MNRAS.362..799M} and later in \citet{2010ApJ...708...58C}. This contains time-averaged AGB spectra combined over a number of years of observations of the same AGB stars covering a reasonable range of AGB classes and temperatures (2,000 -- 4,000 K). This template set is generally considered to be a good compromise for AGB spectra since it captures the temporally extreme pulsations that in practice vary even more rapidly than stellar evolution isochrones can describe. The Lancon AGB spectra was extended beyond the original 5,100 -- 24,900 \AA{} range to cover the total range needed for \progeny{} using a blackbody extension (at the appropriate temperature) at either end.

In terms of parameter coverage, the AGB stellar spectra have the sparsest and least regular coverage, with only 14 templates covering the approximate temperature range 2,000 -- 4,000 K. They also do not have a specified (i.e.\ known) surface gravity, so for the purposes of isochrone parameter interpolation they are placed at a fiducial value of -1 (as can be seen in Figure \ref{fig:atmos_cover}). The consequence of this is mitigated by the `base' and `extend' stellar spectra extending down to a surface gravity of 0, which covers the majority of AGB evolutionary phases in practice. It is only the most extreme TP-AGB pulsations that require the use of the empirical Lancon libraries provided here.

\subsubsection{Hot Stars and White Dwarfs}

The final stellar spectrum variant that must be included for suitable isochrone coverage is `hot' stars including stellar remnants (i.e.\ white dwarfs and neutron stars, although the latter are too dim to be relevant in a SSP). This includes most stars hotter than 50,000 K and ones with a surface gravity 6 $\text{cm}/\text{s}^2$ or greater, i.e.\ extremely compact and hot objects. It is notable from \ref{fig:atmos_cover} that only the MIST isochrones have significant coverage inside this regime, i.e.\ the PARSEC and BaSTI isochrones do not use or require `hot' stellar spectra. The hot remnant stellar spectra are important for producing at least some of the UV upturn for old stellar populations with \progeny{} (see Figure \ref{fig:comp_UV}), but it should be noted that there is still a great deal of uncertainty about what contributes to UV upturn features in elliptical galaxies. For instance, it is not well understood how dominant exposed core helium burning could be for producing UV photons, and the thorough modelling of binary evolution certainly has a significant effect \citep{2018MNRAS.479...75S}.

The fiducial `hot' template spectra comes from \citet{2003ASPC..288...31W} (and earlier works). This library is generated by a one-dimensional line-blanketed non-Local Thermodynamic Equilibrium (i.e.\ non-LTE) code, and is considered appropriate for use with white dwarf spectra in particular. It is a combination of the TMAP2 and TMAP3 Werner data available online at the Spanish Virtual Observatory\footnote{\url{svo2.cab.inta-csic.es/theory/newov2/index.php}}. The library used reaches up to 190,000 K, and covers the range of temperature and surface gravity for the majority of hot stars and white dwarf remnants observed on the MIST isochrones (temperature 20,000 -- 190,000 K and logarithmic surface gravity 5 -- 9. For this reason it works reasonably well for all hot phases (including helium burning) and remnants (including white dwarfs). The very coolest white dwarfs are not quite covered (below 20,000 K), but these barely exist within the current age of the Universe, and are such low luminosity that they contribute very little flux to any reasonable SSP.

We note that the PHOENIX Allard `extend' template spectra are often used for representing reasonably hot (sub 70,000 K) stellar photospheres. Beyond 4,000 K the PHOENIX code used has some limited non-LTE approximations to better represent `hot' spectra. In detail it is not identical to full non-LTE spectra, where we compared matching (in temperate and surface gravity) PHOENIX Allard and PoWR OB stars \citep[as described in][]{2019A&A...621A..85H}. Between 1,000 -- 10,000 \AA (where the differences are most significant) fluxes are within a few percent on average, with the biggest deviations for certain spectral features being tens of per cent. In practice these spectra are mostly used for very young (sub 10 Myr) stellar populations. For most integrated SED fitting work the impact of using PoWR OB stars for hotter spectra (rather than Allard) will be negligible, but for extreme bursts there could be biases in derived properties at the tens of per cent level. For the rest of this work (and the associated ProGeny II companion paper) we use the Allard template spectra without the use of an additional OB template library. In the future development of \progeny{} we will investigate adding dedicated OB stellar spectra as a further option. But as with Wolf-Rayet stars, keen users can add extra OB star spectral templates as desired when running \progeny{} locally.

\subsection{Initial Mass Function}
\label{sec:FL_IMF}

\begin{figure}
\includegraphics[width=\columnwidth]{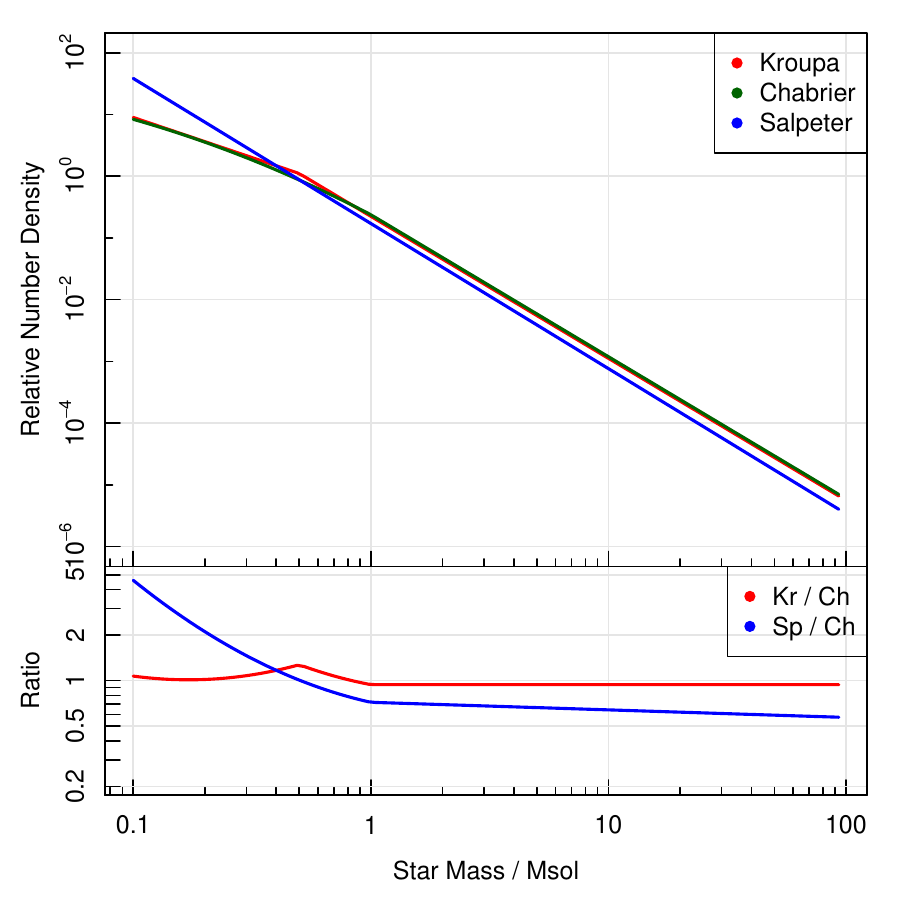}
\caption{Comparison of the three different classic IMFs available as standard within \progeny. In all cases 1 solar mass of stars have been formed by the IMF.  The bottom panel shows the ration of number counts for each IMF relative to the Chabrier IMF. The Chabrier and Kroupa IMFs are in general very similar, as can be seen in the bottom panel showing the number counts ratio of Kroupa and Salpeter relative to Chabrier (red and blue lines respectively). Ch = Chabrier, Kr = Kroupa, Sp = Salpeter.}
\label{fig:IMF_comp_base}
\end{figure}

The last of the critical components for any stellar population library is the initial mass function. Briefly stated, it is the expected distribution of stellar masses in an instantaneous burst of star formation. Over the years it has evolved from a simple power law approximation \citep[e.g.\ the Salpeter IMF][]{1955ApJ...121..161S} to more complex forms involving breaks and/or bends \citep[e.g.\ the Kroupa and Chabrier IMFs:][]{2001MNRAS.322..231K, 2003PASP..115..763C}. Figure \ref{fig:IMF_comp_base} compares these three classic IMFs, where in modern times the Kroupa and Chabrier are the most commonly used. As standard, more stars of lower mass are always formed in a particular burst of star formation, but the precise ratio of low mass to high mass stars finds certain parameterisations labelled as top-heavy (more mass in more massive stars) or bottom-heavy (more mass in less massive stars), usually in reference to a classic fiducial IMF such as the Salpeter.

The IMF might also vary depending on the intensity of local star formation and/or the gas metallicity \citep[see][]{2012MNRAS.422.2246M, 2015ApJ...806L..31M, 2019MNRAS.482..118G}. In these variants, the general consensus is that IMFs might be more top-heavy when star formation is extremely intense (bursty) and/or the gas metallicity is low. An ongoing matter of research and speculation is how universal or not the IMF might be, in particular whether it varies predictably based on local star formation properties and/or cosmic time. Recent work has suggested that it might be appropriate to parameterise the IMF differently when fitting extremely bursty and/or high redshift stellar populations \citep{2013MNRAS.433..170H, 2014ApJ...796...75C, 2016MNRAS.462.3854L}, or even for different galaxy components \citep{2013MNRAS.428.3183D, 2021MNRAS.505..415L}. \progeny{} supports any reasonably specified IMF as a functional input when creating SSPs, but the fiducial build of the SSPs come with pre-made Chabrier, Kroupa, and Salpeter IMFs, and these are the most readily comparable to other SPLs that usually include one of both of these IMFs.

\begin{figure}
\includegraphics[width=\columnwidth]{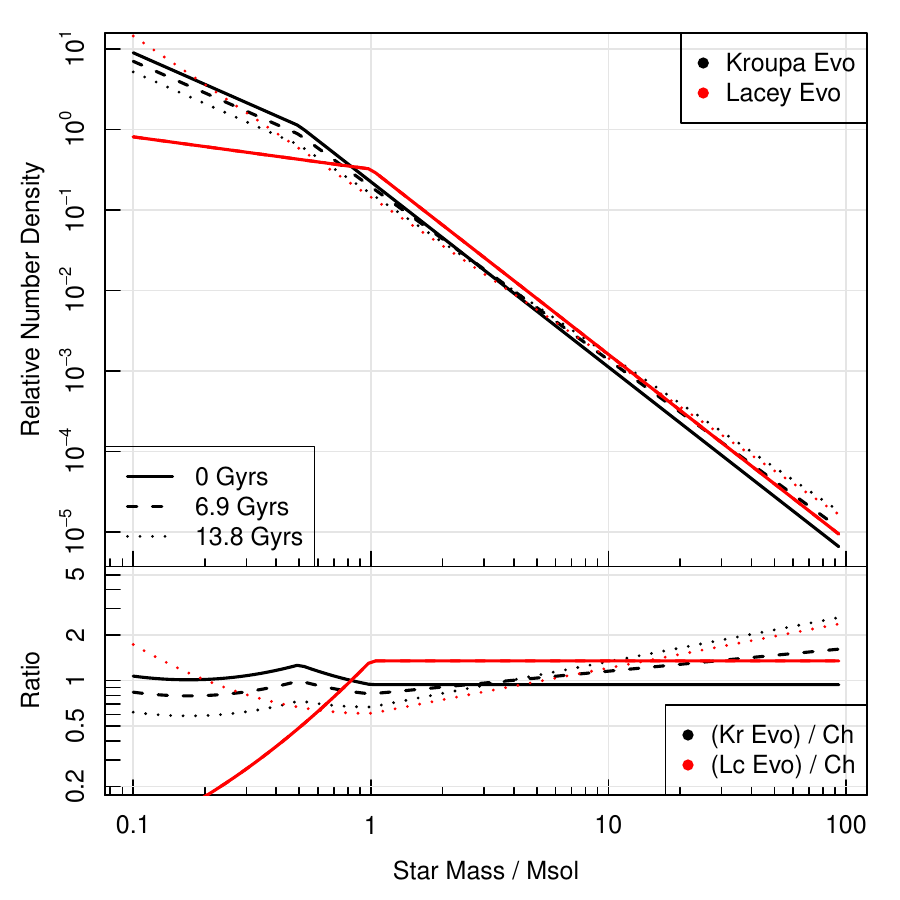}
\caption{Comparison of the different age-evolving IMFs available as standard within \progeny. In all cases 1 solar mass of stars have been formed by the IMF. The bottom panel shows the ration of number counts for each IMF relative to the Chabrier IMF. Ch = Chabrier, Kr = Kroupa, Lc = Lacey.}
\label{fig:IMF_comp_evo}
\end{figure}

For added flexibility it is also possible to encode an evolving IMF with \progeny, either with parameters that vary smoothly over time / metallicity or with sharp transitions between IMF properties. Four reasonable variations of an evolving IMF are included with \progeny: a smoothly evolving form of the Kroupa IMF that adopts classic local Universe values at low redshift (higher metallicity), and top-heavy forms in the early Universe (lower metallicity); and a version of the \citet{2016MNRAS.462.3854L} IMF that switches between a star burst dominated top-heavy IMF at high redshift, and the \citet{1983ApJ...272...54K} form at lower redshifts (the transition being at 10 Gyr lookback time, so $z \sim 1.9$, but note this value is somewhat arbitrary and just selected to be near cosmic noon). Figure \ref{fig:IMF_comp_evo} compares the two age-evolving forms, where we can see both the Lacey and Kroupa functions predict similar top-heavy IMFs for massive stars at the oldest look-back times (when the Universe was youngest), but quite different predictions for the low redshift Universe. This is especially true for the lowest mass stars, where the different IMFs have different cut offs (0.01 versus 0.1 solar masses when comparing Kroupa and Lacey).

It is worth noting that the effect of different IMFs can be highly non-trivial when comparing a broken power law with a pure power law. For one thing, it is possible for an IMF to be both bottom and top-heavy at the same time, i.e.\ when comparing the ancient look-back age IMF of the Lacey model (which is a pure power law) to the low redshift Kennicutt 1983 form we see we expect both more massive stars and more low mass stars. It is probably more accurate to describe the difference between the two in the low redshift regime (i.e.\ the Kennicutt 1983 form) as being relatively `middle-heavy', but such terminology is rarely used.

In total, the pre-built \progeny{} IMFs come in seven different flavours of IMF: Salpeter, Kroupa, Chabrier, age-evolving Kroupa, age-evolving Lacey, metallicity-evolving Kroupa, and metallically evolving Lacey. By following the included examples it is relatively easy for users to encode their own Universal IMFs, or even other evolving (either age or metallicity) variants. An obvious extension might be to parameterise the evolution via a property other than stellar population age. It often makes more sense to think of IMF evolution as being relative to cosmic age rather than age of a particular stellar population (which can be achieved by specifying the look-back time of observation in \progeny), or perhaps relative to the metallicity enrichment history of the star formation event in question. The basic age-evolving IMFs available with \progeny{} are shown in Figure \ref{fig:IMF_comp_evo}, but note the exact implementation presented here is somewhat arbitrary --- there is no well informed model for exactly how the IMF might evolve with age (or metallicity), so the existence of these functions is really to aid user experimentation (i.e.\ the default values should not be treated as especially well justified). Longer term work (beyond the scope of this paper), now that \progeny{} exists with this degree of flexibility, will be to better justify if (and how) these evolving IMF forms can be used for fitting SEDs with \prospect.

\section{Comparisons and Discussion}
\label{sec:litcomp}

Below we make some instructive comparisons of \progeny. First internally (where we vary the isochrones, stellar spectra and IMF), and secondly to well know literature SSPs (that are mixture of similar and dissimilar SPLs).

\subsection{Internal Low-Level Comparisons}

It is important to understand the role that the key components in \progeny{} have when generating spectra at specific ages and metallicities (the core low-level output of any SSP). The most significant variations in spectral output come from the choice of isochrone (MIST, PARSEC, or BaSTI in the case of \progeny), the stellar spectra used (be they empirical or theoretical) and the type of IMF (both fixed and evolving, in our case).

\subsubsection{Impact of Isochrones}

\begin{figure*}
\begin{center}
\includegraphics[width=8.5cm]{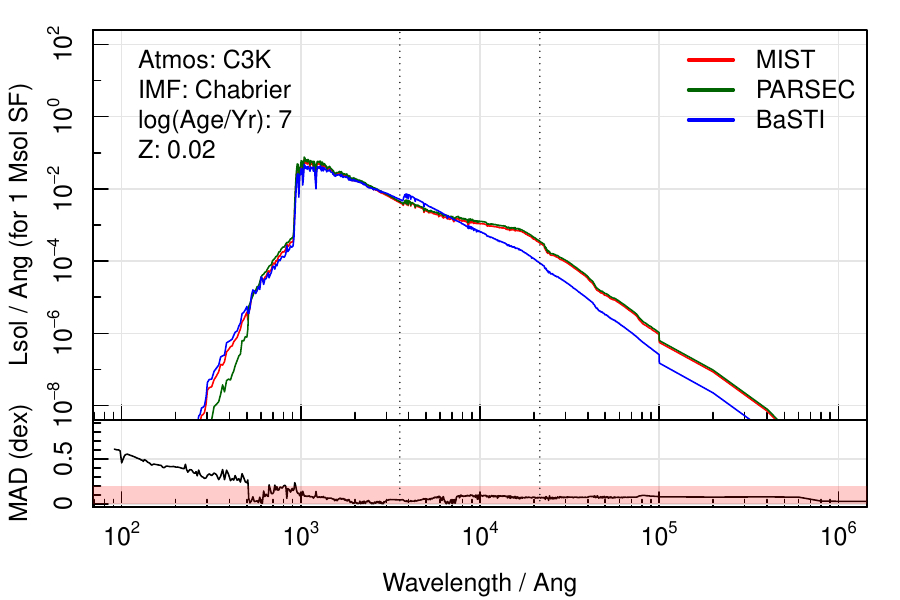}
\includegraphics[width=8.5cm]{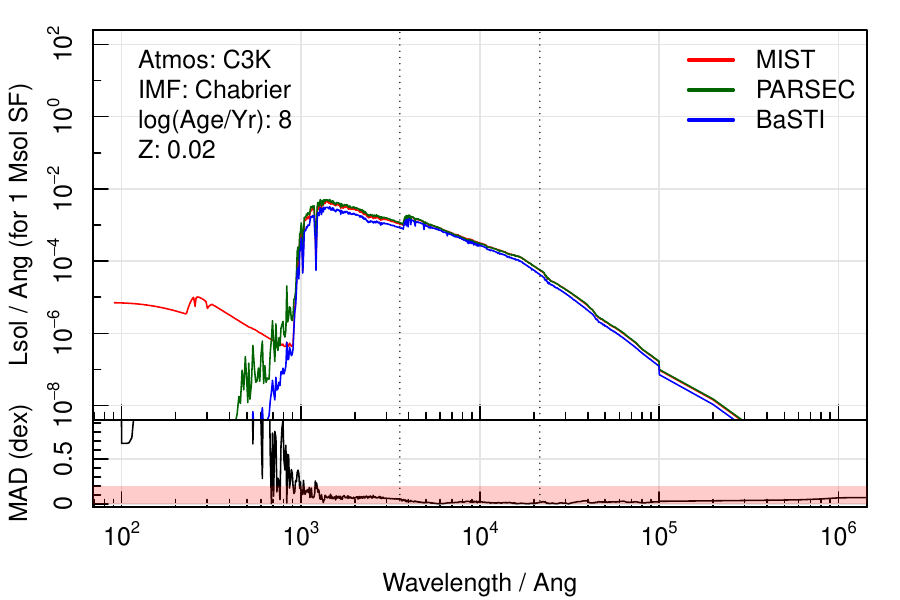}
\\
\includegraphics[width=8.5cm]{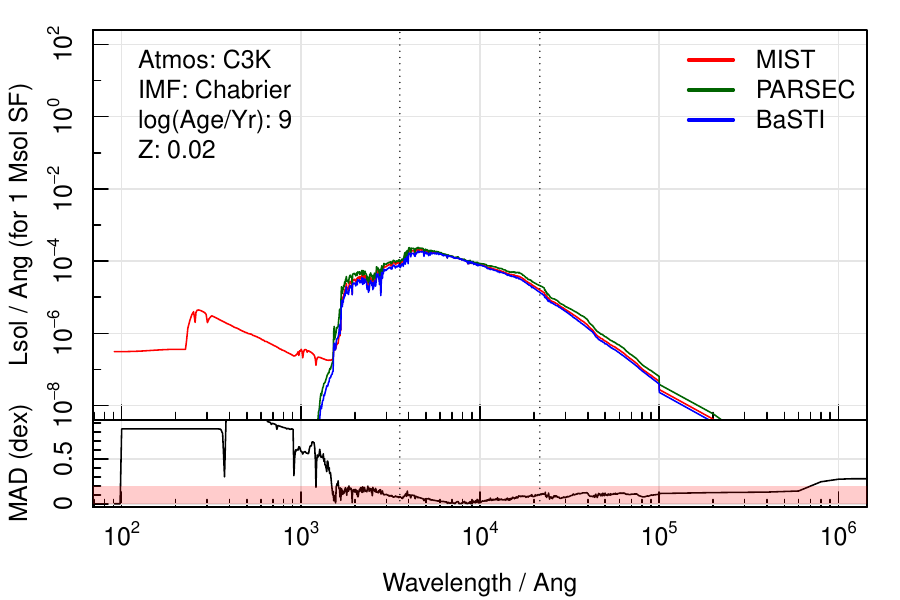}
\includegraphics[width=8.5cm]{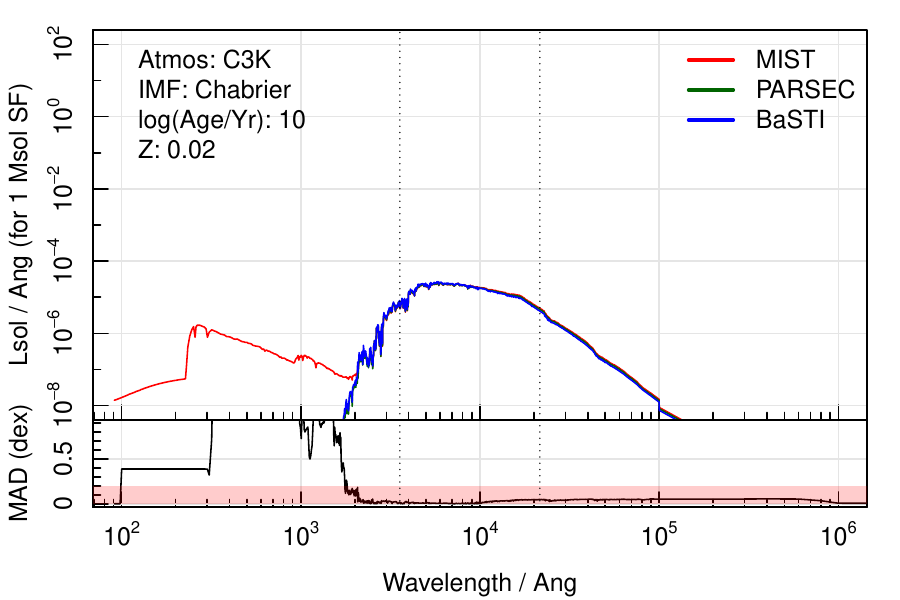}
\\
\caption{Comparison of \Zsol{} spectra at different stellar population ages for different isochrones. The ages presented are log spaced with 1 dex intervals, covering 10 Myr  through to 10 Gyr to allow comparison between all isochrones initially available with \progeny. The vertical dashed lines indicate the $u$ and Ks band central wavelengths, common limits for SED analysis work.}
\label{fig:comp_iso_Z0}
\end{center}
\end{figure*}

In Figure \ref{fig:comp_iso_Z0} we make a direct comparison of the three main isochrones initially available with \progeny{} (MIST, PARSEC and BaSTI), where the other key characteristics of the SSPs are kept fixed (as per the Figure legend). The bottom sub-panels show the median absolute deviation (MAD) of the spectra in dex, where the red-shaded region indicates where the MAD is less than 0.2 dex \citep[a classic threshold of SSP inferred stellar mass accuracy, see][]{2011MNRAS.418.1587T, 2020MNRAS.495..905R}. We note that here we are using the isochrones as provided, without additional attempt at remedy or treatment for `missing' physics.

In general, except for the very youngest stellar populations, the spectra agree much better than 0.2 dex through the optical and NIR regime. The biggest tension in the NIR is clearly for the youngest (10 Myr) sample, where MIST and PARSEC produces enhanced NIR flux compared to BaSTI. Investigating the isochrones, it is clear this is a feature of MIST/PARSEC producing significant pre main-sequence stellar light, which is dominated by cool collapsing photospheres that peak in the NIR. Beyond this feature the generated SSPs tend to be most consistent at $\sim$1 micron.

In the UV the agreement is much poorer for all ages, with very different implied spectra. As discussed earlier in Section \ref{sec:FL_iso}, this is entirely due to the physics incorporated within the isochrone regarding a mixture of remnant stars and horizontal branch stars. It is worth emphasising that the serious disagreement in predicted UV flux occurs short-wards of FUV filter (~1,500 \AA), and for typical SED analysis at low redshift this should not be a concern. When analysing higher redshift sources some care should be taken when including data in this rest-frame regime: it is likely including such observations to an SED fit does more harm than good given the lack of predictive agreement seen here. Note we are not even exploring the additional impact of binary evolution on the UV spectra, which can also have a dramatic impact \citep{2017PASA...34...58E}.

\subsubsection{Impact of Stellar Spectra}

\begin{figure*}
\begin{center}
\includegraphics[width=8.5cm]{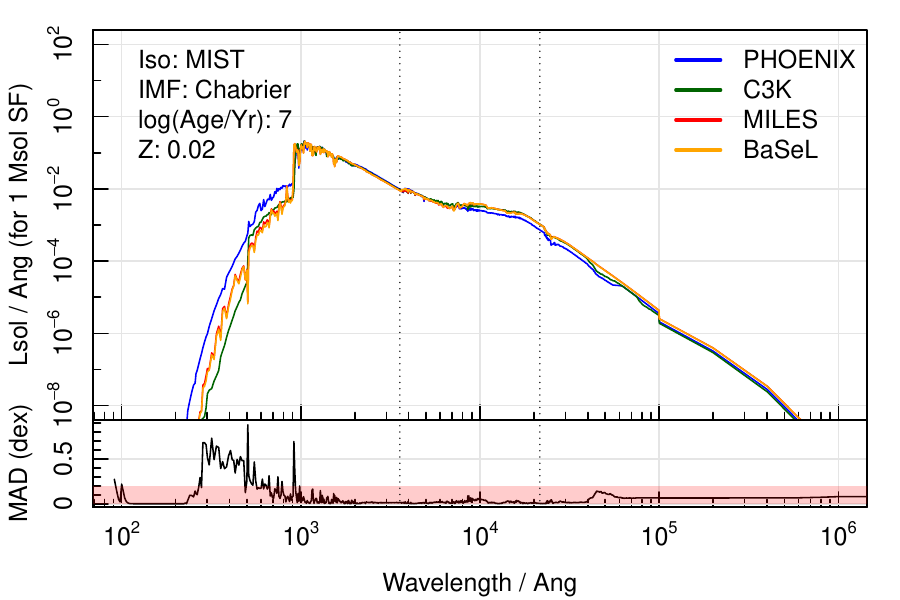}
\includegraphics[width=8.5cm]{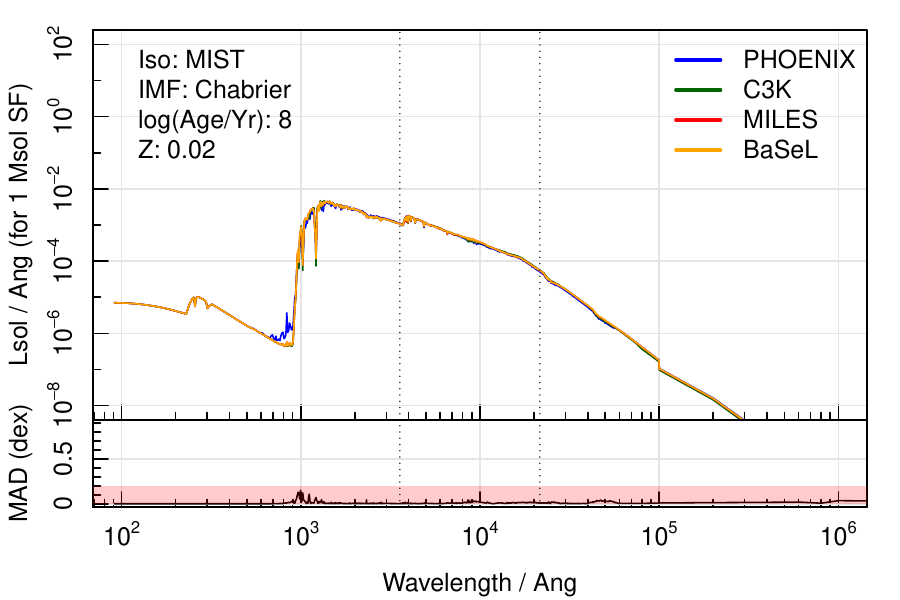}
\\
\includegraphics[width=8.5cm]{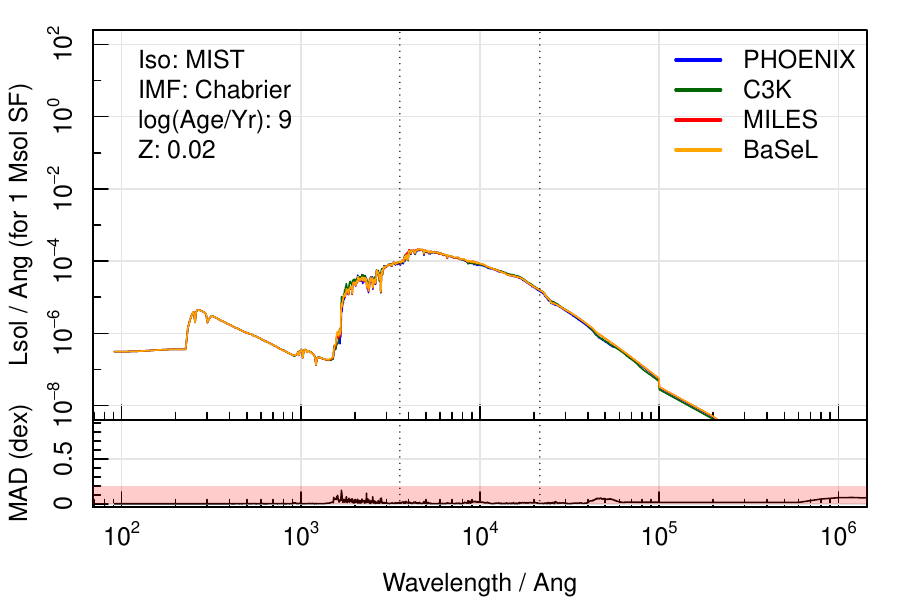}
\includegraphics[width=8.5cm]{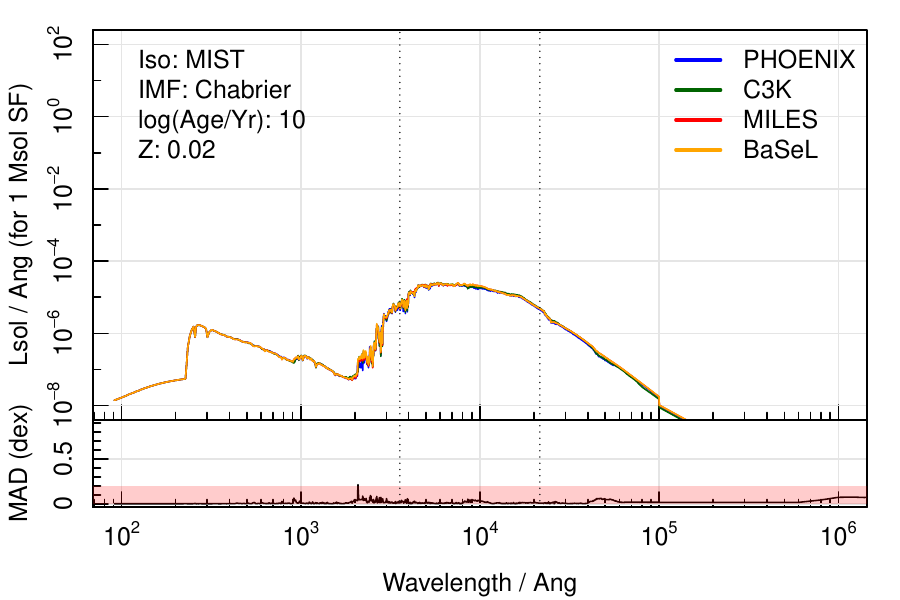}
\\
\caption{Comparison of \Zsol{} metallicity spectra at different stellar population ages for different stellar spectral libraries. The ages presented are log spaced with 1 dex intervals, covering 10 Myr  through to 10 Gyr to allow comparison between all stellar spectra initially available with \progeny. The vertical dashed lines indicate the $u$ and Ks band central wavelengths, common limits for SED analysis work.}
\label{fig:comp_atmos_Z0}
\end{center}
\end{figure*}

Similar to the above, in Figure \ref{fig:comp_atmos_Z0} we make a direct comparison of the four main `base' stellar spectra initially available with \progeny{} (PHOENIX (Husser), C3K, MILES and BaSeL), where the other key characteristics of the SSPs are kept fixed (as per the Figure legend).

Agreement here is much better than that noted between isochrones, with much less the 0.2 dex MAD between the spectral predictions. The only slight discrepancy of note is for youngest extreme UV population (below 1,000 \AA). The main advantages and disadvantages between the different options for stellar spectra are really native wavelength coverage, spectral resolution and atmospheric coverage (i.e.\ the details presented in Table \ref{tab:atmos}). For broadband SED analysis, especially in the optical -- NIR regime, the choice of stellar spectra has almost no tangible impact within \progeny. Because of its dense atmospheric parameter sampling and good temperature coverage, we suggest a weak preference for the C3K stellar spectral library for use within \progeny, but we note the PHOENIX (Husser) library has the potential for much high spectral resolution SSPs (0.01 \AA{} sampling versus 1 \AA{} in C3K).

\subsubsection{Impact of Initial Mass Function}

\begin{figure*}
\begin{center}
\includegraphics[width=8.5cm]{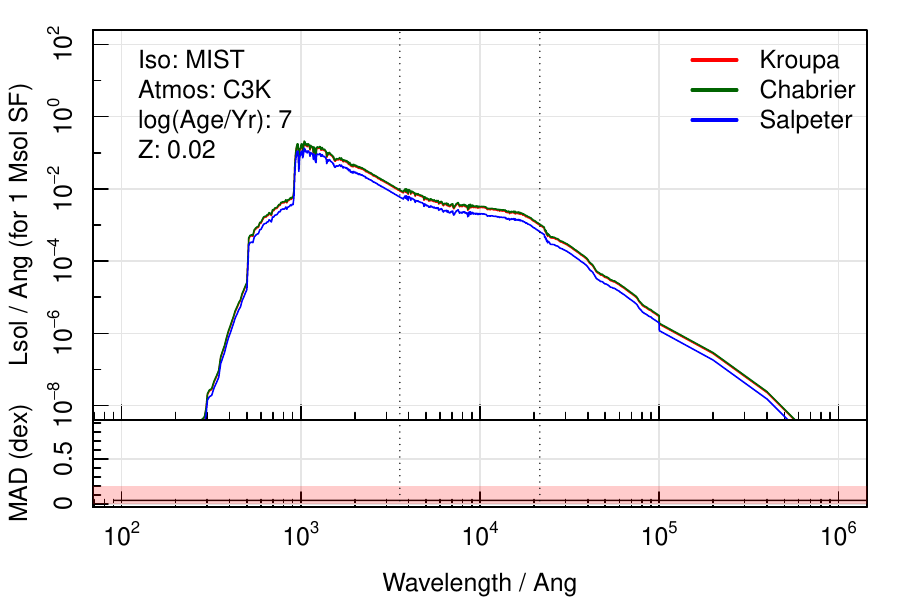}
\includegraphics[width=8.5cm]{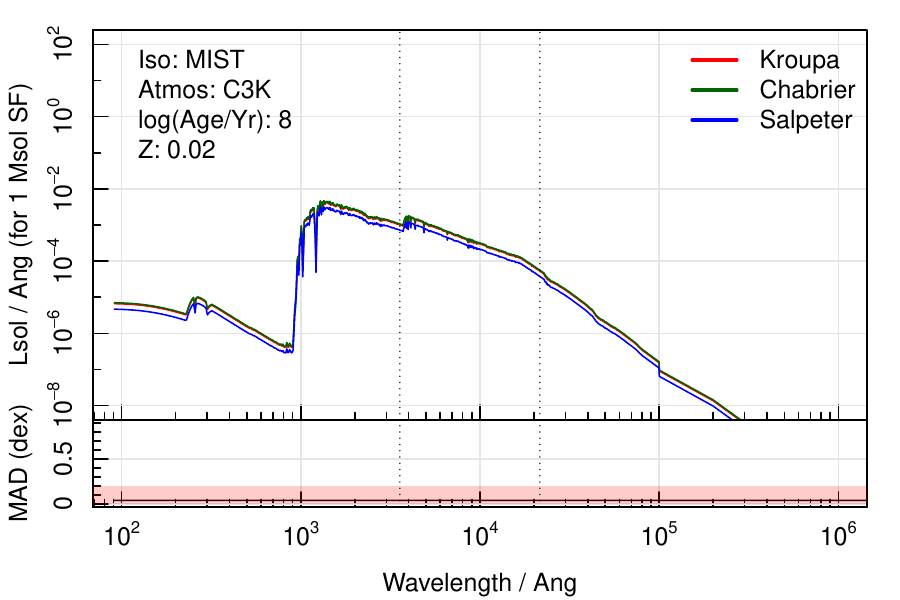}
\\
\includegraphics[width=8.5cm]{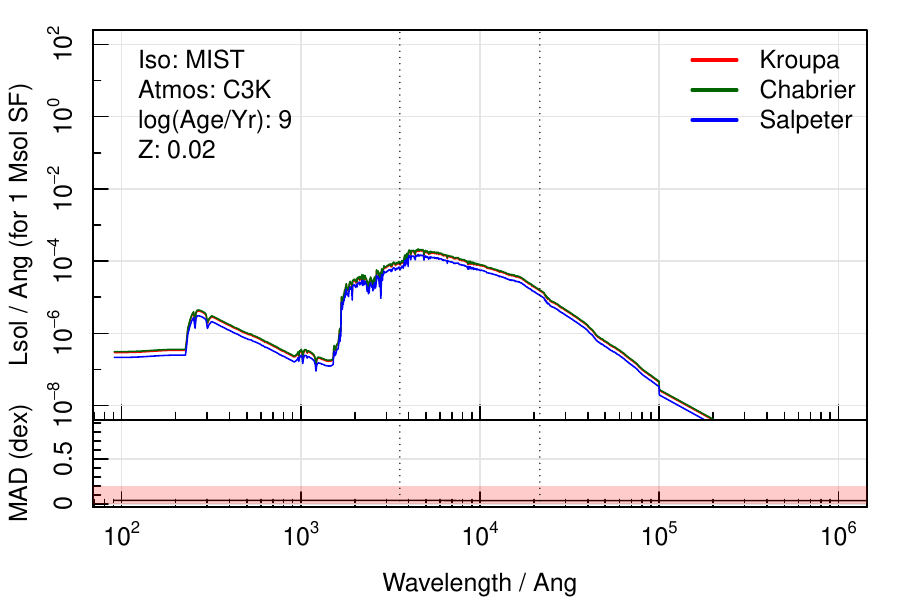}
\includegraphics[width=8.5cm]{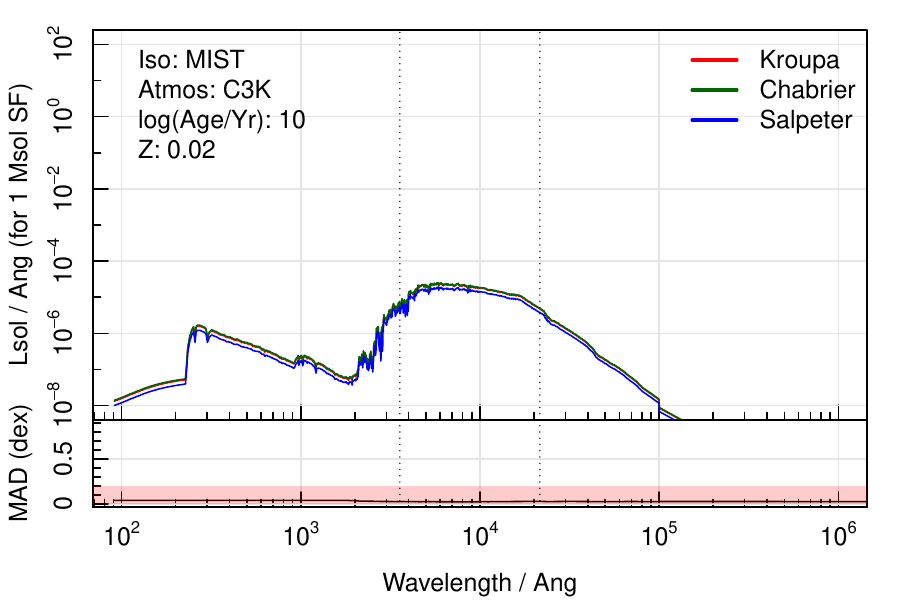}
\\
\caption{Comparison of \Zsol{} spectra at different stellar population ages for different IMFs. The ages presented are log spaced with 1 dex intervals, covering 10 Myr  through to 10 Gyr to allow comparison between all stellar spectra initially available with \progeny. The vertical dashed lines indicate the $u$ and Ks band central wavelengths, common limits for SED analysis work.}
\label{fig:comp_IMF_Z0}
\end{center}
\end{figure*}

Finally, in Figure \ref{fig:comp_IMF_Z0} we make a direct comparison of the three classic IMFs available with \progeny{} (Kroupa, Chabrier and Salpeter), where the other key characteristics of the SSPs are kept fixed (as per the Figure legend).

The differences seen here are negligible, bar the general reduction in spectral normalisation seen for Salpeter. This is a direct consequence of the Salpeter IMF putting proportionally more mass into low luminosity stars. These stars are sufficiently sub-luminous that they have little impact on the spectral shape, so the overall effect is a drop in flux density (i.e.\ an increase in mass-to-light ratio).

\subsubsection{Summary of Low-Level SSP Impact}

From the above analysis it is clear that within a SPL, holding the methodology of how ingredients are combined fixed, the relative impact of isochrones, stellar spectra and IMFs have a clear hierarchy (within the bound of the IMF being a `classic' local Universe variety, not an exotic type). The choice of isochrone has by far the most impact throughout the spectrum, with particularly notable uncertainty in the extreme UV. In this regime various aspects of complicated physics (including, but not limited to, remnant stars, horizontal branch stars and planetary nebulae) become dominant. In some SPLs interventions are made to incorporate some of these physical effects, but \progeny{} relies on the available physics of the isochrones only. Beyond the choice of isochrones, the choice of stellar spectra has the next most impact. The main driver of selecting which one we prefer in \progeny{} is the atmospheric parameter coverage, wavelength coverage and resolution --- in general for similar atmospheric parameters the stellar spectra are highly consistent, with only small deviations in the UV and IR. Finally, the choice of IMF has little impact on the spectral shape, but some notable impact on the normalisation (mass-to-light), where the impact of this is discussed in more detail in the next Section. These general findings are consistent with the literature consensus discussed in the Introduction \citep{1996ApJ...457..625C, 2005MNRAS.362..799M, 2007MNRAS.381.1329M, 2009ApJ...703.1123P, 2009ApJ...699..486C}, but \progeny{} offers an easy route for interested researchers to explore such details (and more subtle behaviour) via its open software and detailed scripts and vignettes.

\subsection{Internal High-Level Comparisons}

\begin{figure}
\includegraphics[width=\columnwidth]{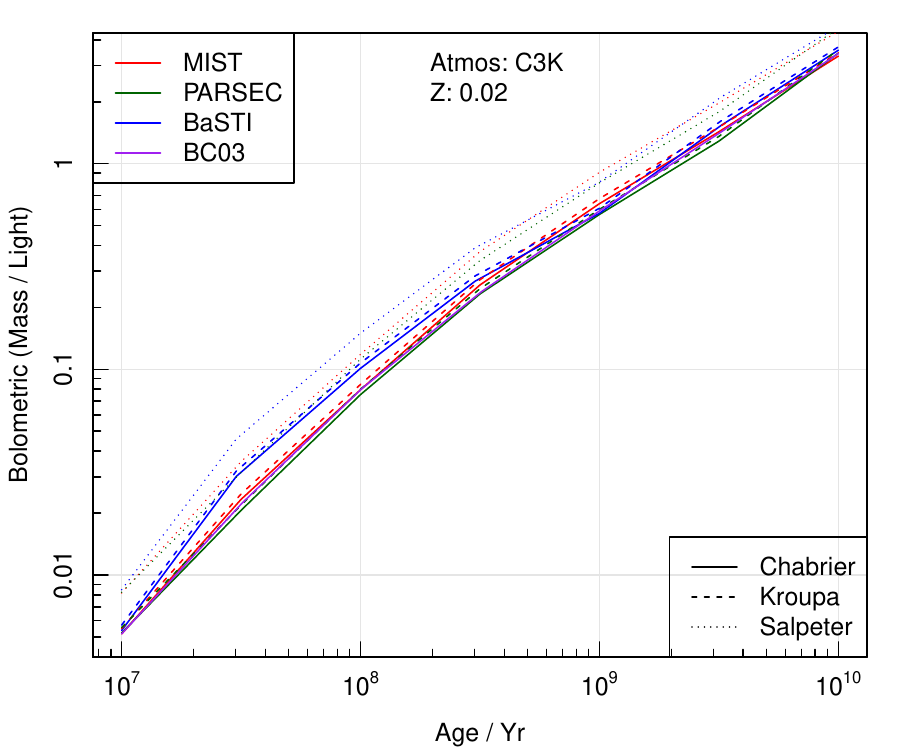}
\includegraphics[width=\columnwidth]{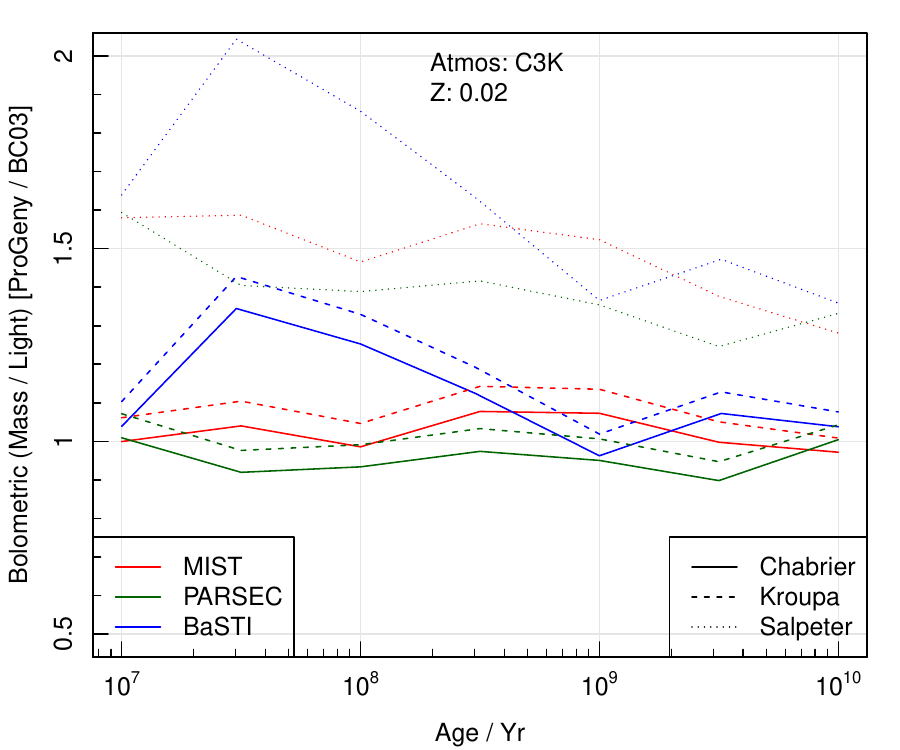}
\caption{Comparison of the bolometric mass-to-light ration for different combinations of isochrones and IMF (there is very little variation due to stellar spectrum choice, so that is left as C3K for this comparison). Top panel shows the measured $M/L$, and bottom panel shows this relative to BC03, where in general \progeny{} SSPs produce more mass-to-light than BC03.}
\label{fig:M2L_comp}
\end{figure}

A useful high-level comparison between different SSPs is to analyse how much mass is required to produce a given bolometric luminosity, i.e.\ $M/L$ where both quantities are computed in solar units. Figure \ref{fig:M2L_comp} compares to $M/L$ for various combinations of \progeny{} isochrones and IMFs (using the C3K stellar spectrum and \Zsol), with BC03 (using a Chabrier IMF) added for reference. The bottom panels divides through by the BC03 SSP $M/L$ to make variations clearer. The main note is that most combinations of \progeny{} isochrones and IMFs produce more mass per unit light, the exception being PARSEC with a Chabrier IMF. In principle that combination should be the most similar to BC03, but we observe that PARSEC is a significant evolution of the older Padova isochrones it was developed from (as used by BC03) and there are other treatments in the BC03 SSP that handle various stellar population aspects differently: remnant stars; planetary nebulae (explicitly added in BC03, and not currently generated in \progeny); AGB phases; stellar spectra (especially for hot stars) etc.

Using the bottom panel from Figure \ref{fig:M2L_comp} it is possible to make approximate corrections to both measured star formation rates and/or stellar mass. When SED fitting the SFR is usually defined as the mass formed over a short recent interval (ranges from 10--100 Myr exist in the literature). Using the upper limit of 100 Myr is reflective of how we define SFR in \prospect, and here we see very small corrections between MIST and PARSEC for Chabrier and Kroupa IMFs, and $\sim 0.2$ dex corrections for Salpeter. BaSTI is the most different in this domain, producing a factor $\sim 0.1$ dex more mass-to-light than either MIST or PARSEC at 100 Myr. To correct for stellar mass the age of stellar populations above 1 Gyr is most relevant at almost all redshifts. Here all three isochrones are very similar for both Chabrier and Kroupa IMFs with about 0.04 dex of scatter between them, and we see a factor $\sim 0.1$ dex correction for Salpeter (also with 0.04 dex of scatter). Given it is difficult to objectively determine which isochrone (if any) is `correct', these mass-to-light shifts should be treated as the absolute lower limit of uncertainty when measuring SFR and/or stellar mass.

\begin{figure*}
\includegraphics[width=8cm]{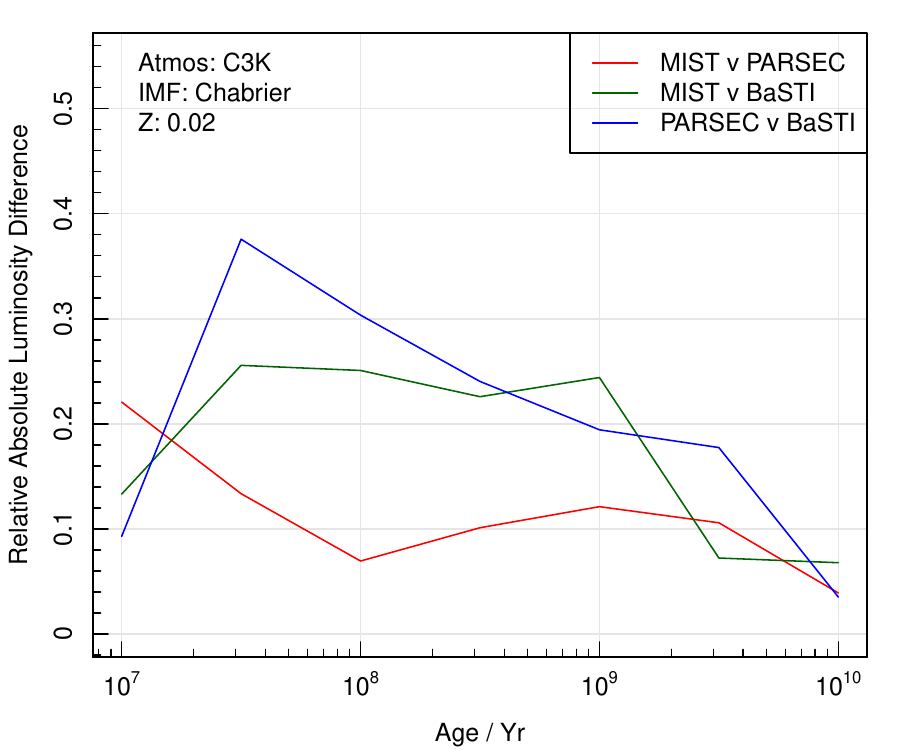}
\includegraphics[width=8cm]{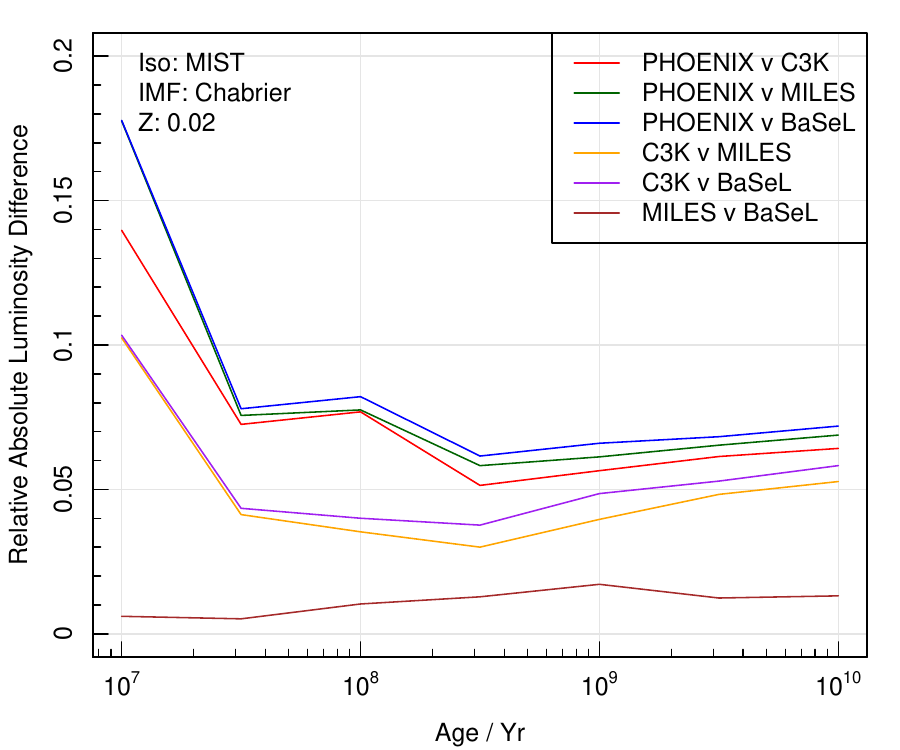}
\includegraphics[width=8cm]{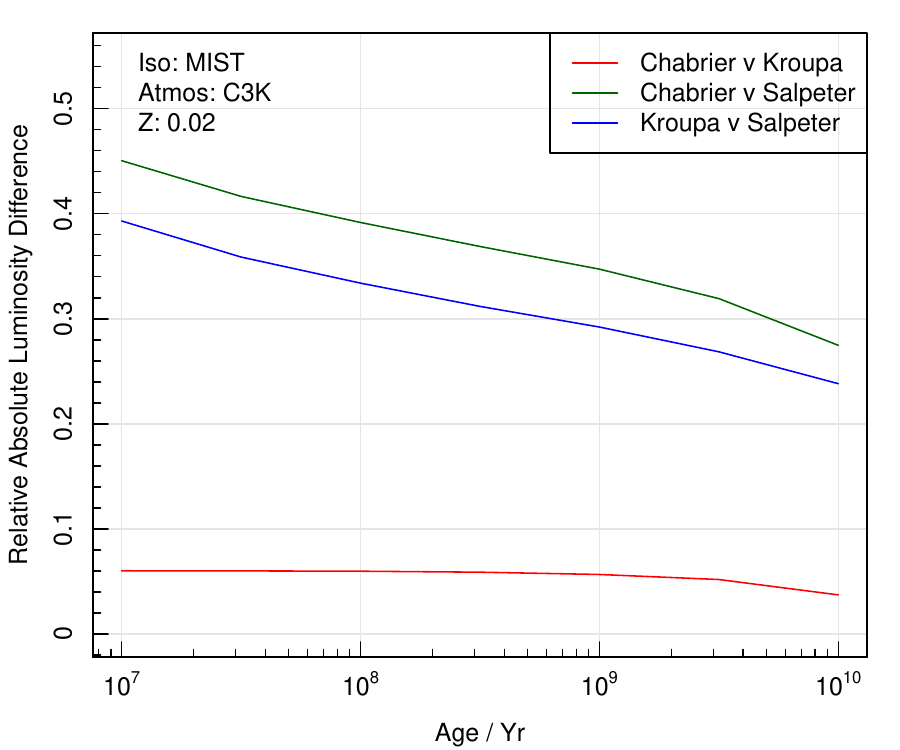}
\caption{SED comparisons of different isochrones, stellar spectra and IMF choices with \progeny. The y-axis is the relative total integrated flux discrepancy when switching isochrones, stellar spectra or IMFs as a function of stellar age, where both relative excesses of flux and deficits are equally penalised. In all cases the major properties kept constant during comparisons are shown in the top left.}
\label{fig:Lum_comp}
\end{figure*}

A core difference between SSPs is how much their spectra differ at a given epoch. One way to capture this information is to compute the relative integrated absolute flux difference between two spectra as a function of time. This has the desired effect of treating the spectra symmetrically, both penalising the relative under and over production of flux at all wavelengths in a consistent manner. For \progeny{} this is presented in Figure \ref{fig:Lum_comp}, where we compare SSPs generated for the MIST, PARSEC, and BaSTI isochrones; C3K, PHOENIX (Husser), MILES, and BaSeL stellar spectra; and also the Chabrier, Kroupa, and Salpeter IMFs commonly used in modern literature.

The immediately notable feature of Figure \ref{fig:Lum_comp} is that our choice of isochrones has a much more dramatic impact than our choice of stellar spectra or IMF at nearly all stellar ages, with particularly dramatic differences seen at periods of stellar evolution that are both extremely luminous and difficult to model. At around 1 Gyr the big spike is dominated by differences in how the various AGB phases are treated. At much older ages we see the relative flux disagreement settles down to around 12-13\%. Since most stellar mass measurements are dominated by stellar population in this age window it suggests fitting error of no less than  $\sim0.05$ dex even in a situation where every other aspect of the process (stellar spectra, interpolation, IMF, SED engine) is identical. In comparison, the impact of the IMF when switching between the two most popular modern variants (Chabrier and Kroupa) is extremely minimal, causing only 6\% relative flux disagreement and 0.03 dex stellar mass variance. The switch to Salpeter does cause a dramatic spectral shift, but this is because it is extremely bottom-heavy (a large fraction of mass in low mass stars), meaning it produces much less light for a given unit of star formation mass. This is consistent with the spectra comparison in Figure \ref{fig:comp_IMF_Z0}.

Considering just the behaviour of the isochrones, overall MIST produces the least variation with the other options. It is particularly consistent with PARSEC overall, with the two producing the smallest spectral differences for almost all stellar ages. PARSEC and BaSTI appear to be much more discrepant with each other in general. Given MIST also has excellent age and metallicity coverage and relatively detailed treatment of stellar remnants, we recommend MIST as our fiducial isochrone in general usage. That said, any detailed study should explore the impact of using different isochrones, and a whole range of SSPs have been pre-generated with \progeny{} using both PARSEC and BaSTI isochrones to accompany this paper and the release of the \progeny{} package.

Assuming we using either a Chabrier or Kroupa IMF, extremely young ($10^7$ Gyr) and old ($10^{10}$ Gyr) stellar population spectra uncertainties are mildly dominated by stellar spectrum choice (but isochrones are not far behind). The scatter between these is minimised at a few hundred Myr, and slowly rises beyond that to a typical 5--7 \% variation. The least disagreement is seen between MILES and BaSeL stellar spectra, likely because both of these are empirically constructed. C3K agrees more with these two empirical libraries than PHOENIX (Husser). Given the excellent parameter coverage of C3K (intrinsically much larger and well sampled than MILES or BaSeL), this result partly informs our decision to recommend C3K as our fiducial `base' stellar spectrum library. As with the isochrones above, SSPs using all of the discussed stellar spectra have been pre-generated with \progeny. Anybody carrying out a detailed study of SEDs with \prospect{} should consider using the other stellar spectrum combinations too.

An obvious issue from an inference perspective is we do not know the truth re isochrones, stellar spectra, and/or IMF, and it is quite possible that a combination of the `wrong' isochrone and `wrong' IMF combine in a manner that provides a better overall fit to our data. The above findings suggest that much better fidelity in isochrones and atmospheric modelling would be needed before strong statements about IMF behaviour could be conclusively demonstrated in any absolute sense. Within a consistent SED software environment (e.g.\ \prospect) relative statements about the impact of the IMF can be made however. The impact of using different SSPs (including our new \progeny{} ones) is explored in depth in our Bellstedt \& Robotham (in prep.) companion paper.

An important component of calculating the stellar mass remaining in a stellar population (almost always what is meant by a `stellar mass', versus the alternative of total mass formed) is the recycling behaviour of an SSP as a function of time. How these behave is almost entirely a function of the isochrones chosen combined with the IMF. Top-heavy IMFs have the effect of distributing much more mass in massive stars that die sooner and feedback mass much more efficiently than less massive stars. Figure \ref{fig:Recycle_comp_ProGeny} shows the impact of varying both the isochrones (MIST, PARSEC, and BaSTI) and IMF (Chabrier, Kroupa, and Salpeter). Here we see a bigger differences when switching IMFs than isochrones, with Chabrier consistently having less mass remaining at all epochs (meaning more mass from stars is being returned to the inter-stellar medium) and Salpeter by far the most. Regarding the isochrones, MIST has a similar bias, with relatively less mass existing within stars at all epochs compared to PARSEC and BaSTI. MIST plus Chabrier is the most aggressive combination in terms of returning material back to the ISM, with only about 54\% of mass locked up in stars at 10 Gyr. BaSTI in comparison locks up much more mass in stars at all times, with BaSTI plus Salpeter producing the most extreme recycling function (77\% of mass locked up in stars at 10 Gyr). When combing either MIST or PARSEC (isochrones) and Chabrier or Kroupa (IMFs) the effects are subtle --- even for the oldest ages with the most feedback we would only see a 10\% difference between MIST + Chabrier versus PARSEC + Kroupa in terms of the stellar mass inferred.

\begin{figure}
\includegraphics[width=\columnwidth]{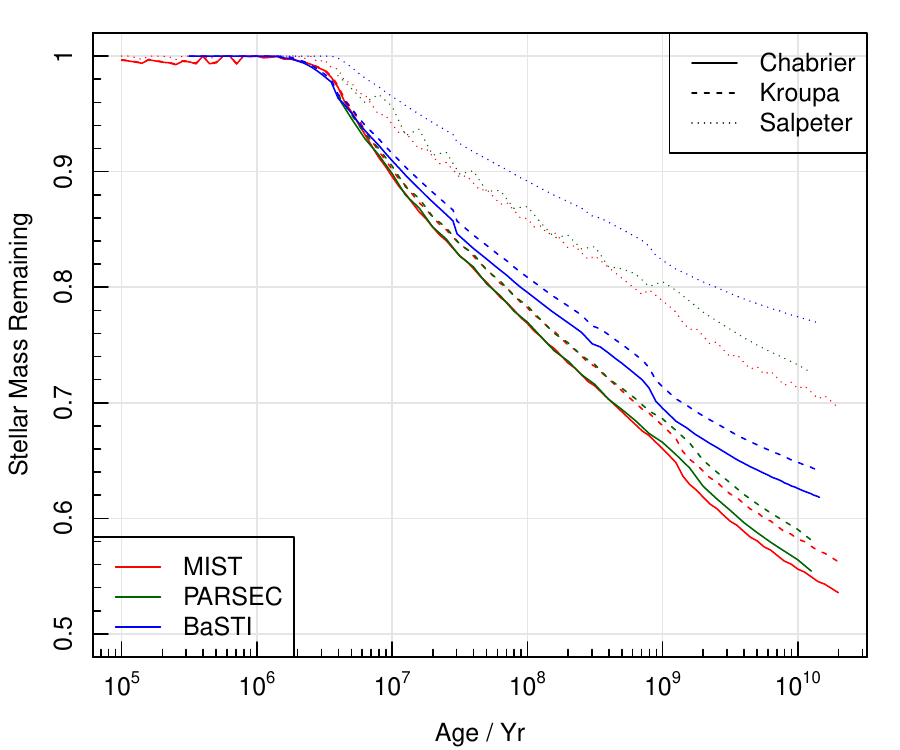}
\caption{Stellar mass remaining comparison of different isochrone and IMF choices with \progeny (using \Zsol). Other versions of this Figure (for different metallicities) can be viewed at \url{rpubs.com/asgr}.}
\label{fig:Recycle_comp_ProGeny}
\end{figure}

\begin{figure}
\includegraphics[width=\columnwidth]{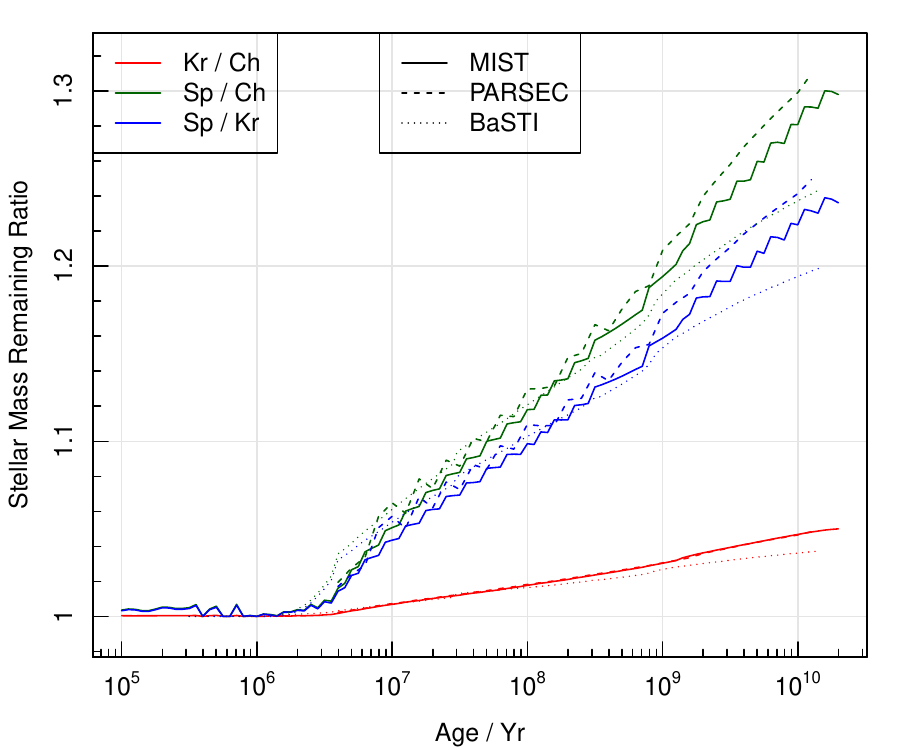}
\caption{Ratio of stellar mass remaining for different isochrone and IMF choices (using \Zsol). Ch = Chabrier, Kr = Kroupa, Sp = Salpeter.}
\label{fig:Recycle_correct}
\end{figure}

Figure \ref{fig:Recycle_correct} highlights another aspect of SED modelling that tends to be ignored or simplified: IMF corrections. The main note here is that the correction between Chabrier and Kroupa IMFs is very small at all ages (at worst $\sim5$\% for the oldest stellar populations). The story is different when comparing either of these to Salpeter. The extreme bottom-heavy nature of the Salpeter IMF means substantially more mass is locked up for a given amount of star formation, and the apparent mass difference grows steadily with age, reaching $\sim30$\% extra mass for the most extreme comparison of Salpeter versus Chabrier for PARSEC isochrones.

Many papers employ simple single factor `corrections' between SFR and stellar mass measurements made assuming different IMFs. In detail it is clear this correction is a strong function of stellar population age and isochrone assumptions (although the effect of the latter is quite minimal in the case or \progeny), and in practice it is almost impossible to fully homogenise IMFs without having the full SFH of the stellar population being corrected. On the positive side, such corrections are rarely the largest error in the process.

\subsection{Literature Comparisons}

It is instructive to compare the spectra and other aspects of the new \progeny{} SSPs against well known and used literature examples. Table \ref{tab:SSPs} summarises the SSPs we compare against in the following sections. While we list the dominant isochrone and atmospheric spectra, it should be noted that these are usually supplemented to achieve better coverage (e.g.\ stellar remnant evolution and AGB phases etc). To read more details of the various SPLs (and these specific SSPs) we guide the interested reader to \citet{2021MNRAS.505..540T}, where a comprehensive accounting is made for the various components used for many classic SPLs.

\begin{table*}
\begin{center}
\caption{Basic information regarding the literature SSPs we compare against. Note for the stellar spectra only the dominant library is listed, nearly all SSPs use a complex mixture of spectra to achieve adequate temperature and surface gravity parameter coverage. Note M05 only has a Kroupa IMF variant.}
\begin{tabular}{|c|c|c|c|c|c|c|c|c|}
Name & Isochrones & Spectra & IMF & log(Age/Myr) & N (Age) & log(Z/Z$_\odot$) & N (Z) & Reference \\
\hline
BC03 & Padova & STELIB & Chabrier & 0 -- 10.3 & 221 & -2.3 -- 0.4 & 6 & \citet{2003MNRAS.344.1000B} \\
CB19 & PARSEC & MILES & Chabrier & 0 -- 10.15 & 221 & -2.3 -- 0.5 & 15 & \citet{2019MNRAS.490..978P} \\
BPASS & STARS & C3K/BaSeL & Chabrier & 6 -- 11 & 51 & -3.3 -- 0.3 & 13 &  \citet{2018MNRAS.479...75S} \\
M05 & Maraston & Mixture & Kroupa & 3 -- 10.18 & 67 & -1.3 -- 0.3 & 4 &  \citet{2005MNRAS.362..799M} \\
FSPS & MIST & Padova & Chabrier & 5.5 -- 10.15 & 94 & -2 -- 0.2 & 22 & \citet{2009ApJ...699..486C} \\
EMILES & BaSTI / Padova & MILES & Chabrier  & 7.48 -- 10.15 & 53 & 2.3 -- 0.3 & 12 & \citet{2016MNRAS.463.3409V}
\end{tabular}
\end{center}
\label{tab:SSPs}
\end{table*}%

\begin{figure*}
\begin{center}
\includegraphics[width=8.5cm]{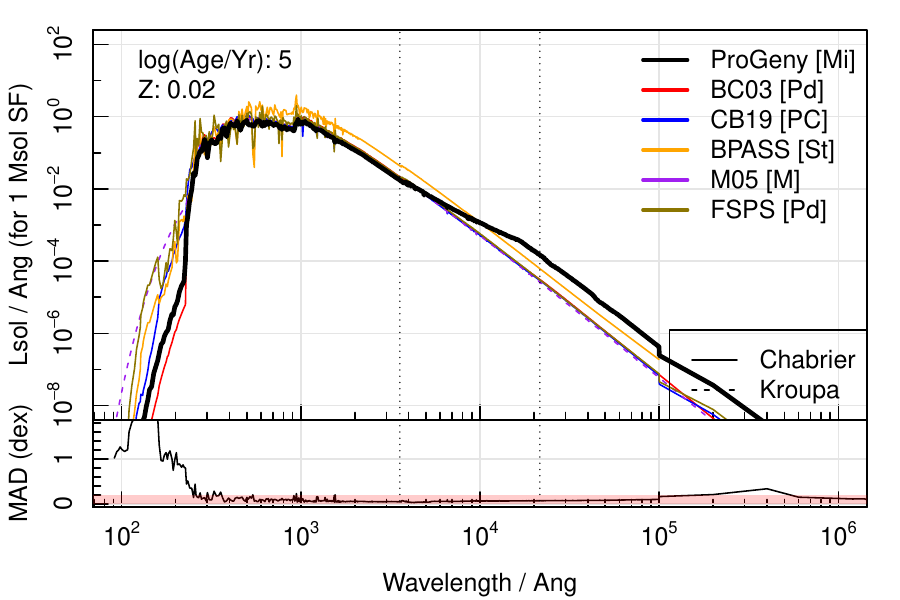}
\includegraphics[width=8.5cm]{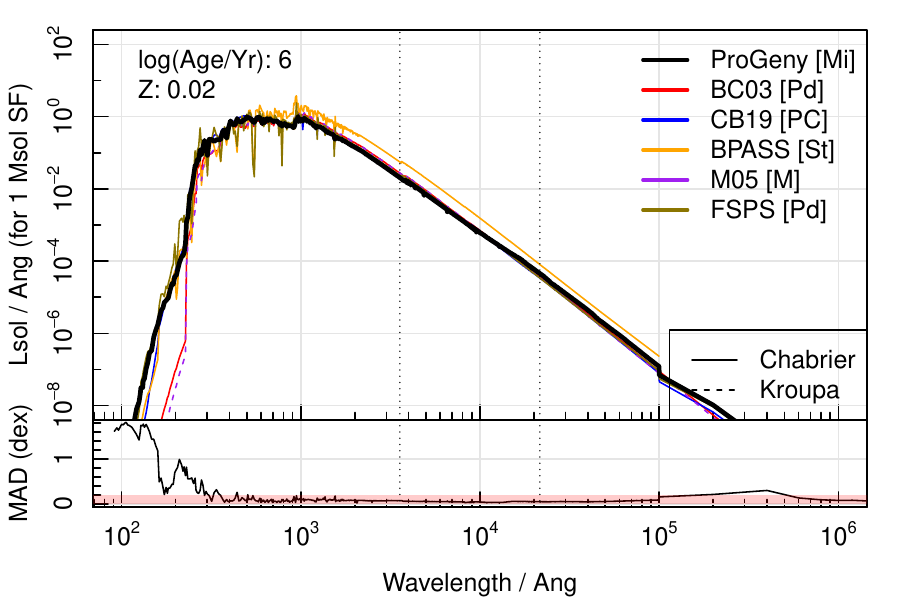}
\\
\includegraphics[width=8.5cm]{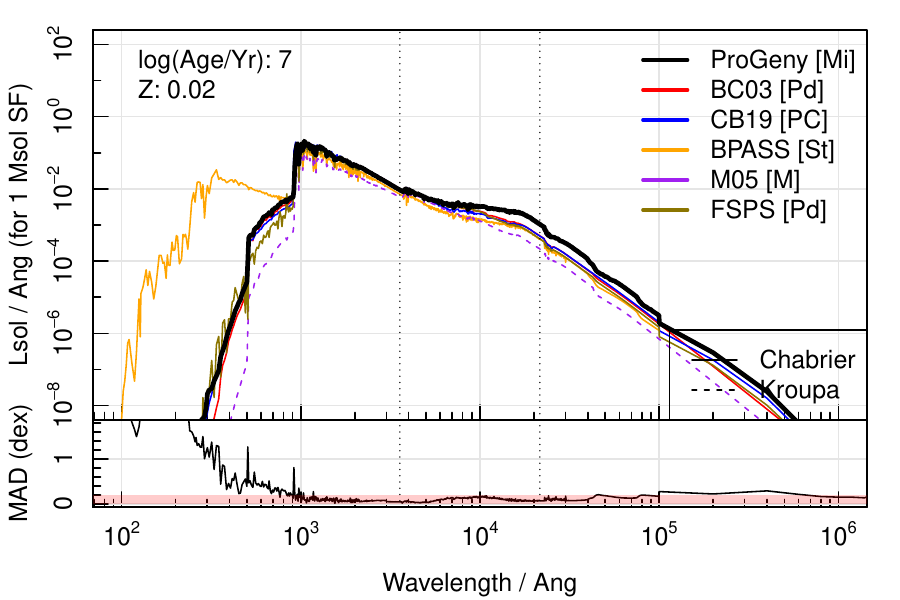}
\includegraphics[width=8.5cm]{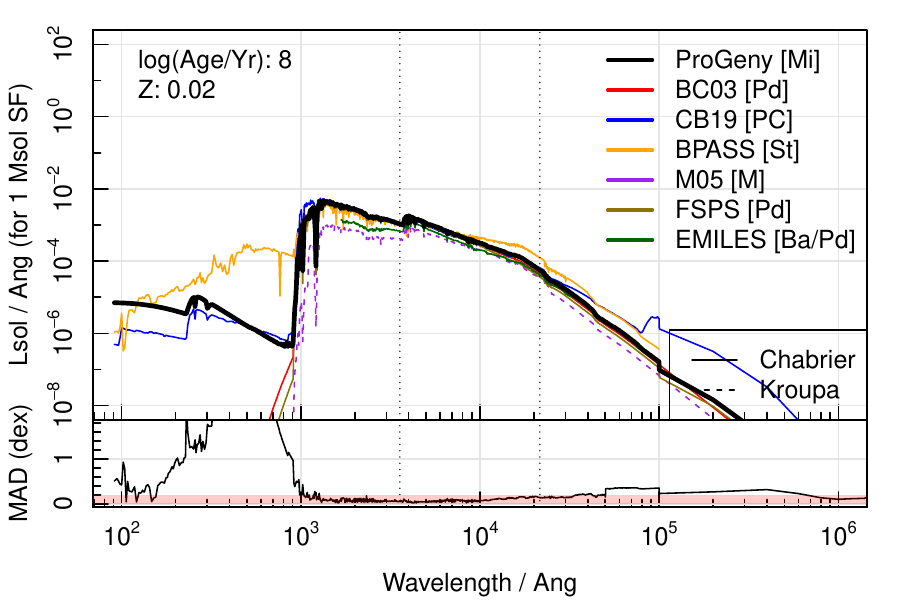}
\\
\includegraphics[width=8.5cm]{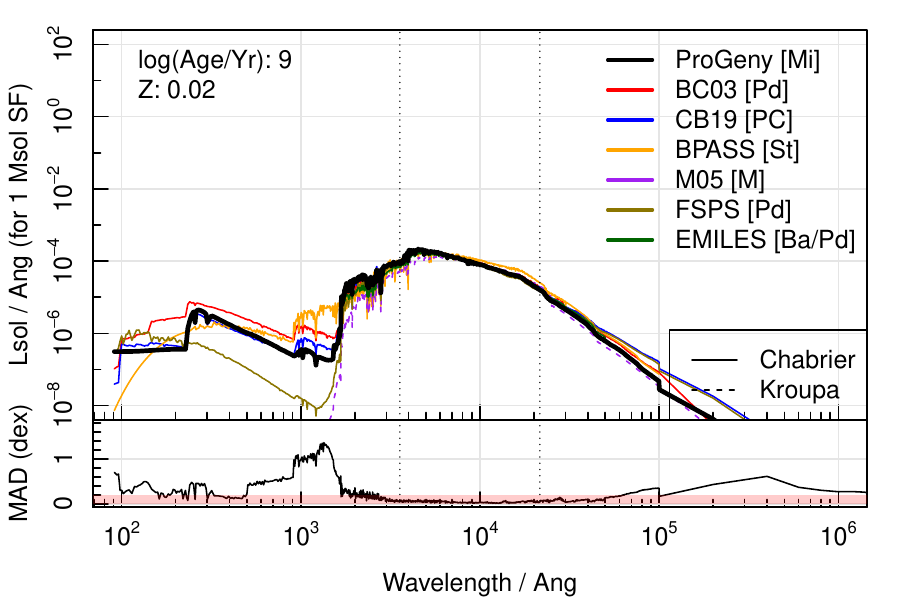}
\includegraphics[width=8.5cm]{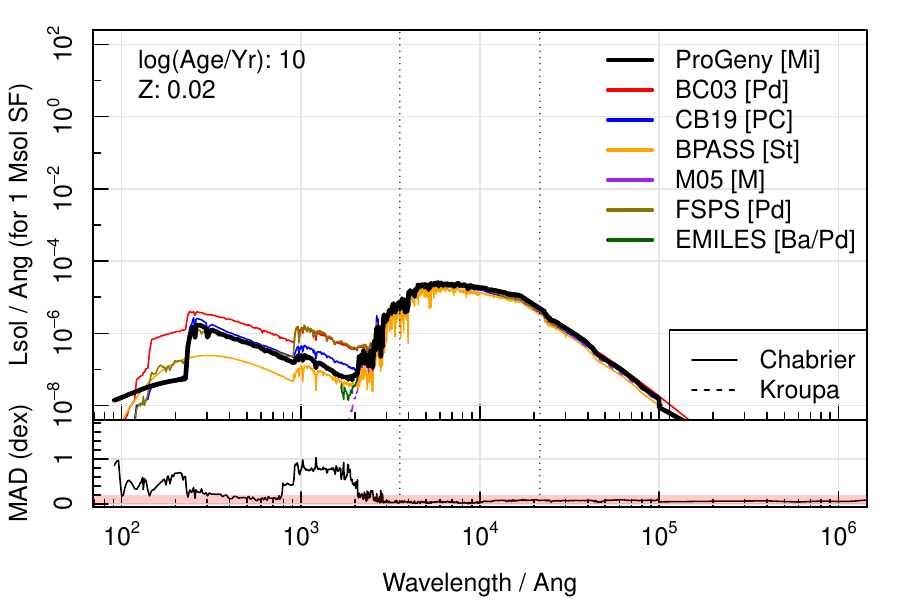}
\\
\caption{Comparison of \Zsol{} spectra at different stellar population ages for different SSPs. The ages presented are log spaced with 1 dex intervals, covering 100,000 yr through to 10 Gyr. Zoomed in versions of the Figure (covering 1,000 -- 10,000 \AA), and alternative metallicity versions of this Figure can be viewed in Appendix \ref{sec:SSP_extra}. The vertical dashed lines indicate the $u$ and Ks band central wavelengths, common limits for SED analysis work. The dominant isochrone used is listed in square brackets, where Mi = MIST, Pd = Padova, PC = PARSEC, St = STARS, M = Maraston, Ba = BaSTI.}
\label{fig:comp_spec_Z0}
\end{center}
\end{figure*}

The most fundamental low-level comparison we can make between SSPs is by analysing the actual spectra produced for specific combinations of metallicity and stellar evolutionary age. Figue \ref{fig:comp_spec_Z0} presents such spectra for solar metallicities, covering 100,000 yr through to 10 Gyr. For reasons not entirely clear, FSPS when run with MIST isochrones produces very unusual spectra for ages below $10^7$ years, with a much flatter spectral shape than seen for any other SSP. These unusual features are not present when running in the same manner but specifying Padova isochrones with FSPS (using the original Fortran software), so it would seem to be a specific issue with the younger MIST isochrones. For that reason we prefer to present FSPS using Padova isochrones for the following comparisons.

A consistent observation is that predictions for SSP spectra below 1,000 \AA{} are highly variable between specific SSPs for age much above a million years, with sometimes many dex of disagreement re the detail of spectra in this regime. The reason for this is complex, but the general issue is the differing treatment of hot remnant stars, horizontal branch (HB) stars, planetary nebulae (PN), and helium burning phases by different isochrones and/or SPLs. This is clear when observing the M05 SSP that has no remnant stars and different treatment of HB stars to standard isochrones: we see a sharp truncation of UV light for all SSPs $10^8$ years and older with no evidence of any UV upturn at all. It is relatively rare that these parts of the SED are interrogated with lower redshift surveys since even UV facilities like GALEX only probe down to 1350 \AA{} in the FUV. However, by redshift 0.3 the rest frame GALEX FUV filter starts to cover the 1,000 \AA{} feature seen here, and much beyond this even typical optical filters will be measuring flux in this region. 

Redder than this 1000 \AA{} feature the SSPs are remarkably consistent over 5 dex in age, with the M05 SSPs being the most obvious outlier below the 4000 \AA{} break. Beyond that observation, the next most significant variation is seen in the near infrared (1 micron and redder) where the various AGB phases become important for ages older than $\sim 10^7$ years (especially above $10^8$ years) and have complex modelling differences at the isochrones level (as noted in Section \ref{sec:FL_iso}) and decisions regarding the best (or perhaps more accurately, least bad) method for matching these stellar evolution phases to the most appropriate stellar spectra. Given the huge wealth of data, it is not surprising that models collectively agree best in the optical regime for giga-year and older stellar populations. This represents the dominant spectral feature in a huge number of passive galaxies and globular clusters, and is a common target when checking the validity of a given SSP.

An additional difference we note between \progeny{} (with MIST isochrones) and other SSPs can be seen in the lowest age (100,000 yr) top-left panel. Red-wards of about 1 micron there is a significant (almost 1 dex) excess to the classic black-body power-law slope you might expect (and as is seen for other SSPs). This feature is entirely cause by the tracking of pre-main sequence stars in MIST, which produces an additional cool ($\sim2,000$ K) spectrum representing the thermal emission from collapsing molecular clouds and proto-stars. This cool spectrum is added onto the dominate spectrum of extremely hot O and B stars on the main sequence which have typical temperatures $\sim50,000$ K. While it looks dramatic when plotted like this, the actual flux contribution in a typical galaxy SED is very small, and this difference in behaviour has little impact on composite (or complex) stellar population spectra in general.

\begin{figure}
\includegraphics[width=\columnwidth]{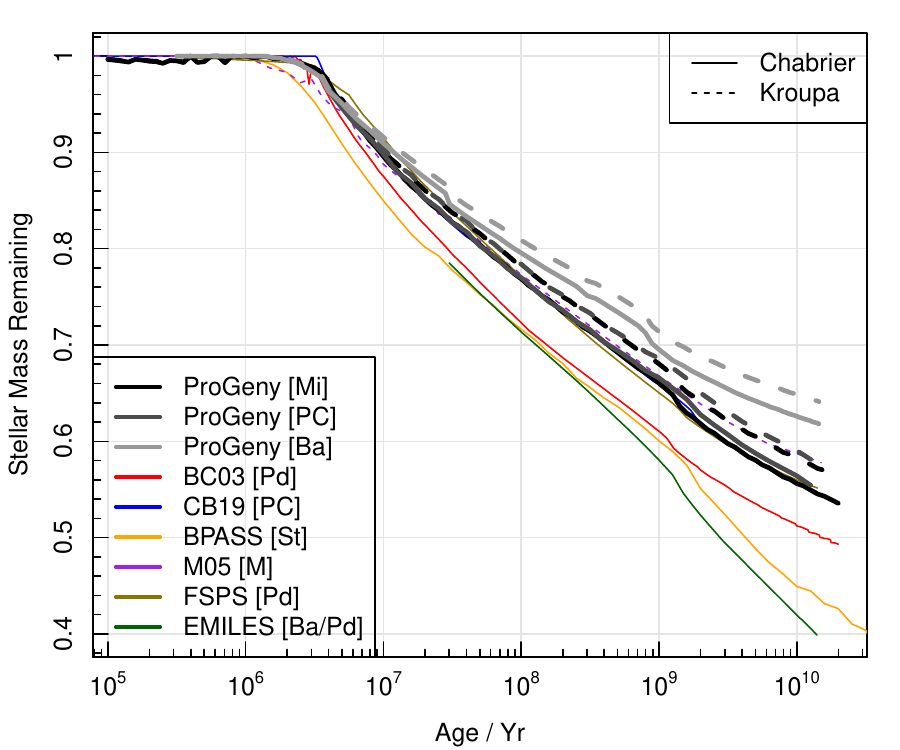}
\caption{Stellar mass remaining comparison of different isochrone and IMF choices with \progeny versus a compendium of literature SSPs (using \Zsol). The dominant isochrone used is listed in square brackets, where Mi = MIST, Pd = Padova, PC = PARSEC, St = STARS, M = Maraston, Ba = BaSTI}
\label{fig:Recycle_comp_ProGeny}
\end{figure}

Another key comparison to make is the variation between SPLs of the stellar mass remaining as a function of time. The main driver of any differences in this regard is clearly the precise isochrone modelling, since complex events like stellar winds and AGB phases must be accounted for. Also, any variations in stellar rotation and binary evolutionary nature (in the case of BPASS in particular) can produce significant differences after a Gyr. Figure \ref{fig:Recycle_comp_ProGeny} compares \progeny{} using both MIST, PARSEC, and BaSTI isochrones with both Kroupa and Chabrier IMFs (since this combination was the most similar in Figure \ref{fig:Recycle_comp_ProGeny}), and compares these to other popular SPLs (which use a mixture of inputs, as per Table \ref{tab:SSPs}).

It is immediately notable that in general SSPs using Padova track isochrones have systematically less mass remaining as a function of time compared to MIST and PARSEC. A key difference between MIST and PARSEC compared to Padova (where we note PARSEC is the successor isochrone to Padova) is the treatment of late time winds and remnant stars. In the older Padova isochrones remnant stars were not tracked, and additional models had to be bolted on to the isochrones to account for the correct evolution of remnant stars. It is clear that FSPS (even though also using Padova 2007 tracks in this case) has different treatments when extending the isochrones to model remnants, in general keeping more mass locked up in stars than BC03. Not surprisingly, CB19 (which uses PARSEC isochrones predominantly) has very similar behaviour to the PARSEC based \progeny{} SSPs. Interestingly, BPASS produces amongst the least mass locked up in stars over time, highlighting that binary feedback produces significant additional mass loss on average. Broadly speaking, this would suggest that there is a 0.1 dex shift in implied stellar mass remaining when switching between more modern non-binary SPLs and binary SPLs for population older than a Gyr (where binary populations would be less massive for a given amount of total stars formed).

\section{Conclusions and Future Outlook}
\label{sec:conclusions}

Here we have, in brief, introduced the new \R{} based \progeny{} SPL package that allows for the rapid and flexible generation of SSPs for use in particular with \prospect, but potentially any other SED generating or fitting software. The package is thoroughly documented, with relevant examples for all user-level functions, and longer form vignettes demonstrating sophisticated types of analysis and comparison\footnote{\url{rpubs.com/asgr}}.

Basic comparisons show that \progeny{} is spectrally consistent with other popular SPLs when using similar isochrones and stellar spectra, with understandable variations in a few key domains (e.g.\ UV upturn) due to our \progeny{} isochrones incorporating different physical processes to some published SPLs, even nominally identical ones. This is not surprising since isochrones with the same name (e.g.\ MIST, PARSEC, BaSTI etc) receive regular updates to their physics over time, and the literature is lacking in clear version control for SPLs and/or SSPs. The largest variations in spectral output are noted when switching isochrones ($\sim 60$\%), followed by the stellar spectra ($\sim 20$\%) and finally IMFs ($\sim 10$\%, between Chabrier and Kroupa variants). This is consistent with previous literature on the topic.

To create a fiducial set of stellar spectra using the recommended combination of MIST for the isochrones, C3K (for the base stellar spectra) and a Chabrier IMF takes 5s at low resolution (matching BC03) and 18s at high resolution (matching CB19) using a modern Apple MacBook. These speeds make it realistic to create SSPs on-the-fly, and for many use cases (since the fitting time will dominate the processing time with \prospect) there is no need to pre-generate static SSPs. This is particularly useful when exploring variations of the input ingredients (like the IMF) since exhaustively generating all potential combinations is onerous in terms of disk space and book-keeping.

As part of this work, we introduce a new standardised FITS based format for defining all the characteristics of stellar spectra to be incorporated into a new SPL/SSP. We noted while constructing the default input stellar spectra that every single library used its own format, ranging from pure binaries, ASCII text files, and FITS files of many different formats. While it is true we have added `yet another standard', the one we propose is a minimal combination of the best characteristics of the data we processed in a highly open and well used binary (FITS) format.

A companion paper (Bellstedt \& Robotham in prep.) investigates the impact of varying the available \progeny{} isochrones, stellar spectra, and IMFs when using the SSPs with \prospect{} to fit GAMA galaxies. Longer term the aim is to establish the most plausible fiducial model, i.e.\ what combination of isochrones, stellar spectra and IMF (including evolving varieties) best reproduce a self consistent model of galaxy evolution in the Universe. Key to this will be investigating the implied cosmic star formation history and metallicity evolution implied by fitting galaxies at low (GAMA) and moderate (WAVES) redshift, and comparing this to `in-situ' measures at earlier epochs \citep[heavily inspired by][]{2020MNRAS.498.5581B}.

We also plan to investigate the impact of using evolving and component dependent IMFs in the Shark semi analytic model \citep{2018MNRAS.481.3573L}. In principle we can couple the IMF to the stellar age and/or the component metallicity when using \prospect{} in its generative SED mode. \citet{2016MNRAS.462.3854L} advocated for the use of top-heavy IMF for bulge components to get the correct sub-mm number counts, but interestingly \citet{2020MNRAS.499.1948L} required no such explicit treatment, using a constant Chabrier IMF for all SEDs. Understanding better the impact of a non-universal IMF will help resolve this tension.

As with any SPL, \progeny{} has certain limitations in its current form. As discussed in the main paper, we do not have explicit treatment for certain evolutionary phases that are not always well captured by traditional isochrones. The most notable features lacking being detailed modelling of the horizontal branch phases (where spectral outputs from both the core and the outer envelope become important), and planetary nebulae (where the PN can be considered as a {\it very} extended photosphere). \progeny{} is not dramatically limited despite these exclusions, producing SSPs that are largely consistent with SPLs that do have such additional handling \citep[most notably][]{2003MNRAS.344.1000B, 2019MNRAS.490..978P}. In any case, these limitations could be overcome by \progeny{} tracking multiple stellar components when relevant. We note that MESA/MIST does explicitly track the properties of the core also, the limitation then being the adoption of appropriate spectra.

Near-term planned enhancements for \progeny{} include the separate treatment of remnant stars in order to make better use of isochrones that do not track these phases in detail. Until these additional tracks have been incorporated, general purpose SED modelling should probably use the MIST isochrones only. We also plan to incorporate a wider variety of isochrone libraries, where the impact of stellar rotation, $\alpha$/Fe enhancement and potentially binary evolution \citep[with BPASS;][]{2018MNRAS.479...75S} can be explored in systematic detail. To self-consistently account for more exotic forms of stellar rotation and $\alpha$/Fe enhancement the appropriate stellar spectra would also need to be added, which means such extensions are likely to happen slowly (multiple years). The ultimate hope is many of these extensions will come from the wider community, with all aspects of \progeny{} being available under a permissive L-GPL3 license.

\section*{Data Availability}
\label{sec:data}

This paper releases multiple aspects of the new \progeny{} stellar population library:

\begin{itemize}
\item The open source (L-GPL3) \R{} package that allows the flexible generation and modification of all SSPs discussed in this work (and many more) available at \url{github.com/asgr/ProGeny}.
\item Long-form examples (vignettes) that generate many of the Figures presented in this work available at ASGR's RPubs\footnote{\url{rpubs.com/asgr}}.
\item Script to generate all 45 SSPs used in this work, also at ASGR's RPubs.
\item Seven pre-generated \prospect{} format (FITS) SSPs that use MIST isochrones, C3K base stellar spectra, and the following IMFs:
	\begin{itemize}
	\item Chabrier (0.1 -- 100 \msol)
	\item Kroupa (0.1 -- 100 \msol)
	\item Salpeter (0.1 -- 100 \msol)
	\item Age-Evolving Kroupa (0.1 -- 100 \msol; default \progeny{} evolution arguments)
	\item Metallicity-Evolving Kroupa (0.1 -- 100 \msol; default \progeny{} evolution arguments)
	\item Age-Evolving Lacey (0.1 -- 100 \msol; default \progeny{} arguments)
	\item Metallicity-Evolving Lacey (0.1 -- 100 \msol; default \progeny{} arguments)
	\end{itemize}
	
	These are available through the \prospect{} SSP download function, and directly at \url{tinyurl.com/prospect-speclib/}.
\end{itemize}

\section*{Acknowledgements}

ASGR acknowledges funding by the Australian Research Council (ARC) Future Fellowship scheme (FT200100375). SB acknowledges funding by the Australian Research Council (ARC) Laureate Fellowship scheme (FL220100191).

All Figures were created with the \R{} {\sc magicaxis} package \citep{2016ascl.soft04004R}. Basic astronomy calculations (including cosmological) used the \R{} {\sc celestial} package \citep{2016ascl.soft02011R}.

Minor typos, grammar and spelling mistakes were identified with the assistance of ChatGPT-4o\footnote{\url{openai.com}} when preparing this document. No passages of text or structural outlines for this paper were created with the help of any large language models.



\bibliographystyle{mnras}
\bibliography{ProGeny} % if your bibtex file is called example.bib

%\begin{thebibliography}{99}
%\bibitem[\protect\citeauthoryear{Author}{2013}]{author2013}
%\bibitem[\protect\citeauthoryear{Jones}{2015}]{jones2015}
%Jones C.~D., 2015, Journal of Interesting Stuff, 17, 198
%\bibitem[\protect\citeauthoryear{Smith}{2014}]{smith2014}
%Smith A.~B., 2014, The Example Journal, 12, 345 (Paper I)
%\end{thebibliography}

%%%%%%%%%%%%%%%%%%%%%%%%%%%%%%%%%%%%%%%%%%%%%%%%%%

%%%%%%%%%%%%%%%%% APPENDICES %%%%%%%%%%%%%%%%%%%%%

\appendix

\section{Isochrone Regions}
\label{sec:iso_regions}

\begin{figure}
\begin{center}
\includegraphics[width=\columnwidth]{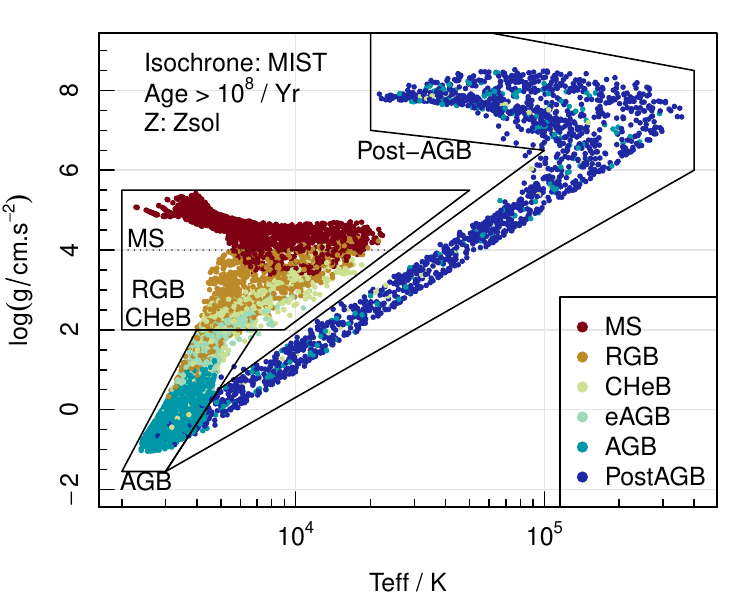}
\label{fig:iso_regions}
\caption{MIST isochrone stellar phase labels with approximate regions overlaid. MS is main sequence; RGB is red giant branch; CHeB is core helium burning; eAGB is early asymptotic giant branch; AGB is asymptotic giant branch (including thermally pulsating TP-AGB stars); PostAGB is post early asymptotic giant branch, covering the evolution of remnants stars and ending with white dwarfs (horizontal feature where logG > 6).}
\end{center}
\end{figure}

Figure \ref{fig:iso_regions} justifies the origin of the isochrone region labelling used in the main paper for Figure \ref{fig:comp_iso}. The differential between main sequence and red giant branch stars is poorly defined (hence the dashed line), and core helium burning and early AGB phases overlap a lot with later AGB phases (including TP-AGB).

\section{Additional Colour Figures}
\label{sec:colour_extra}

Additional versions of Figure \ref{fig:atmos_colour} for different isochrone (MIST / PARSEC / BaSTI) and metallicity combinations. In cases where metallicities do no exist on the exact gridding in the Figure (which correspond to value available in BC03), we use the nearest match in logarithmic space.

\begin{figure}
\includegraphics[width=\columnwidth]{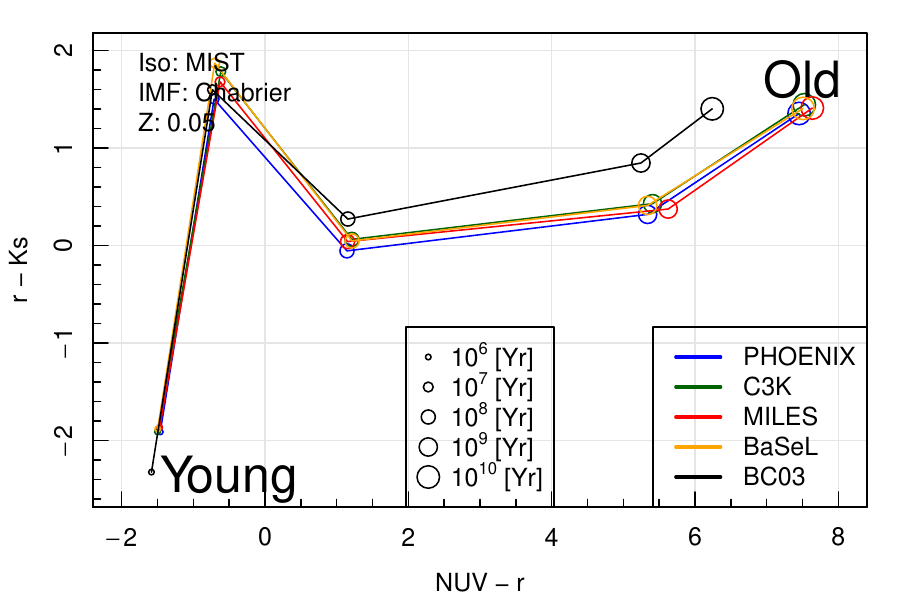}
\caption{Colour-colour comparison of different stellar spectra as a function of age for $Z=0.05$ using MIST isochrones. See Figure \ref{fig:atmos_colour} for additional context.}
\end{figure}

\begin{figure}
\includegraphics[width=\columnwidth]{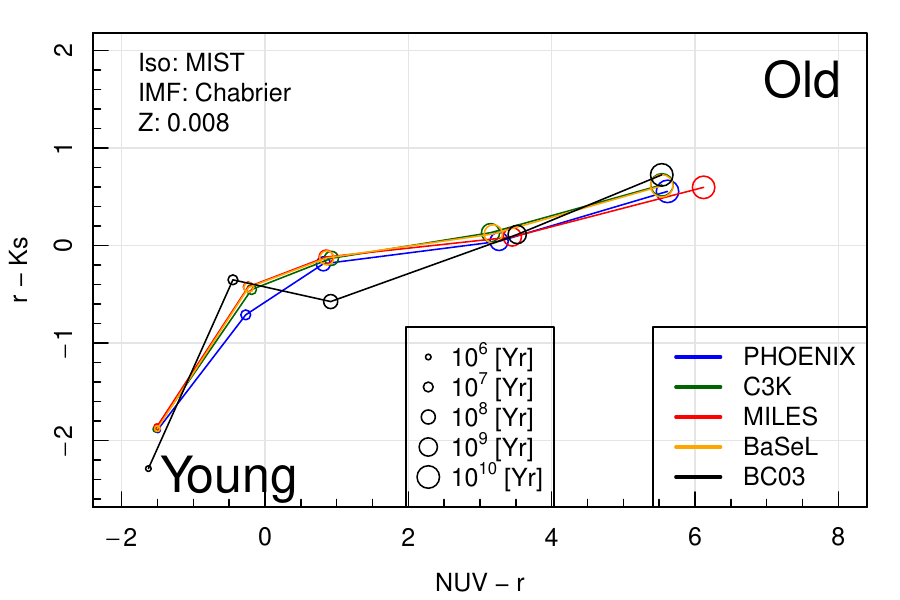}
\caption{Colour-colour comparison of different stellar spectra as a function of age for $Z=0.008$ using MIST isochrones. See Figure \ref{fig:atmos_colour} for additional context.}
\end{figure}

\begin{figure}
\includegraphics[width=\columnwidth]{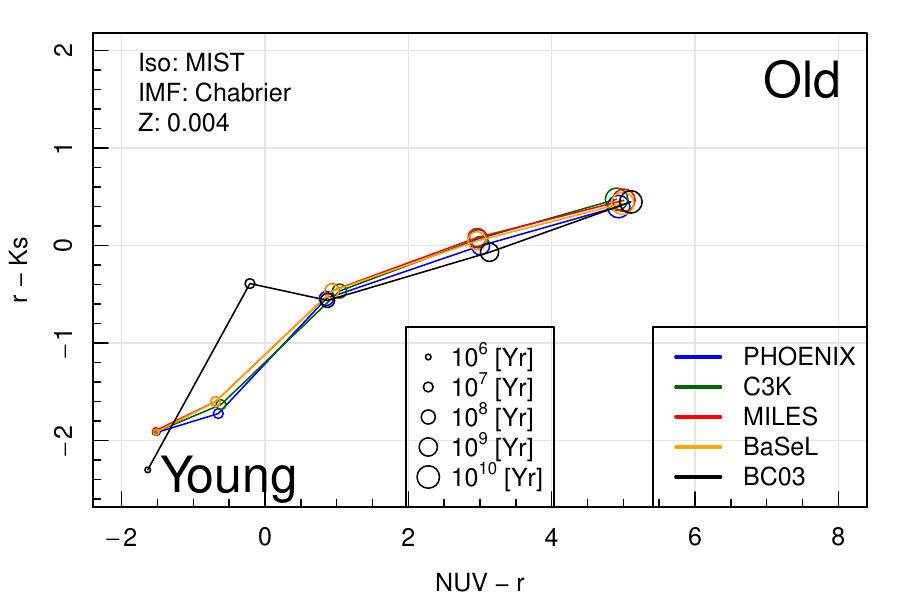}
\caption{Colour-colour comparison of different stellar spectra as a function of age for $Z=0.004$ using MIST isochrones. See Figure \ref{fig:atmos_colour} for additional context.}
\end{figure}

\begin{figure}
\includegraphics[width=\columnwidth]{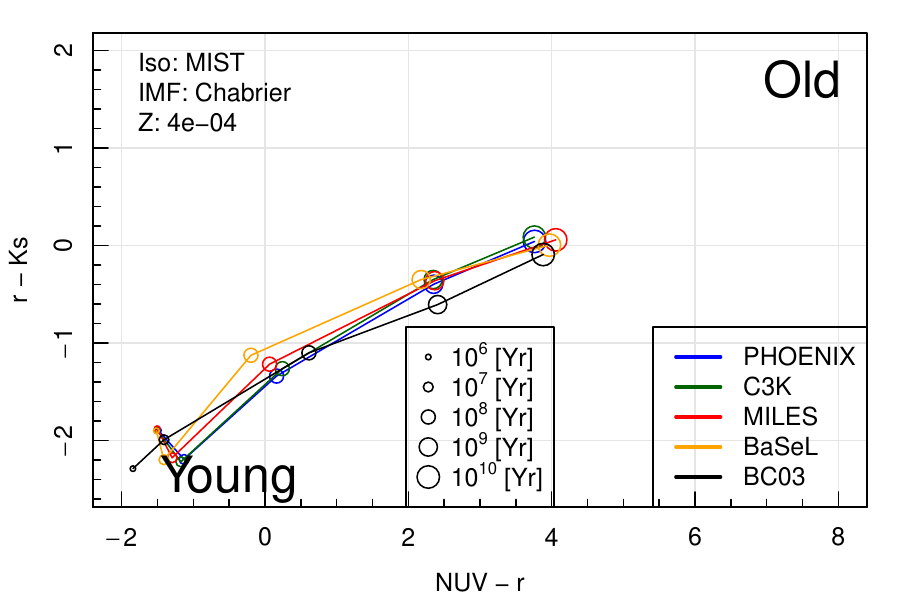}
\caption{Colour-colour comparison of different stellar spectra as a function of age for $Z=0.0004$ using MIST isochrones. See Figure \ref{fig:atmos_colour} for additional context.}
\end{figure}

\begin{figure}
\includegraphics[width=\columnwidth]{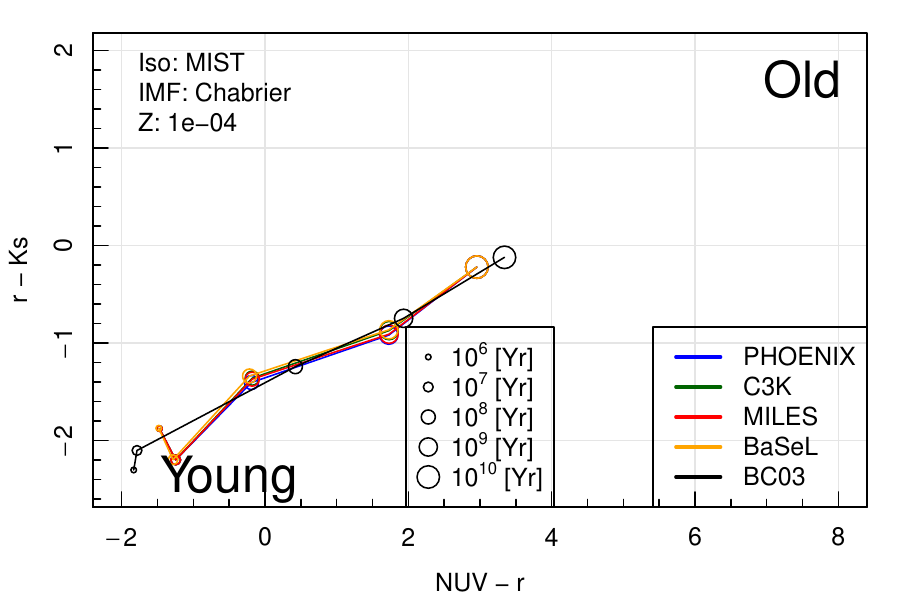}
\caption{Colour-colour comparison of different stellar spectra as a function of age for $Z=0.0001$ using MIST isochrones. See Figure \ref{fig:atmos_colour} for additional context.}
\end{figure}

\begin{figure}
\includegraphics[width=\columnwidth]{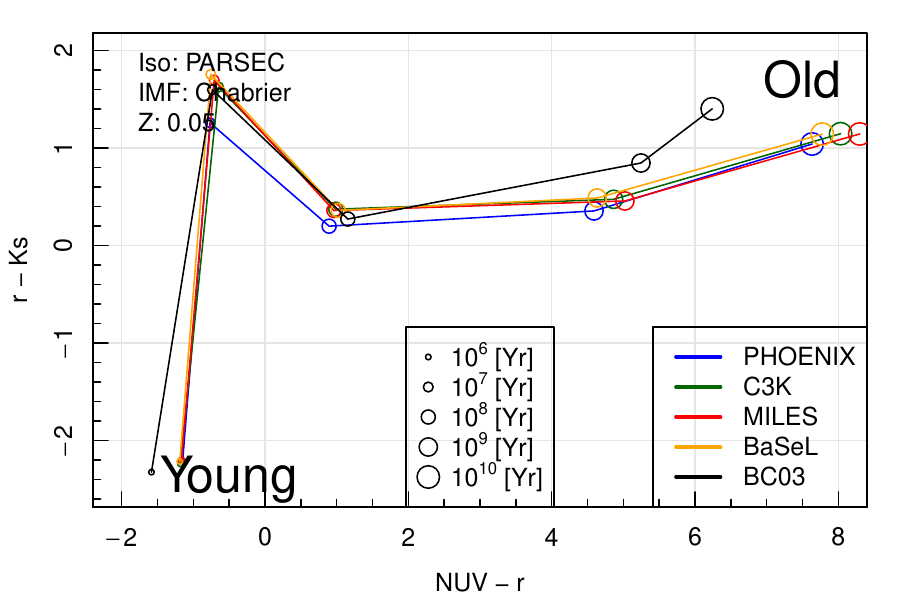}
\caption{Colour-colour comparison of different stellar spectra as a function of age for $Z=0.05$ using PARSEC isochrones. See Figure \ref{fig:atmos_colour} for additional context.}
\end{figure}

\begin{figure}
\includegraphics[width=\columnwidth]{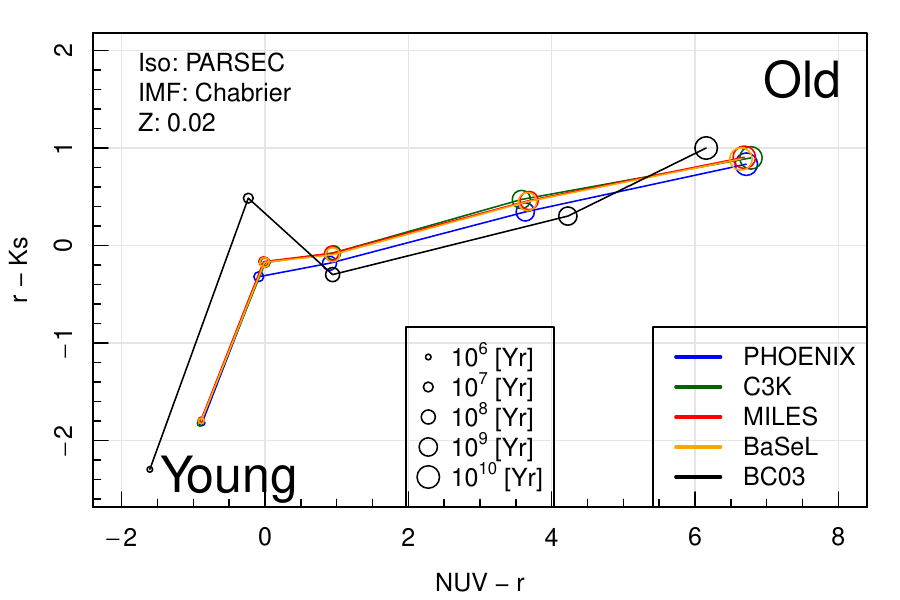}
\caption{Colour-colour comparison of different stellar spectra as a function of age for $Z=0.02$ (solar metallicity) using PARSEC isochrones. See Figure \ref{fig:atmos_colour} for additional context.}
\end{figure}

\begin{figure}
\includegraphics[width=\columnwidth]{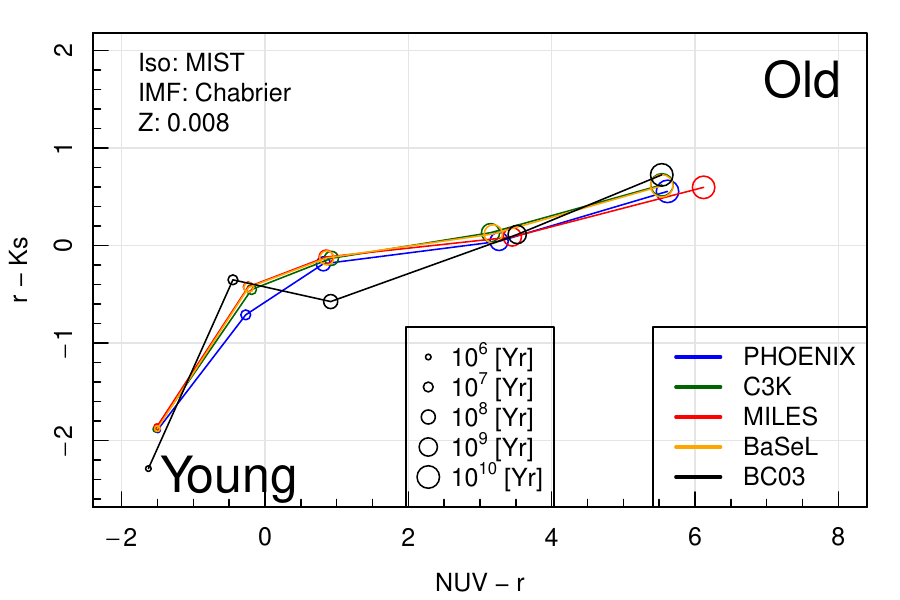}
\caption{Colour-colour comparison of different stellar spectra as a function of age for $Z=0.008$ using PARSEC isochrones. See Figure \ref{fig:atmos_colour} for additional context.}
\end{figure}

\begin{figure}
\includegraphics[width=\columnwidth]{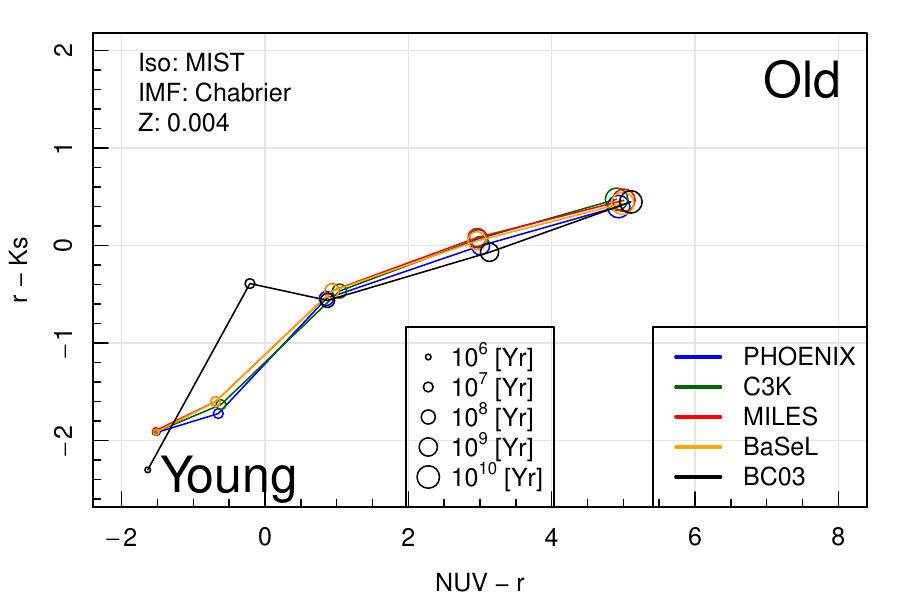}
\caption{Colour-colour comparison of different stellar spectra as a function of age for $Z=0.004$ using PARSEC isochrones. See Figure \ref{fig:atmos_colour} for additional context.}
\end{figure}

\begin{figure}
\includegraphics[width=\columnwidth]{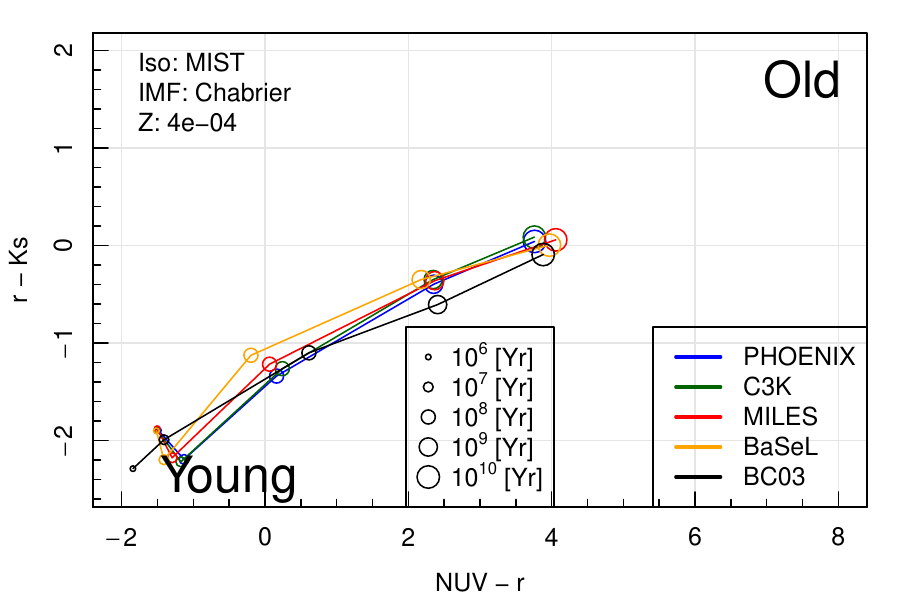}
\caption{Colour-colour comparison of different stellar spectra as a function of age for $Z=0.0004$ using PARSEC isochrones. See Figure \ref{fig:atmos_colour} for additional context.}
\end{figure}

\begin{figure}
\includegraphics[width=\columnwidth]{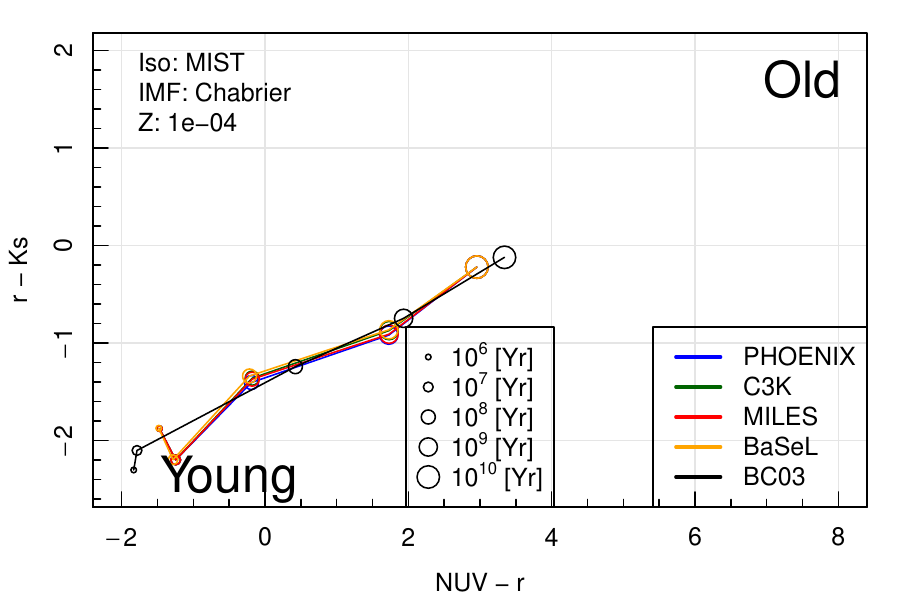}
\caption{Colour-colour comparison of different stellar spectra as a function of age for $Z=0.0001$ using PARSEC isochrones. See Figure \ref{fig:atmos_colour} for additional context.}
\end{figure}

\begin{figure}
\includegraphics[width=\columnwidth]{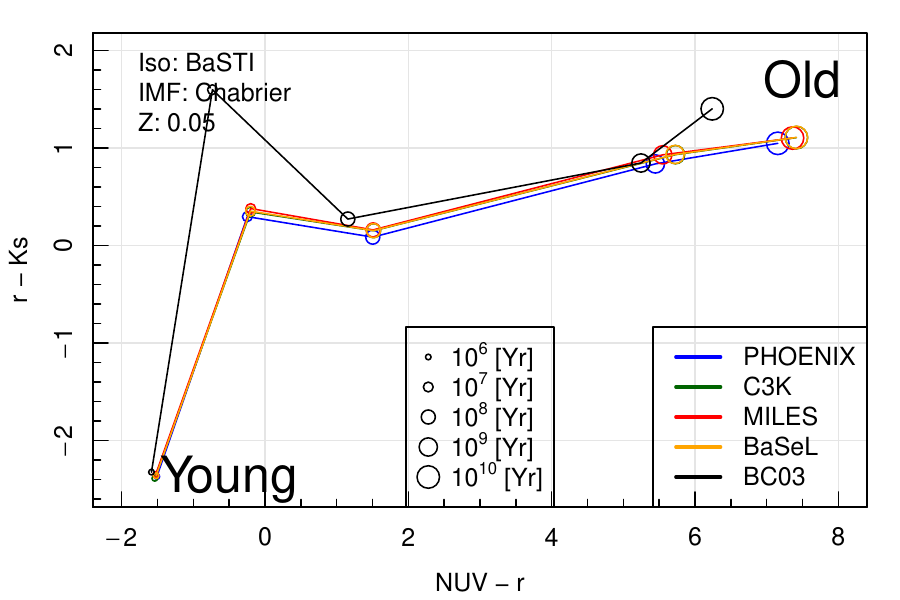}
\caption{Colour-colour comparison of different stellar spectra as a function of age for $Z=0.05$ using BaSTI isochrones. See Figure \ref{fig:atmos_colour} for additional context.}
\end{figure}

\begin{figure}
\includegraphics[width=\columnwidth]{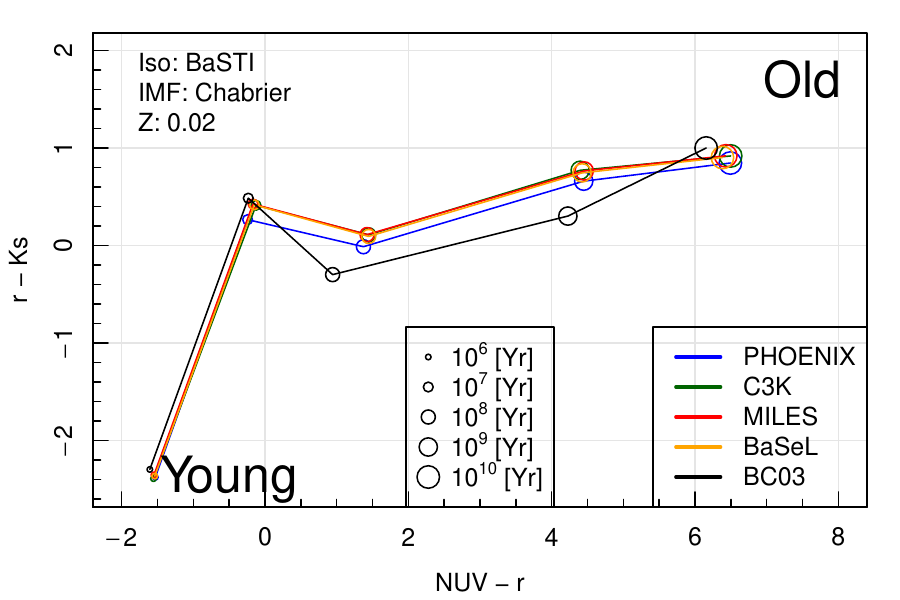}
\caption{Colour-colour comparison of different stellar spectra as a function of age for $Z=0.02$ (solar metallicity) using BaSTI isochrones. See Figure \ref{fig:atmos_colour} for additional context.}
\end{figure}

\begin{figure}
\includegraphics[width=\columnwidth]{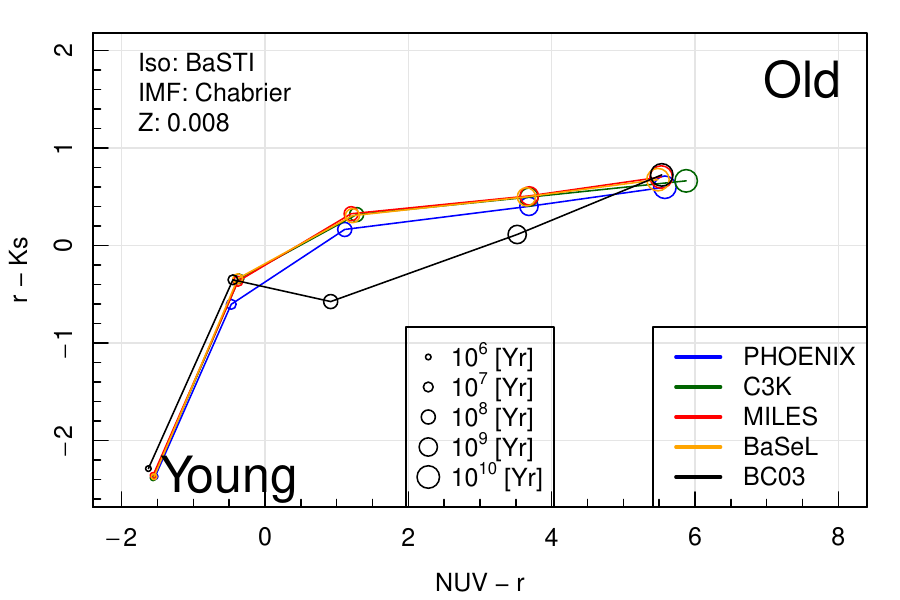}
\caption{Colour-colour comparison of different stellar spectra as a function of age for $Z=0.008$ using BaSTI isochrones. See Figure \ref{fig:atmos_colour} for additional context.}
\end{figure}

\begin{figure}
\includegraphics[width=\columnwidth]{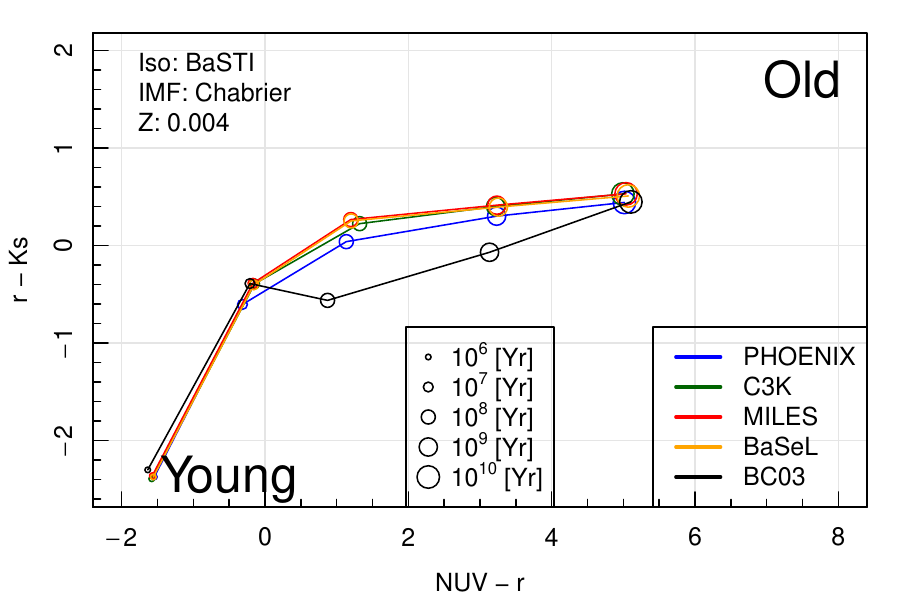}
\caption{Colour-colour comparison of different stellar spectra as a function of age for $Z=0.004$ using BaSTI isochrones. See Figure \ref{fig:atmos_colour} for additional context.}
\end{figure}

\begin{figure}
\includegraphics[width=\columnwidth]{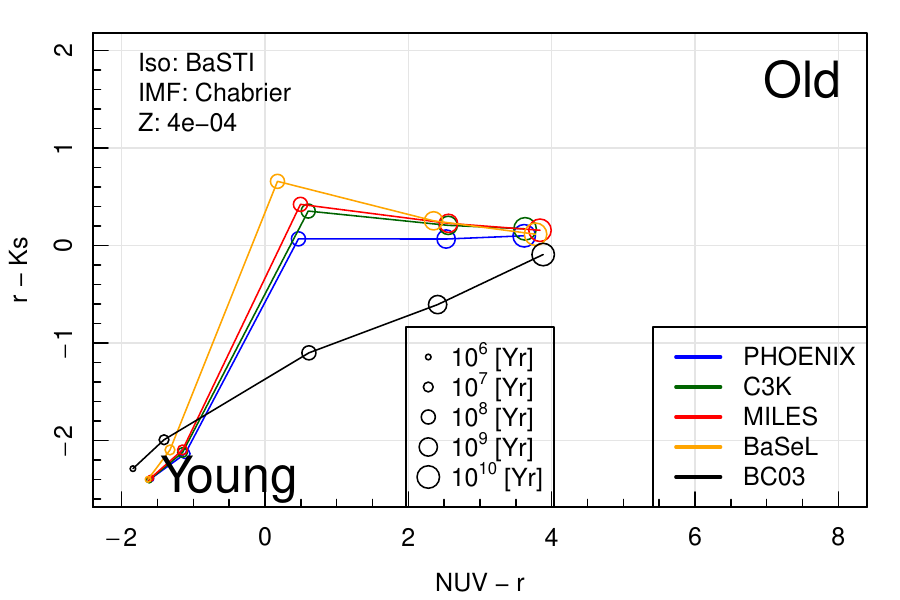}
\caption{Colour-colour comparison of different stellar spectra as a function of age for $Z=0.0004$ using BaSTI isochrones. See Figure \ref{fig:atmos_colour} for additional context.}
\end{figure}

\begin{figure}
\includegraphics[width=\columnwidth]{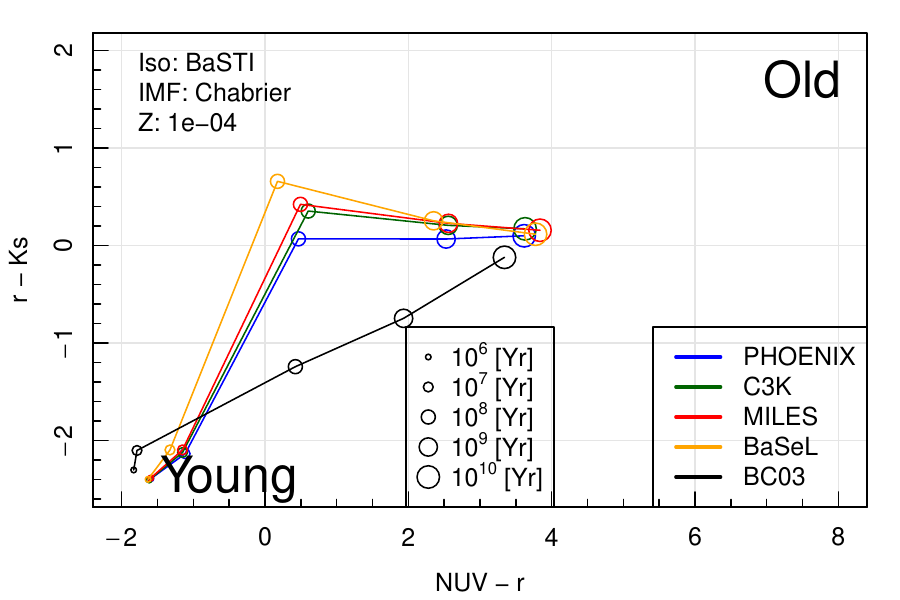}
\caption{Colour-colour comparison of different stellar spectra as a function of age for $Z=0.0001$ using BaSTI isochrones. See Figure \ref{fig:atmos_colour} for additional context.}
\end{figure}

\section{Additional Spectra Figures}
\label{sec:SSP_extra}

A version of Figure \ref{fig:comp_spec_Z0} zoomed in to the region covering 1,000 -- 10,000 \AA. For clarity almost all labels have been removed, and readers should refer to legends in Figure \ref{fig:comp_spec_Z0} to identify the relevant SSPs etc.

\begin{figure*}
\begin{center}
\includegraphics[width=8.5cm]{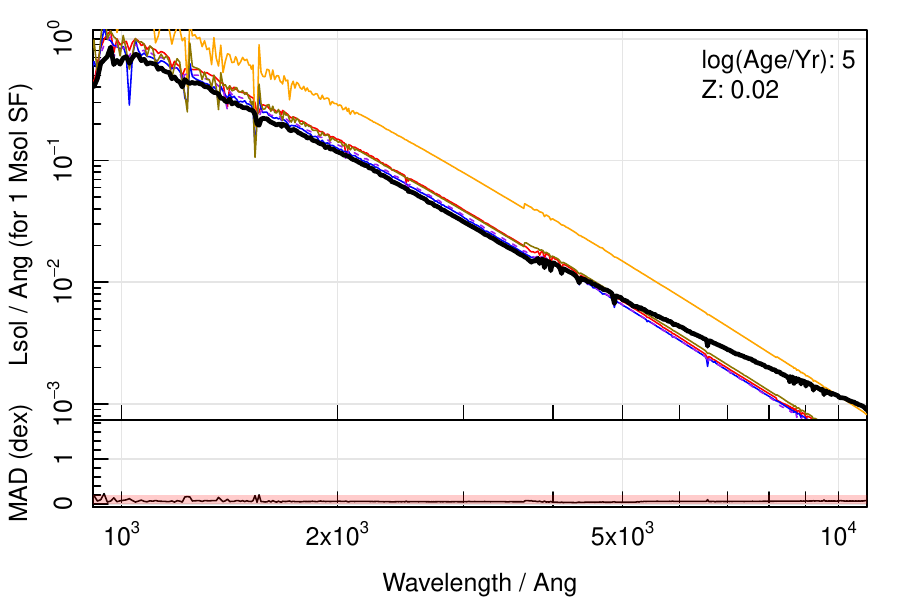}
\includegraphics[width=8.5cm]{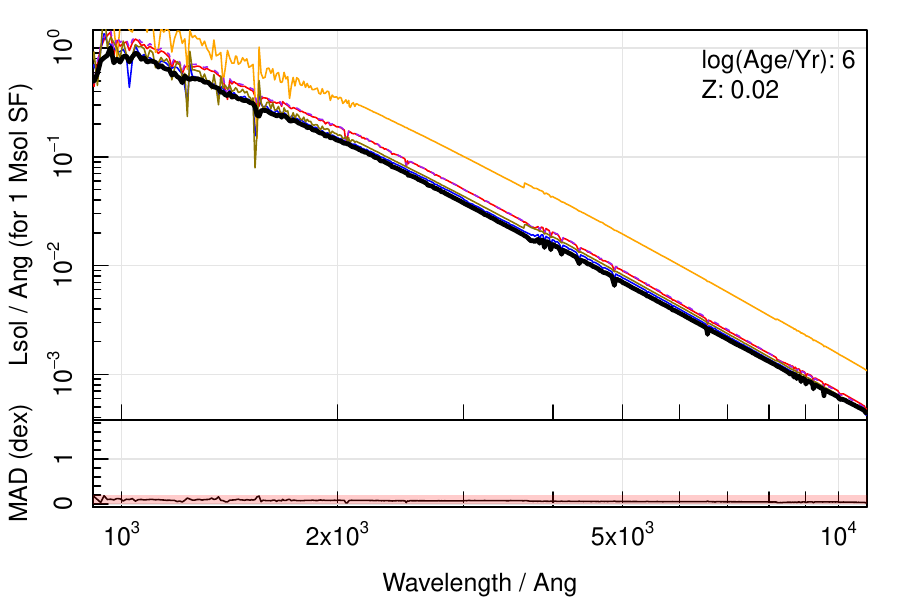}
\\
\includegraphics[width=8.5cm]{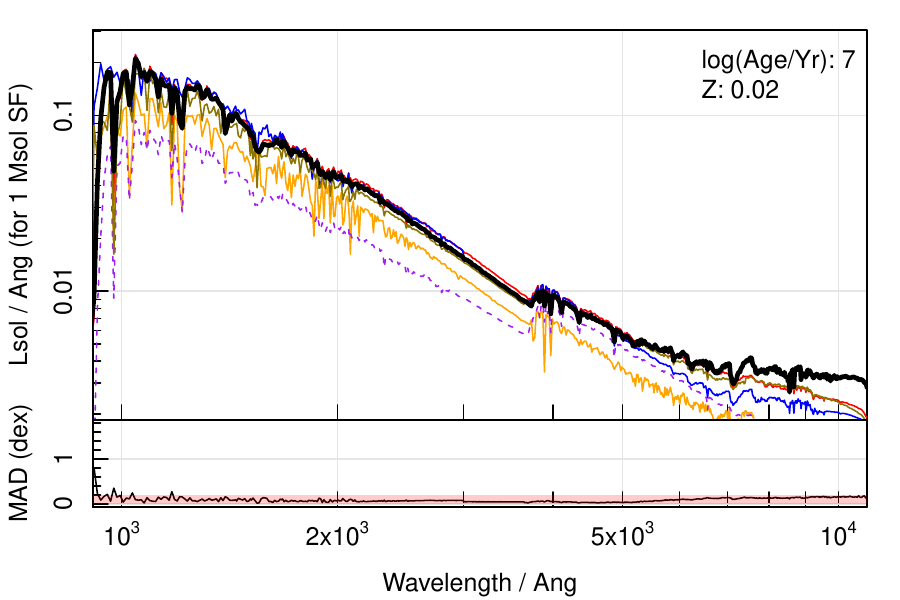}
\includegraphics[width=8.5cm]{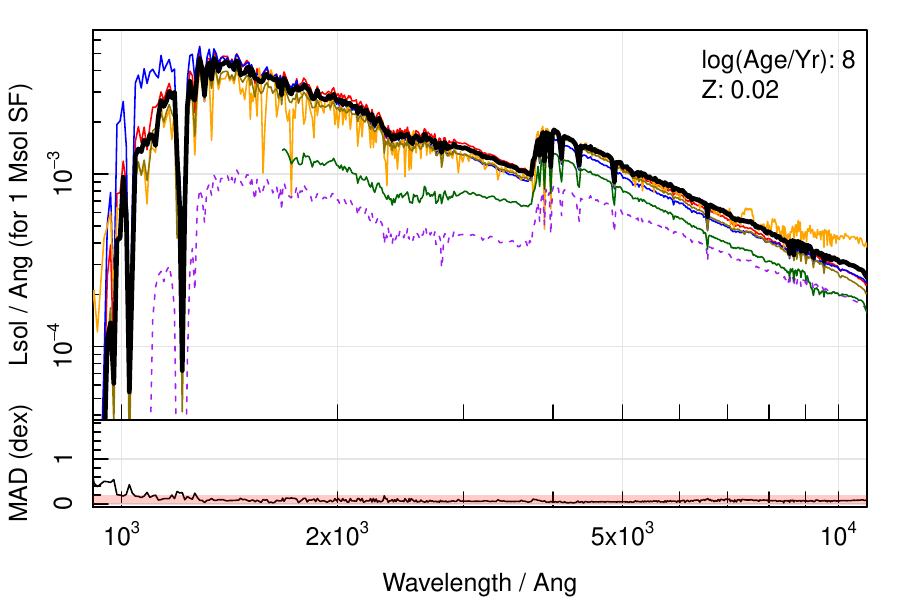}
\\
\includegraphics[width=8.5cm]{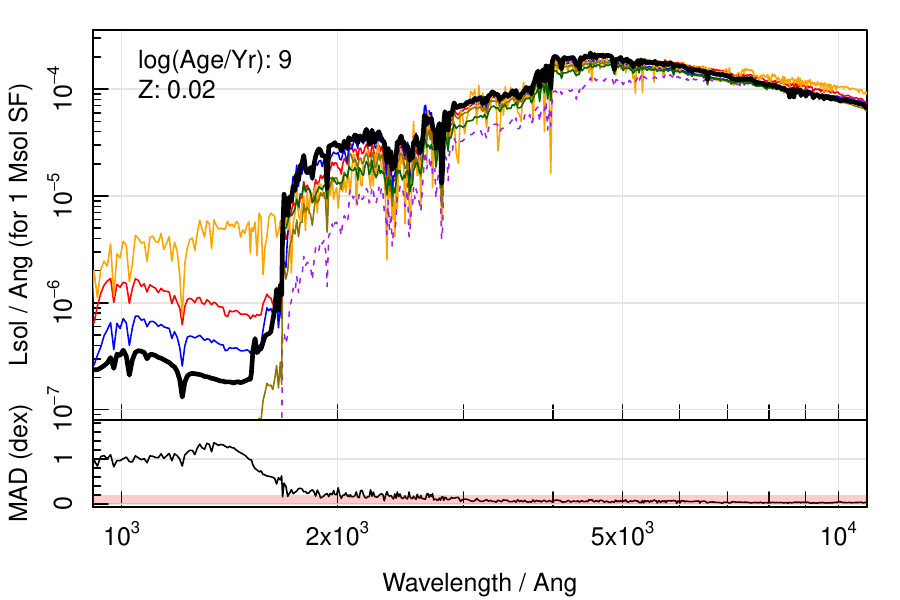}
\includegraphics[width=8.5cm]{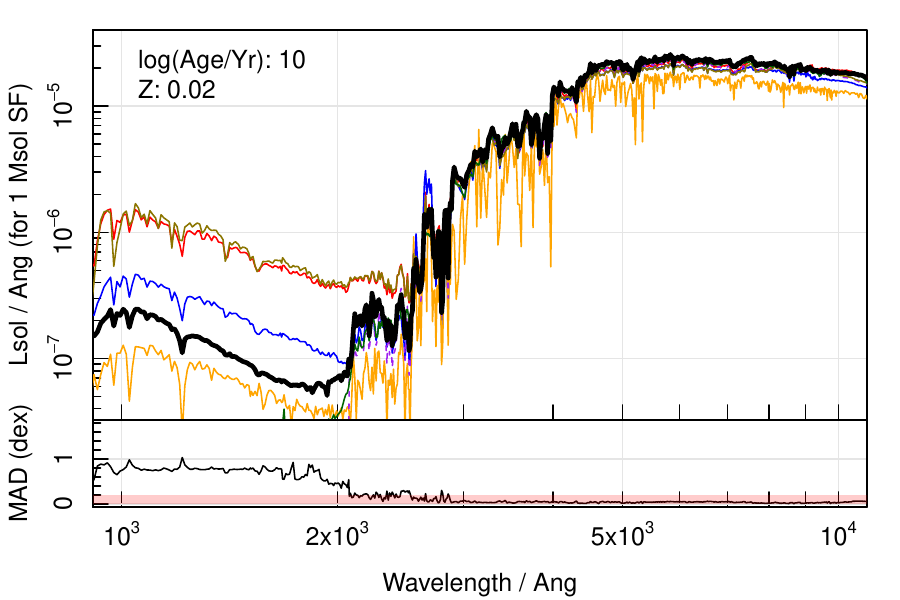}
\\
\caption{Comparison of \Zsol{} spectra at different stellar population ages for different SSPs. The ages presented are log spaced with 1 dex intervals, covering 100,000 yr through to 10 Gyr.}
\label{fig:comp_spec_Z0_zoom}
\end{center}
\end{figure*}

Additional versions of Figure \ref{fig:comp_spec_Z0} for different metallicity combinations. In cases where metallicities do not exist on the exact gridding specified in the Figure, we use the nearest match in logarithmic space.

\begin{figure*}
\begin{center}
\includegraphics[width=8.5cm]{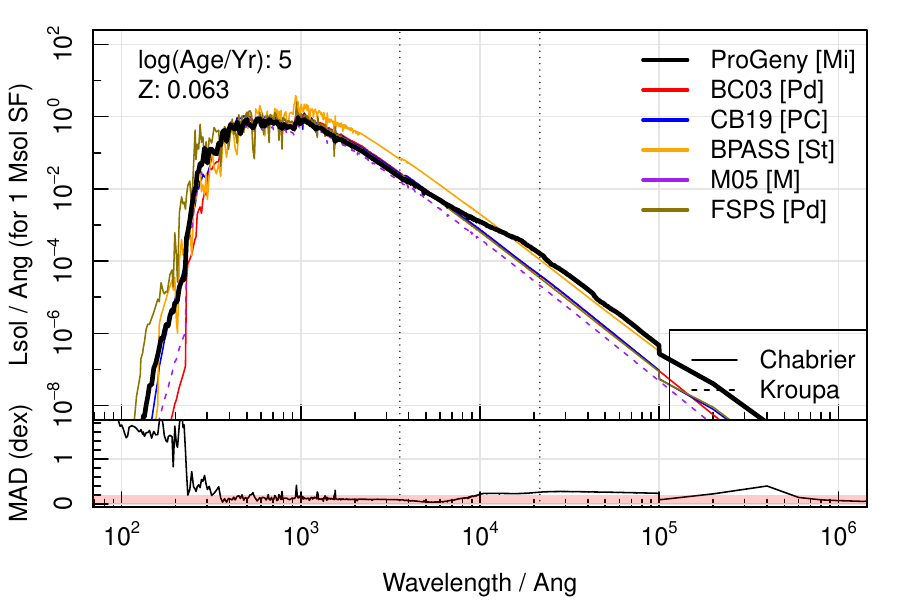}
\includegraphics[width=8.5cm]{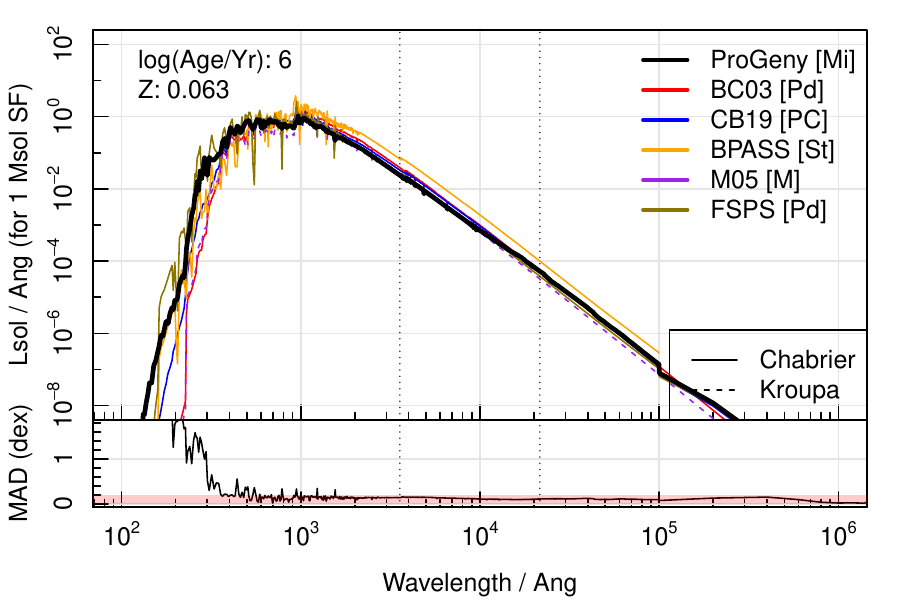}
\\
\includegraphics[width=8.5cm]{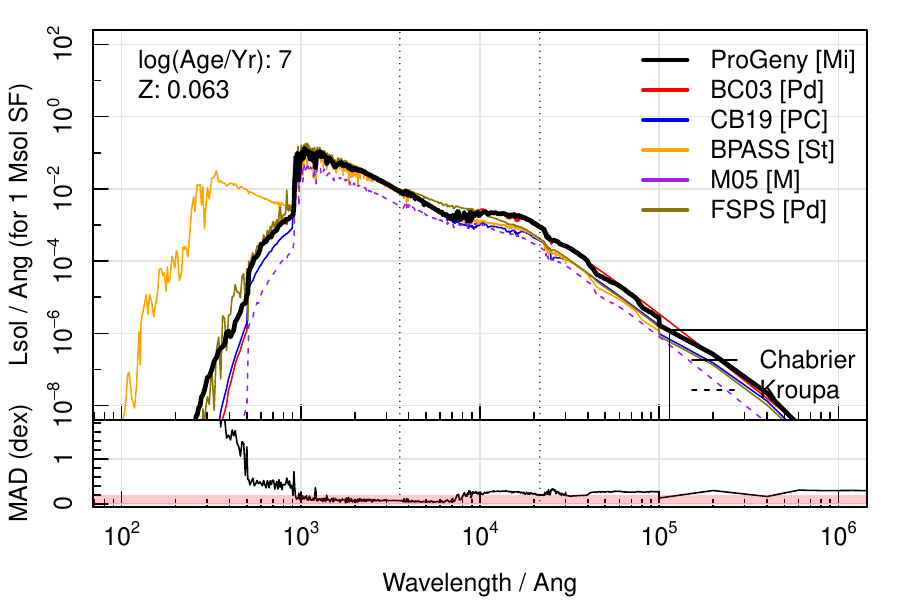}
\includegraphics[width=8.5cm]{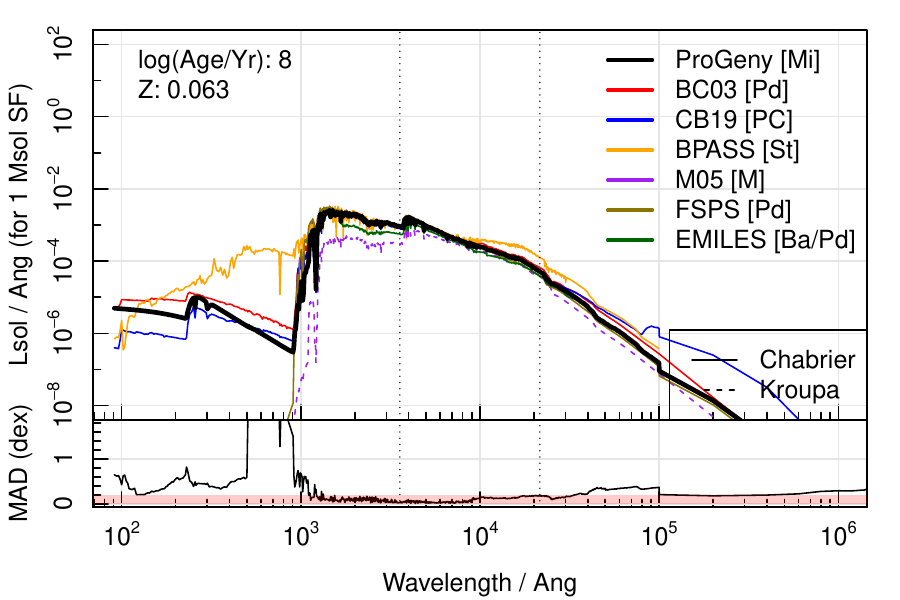}
\\
\includegraphics[width=8.5cm]{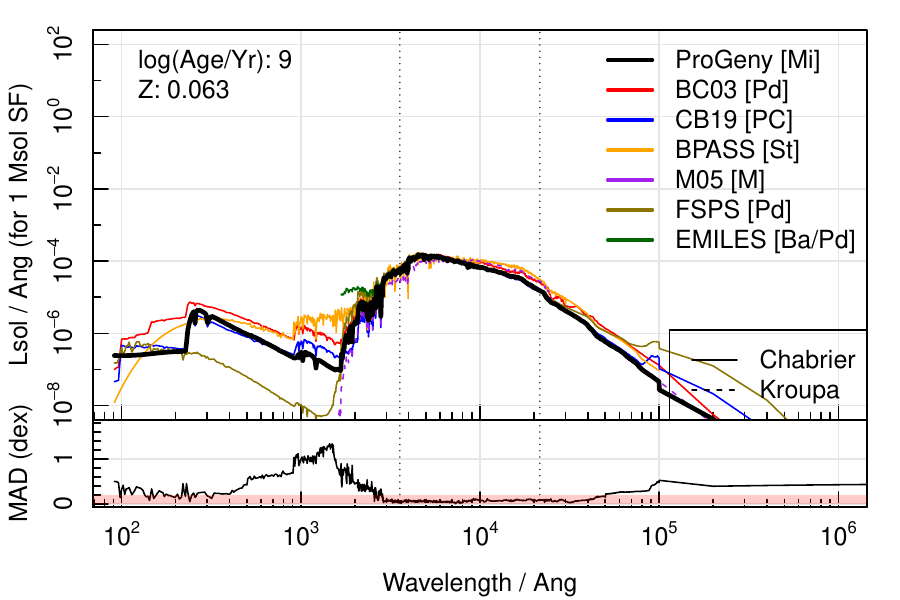}
\includegraphics[width=8.5cm]{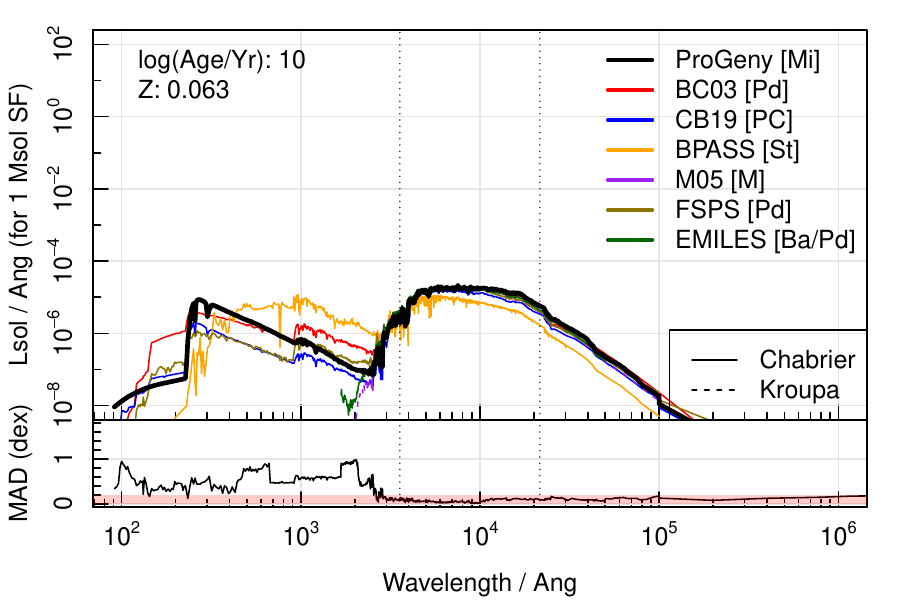}
\\
\caption{Comparison of \Zsol{} +0.5 dex metallicity spectra at different stellar population ages for different SSPs. See Figure \ref{fig:comp_spec_Z0} for additional context.}
\end{center}
\end{figure*}

\begin{figure*}
\begin{center}
\includegraphics[width=8.5cm]{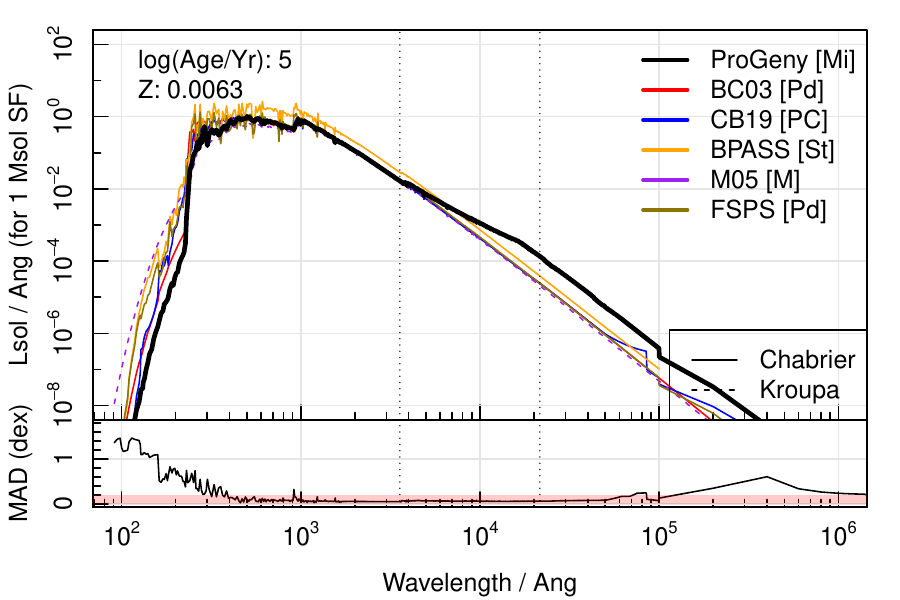}
\includegraphics[width=8.5cm]{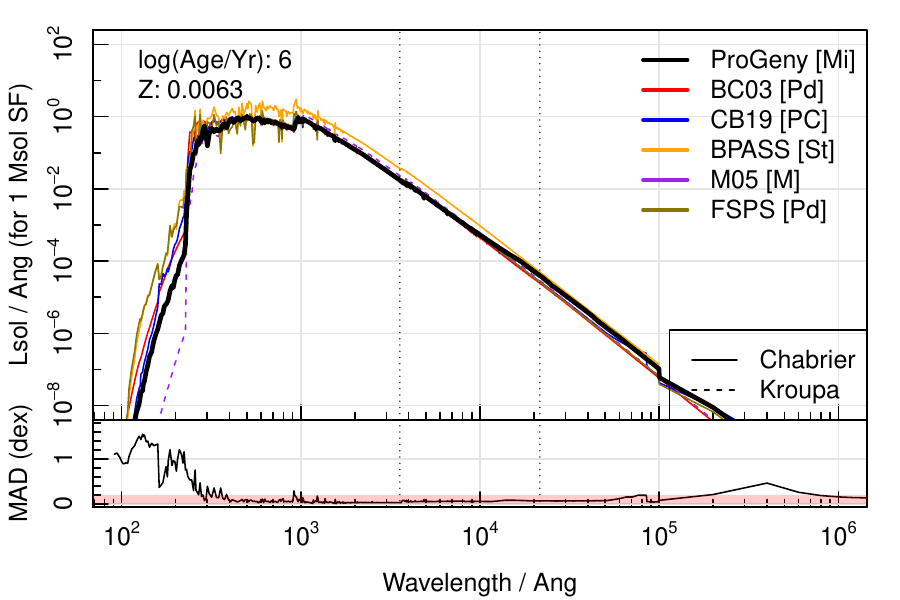}
\\
\includegraphics[width=8.5cm]{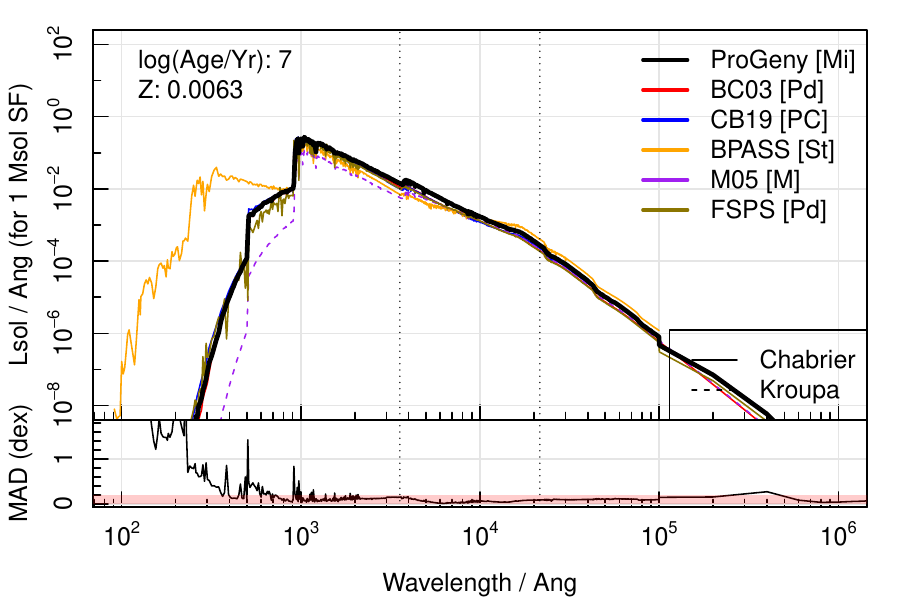}
\includegraphics[width=8.5cm]{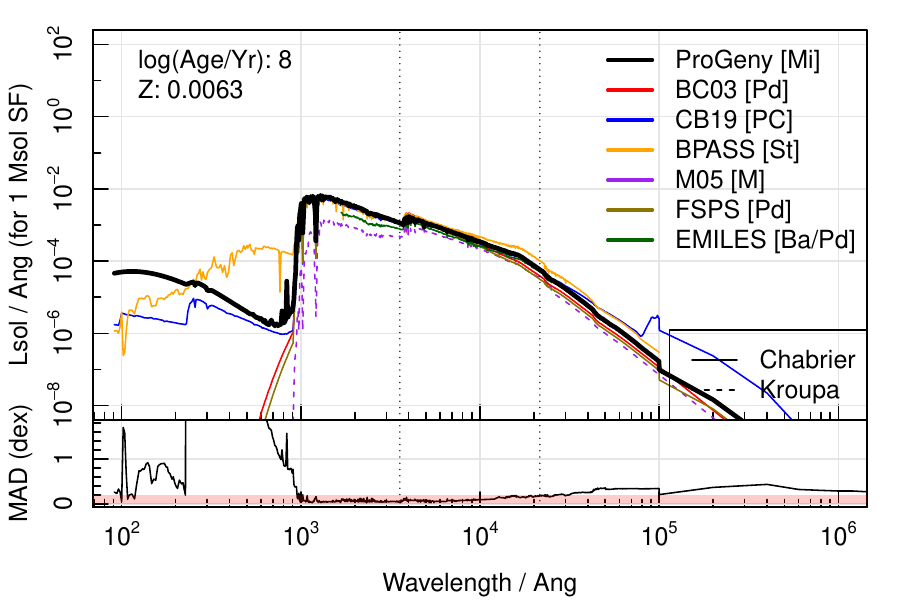}
\\
\includegraphics[width=8.5cm]{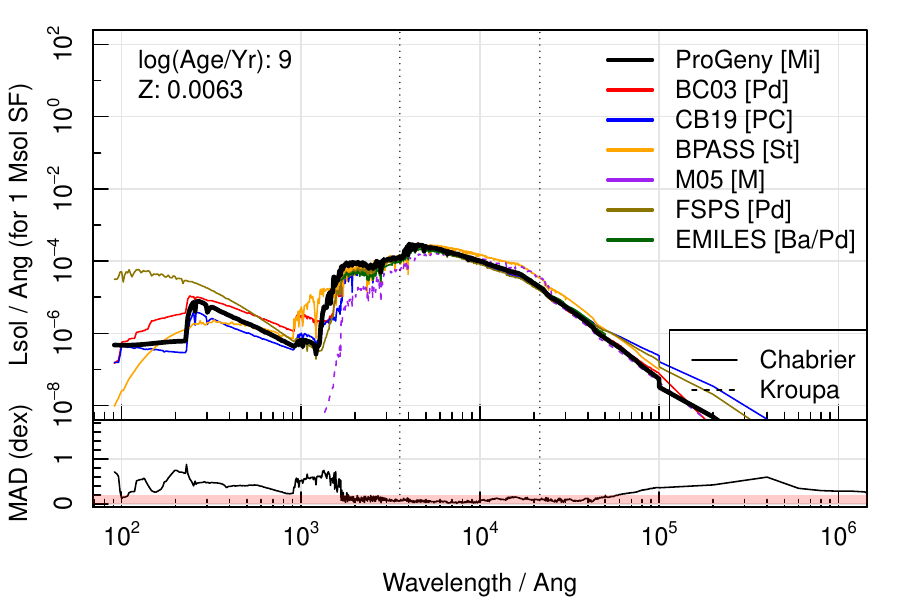}
\includegraphics[width=8.5cm]{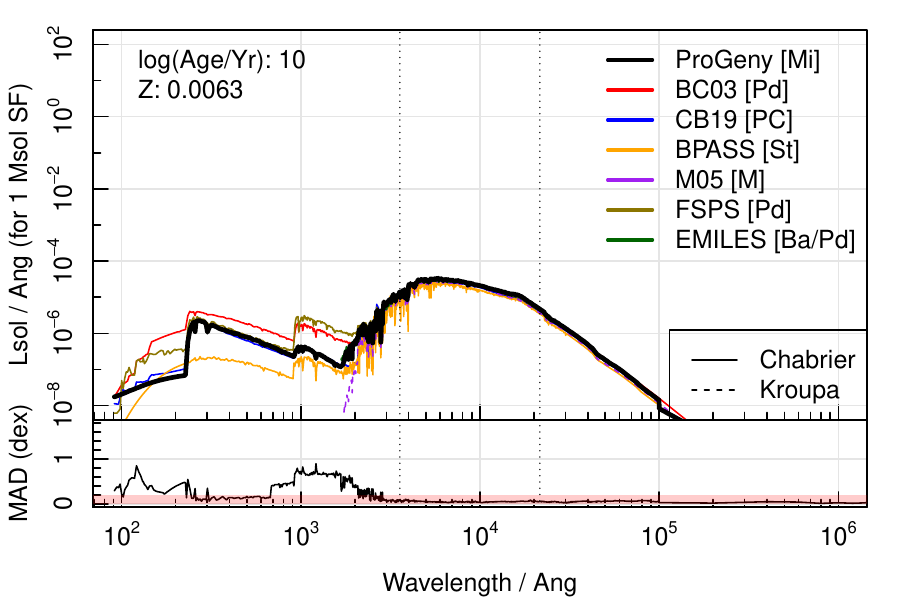}
\\
\caption{Comparison of \Zsol{} -0.5 dex metallicity spectra at different stellar population ages for different SSPs. See Figure \ref{fig:comp_spec_Z0} for additional context.}
\end{center}
\end{figure*}

\begin{figure*}
\begin{center}
\includegraphics[width=8.5cm]{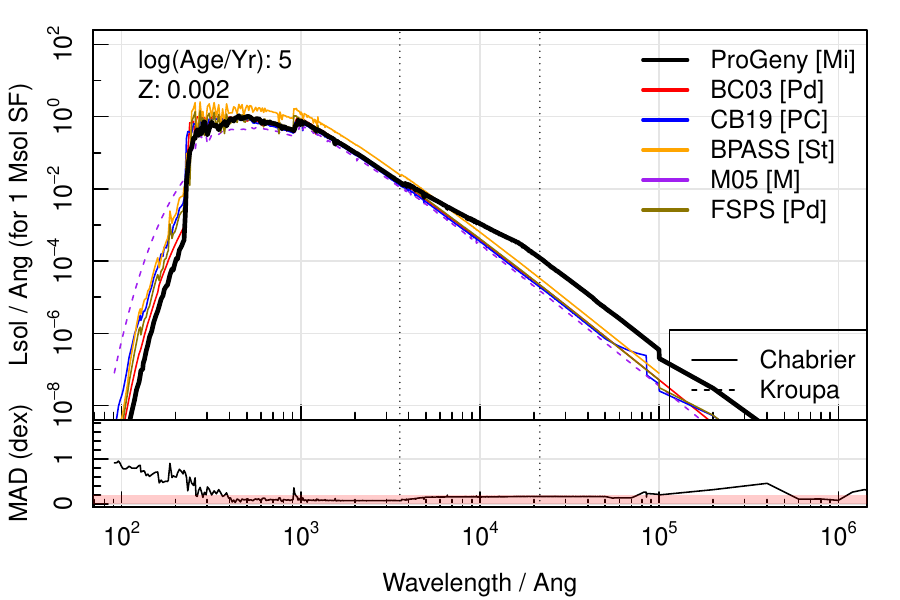}
\includegraphics[width=8.5cm]{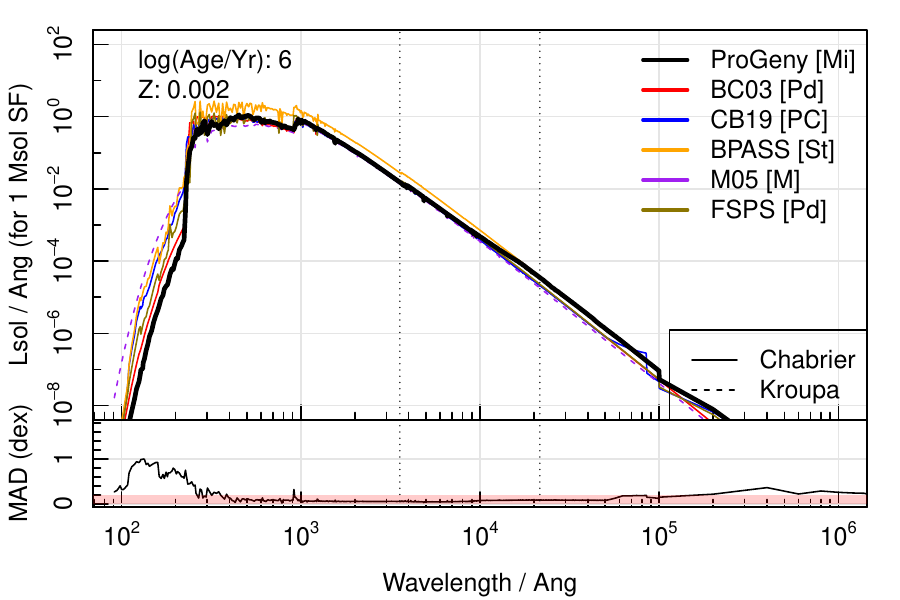}
\\
\includegraphics[width=8.5cm]{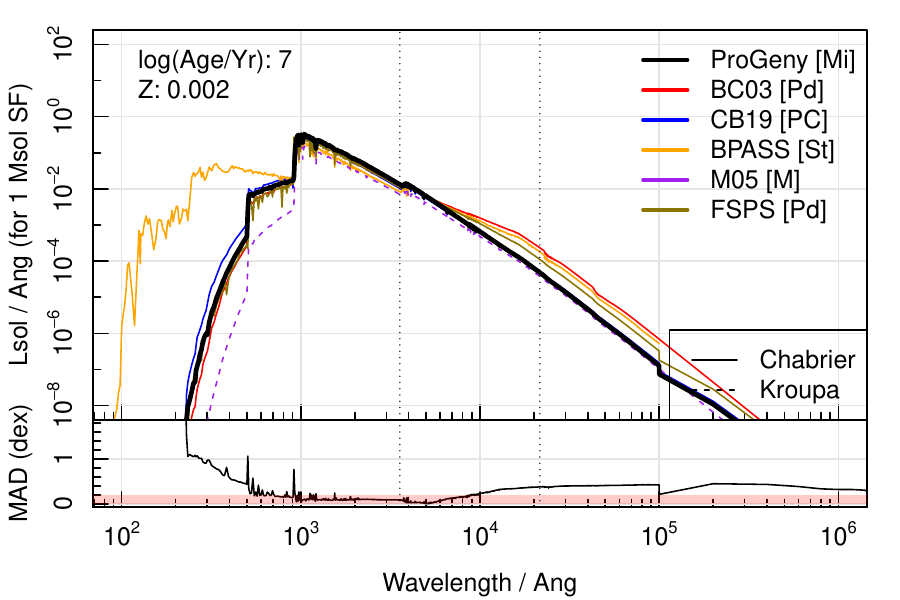}
\includegraphics[width=8.5cm]{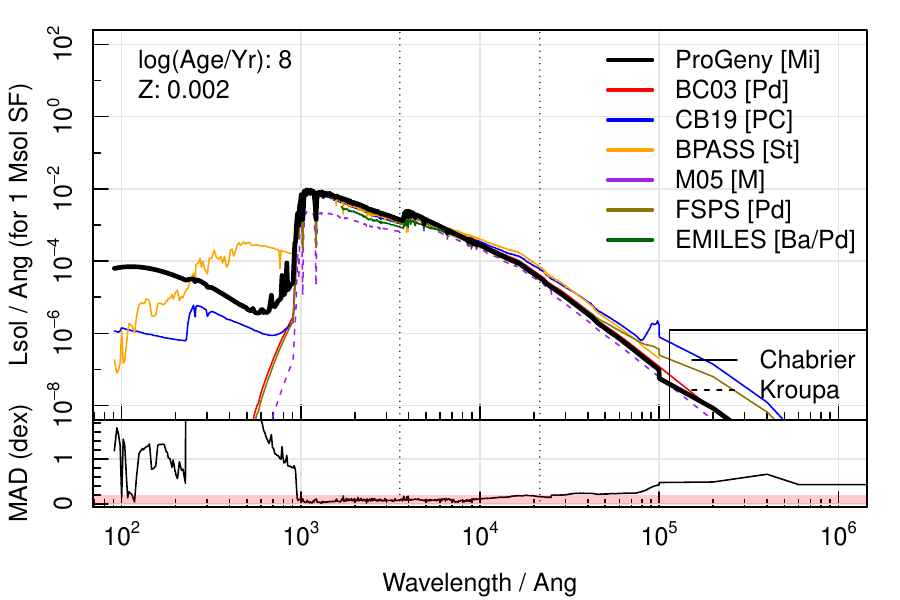}
\\
\includegraphics[width=8.5cm]{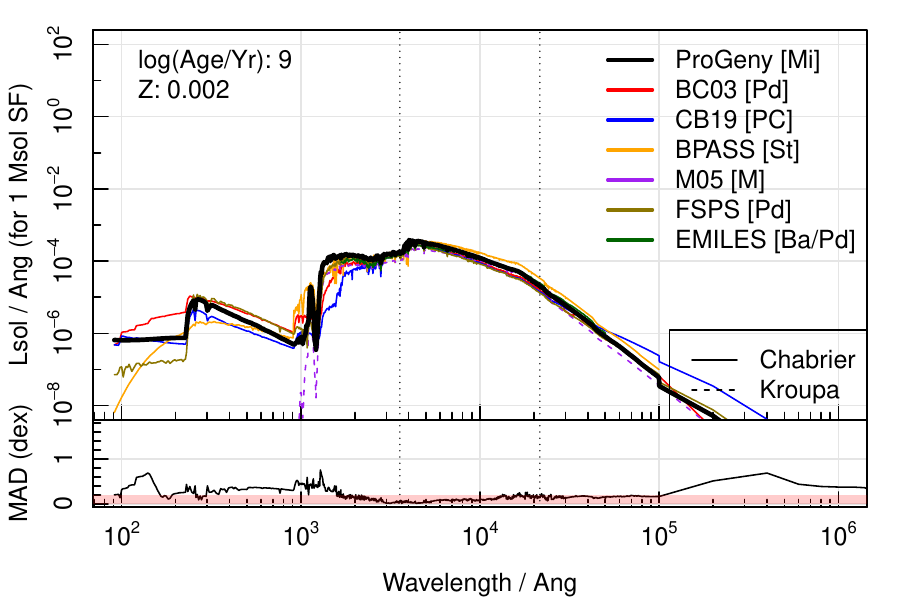}
\includegraphics[width=8.5cm]{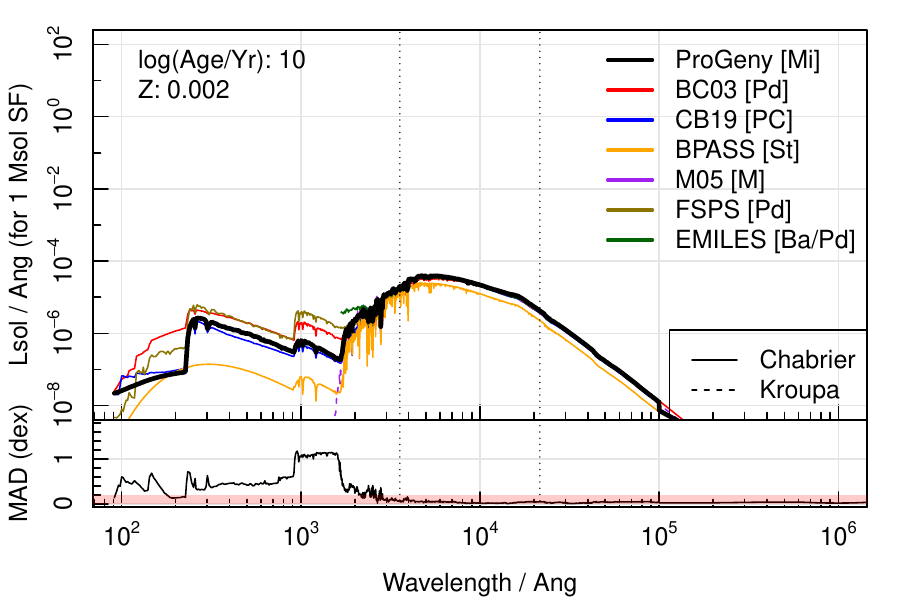}
\\
\caption{Comparison of \Zsol{} -1 dex metallicity spectra at different stellar population ages for different SSPs. See Figure \ref{fig:comp_spec_Z0} for additional context.}
\end{center}
\end{figure*}

\begin{figure*}
\begin{center}
\includegraphics[width=8.5cm]{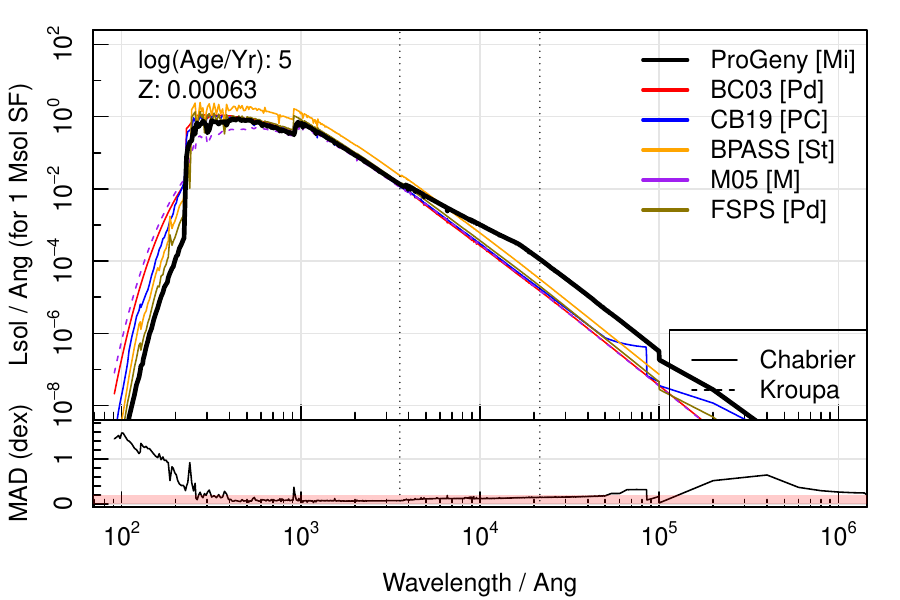}
\includegraphics[width=8.5cm]{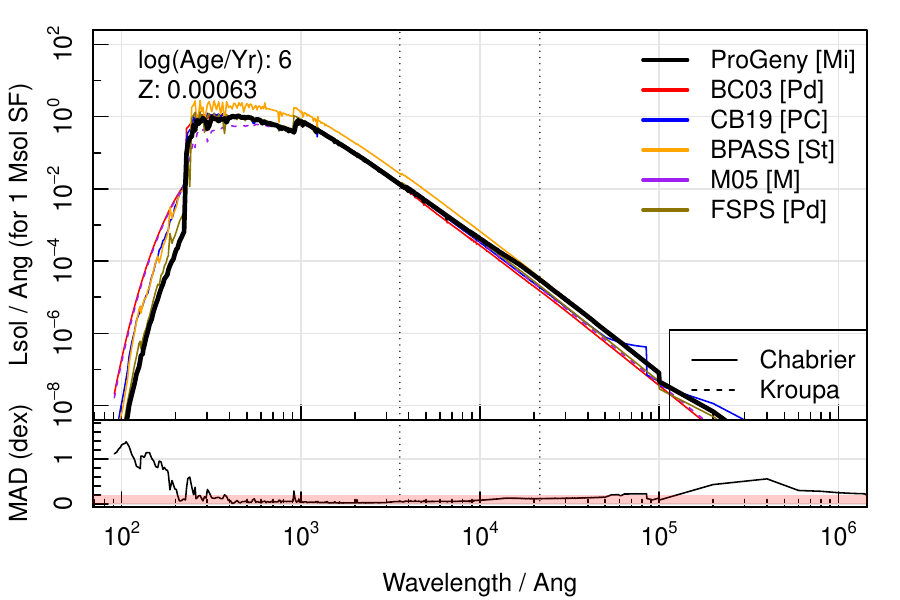}
\\
\includegraphics[width=8.5cm]{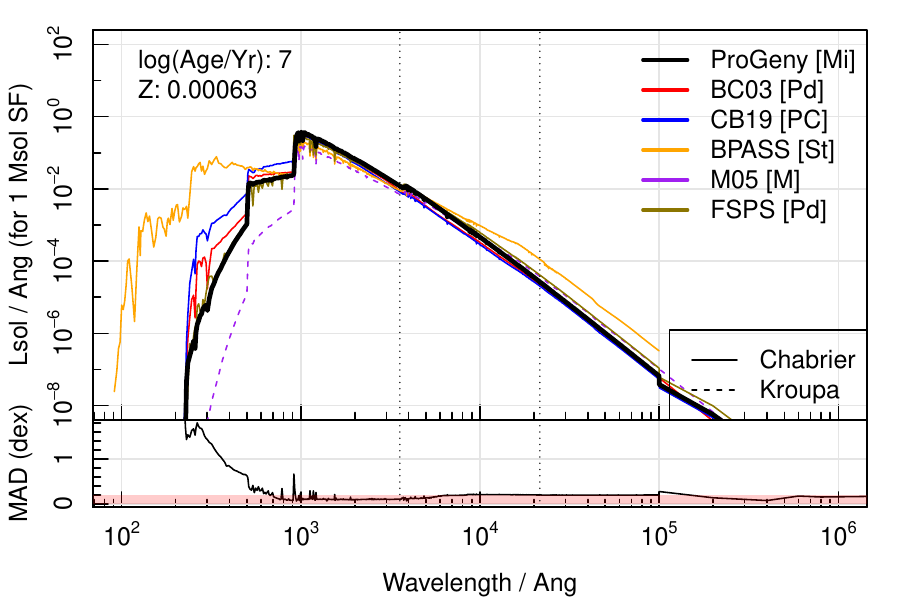}
\includegraphics[width=8.5cm]{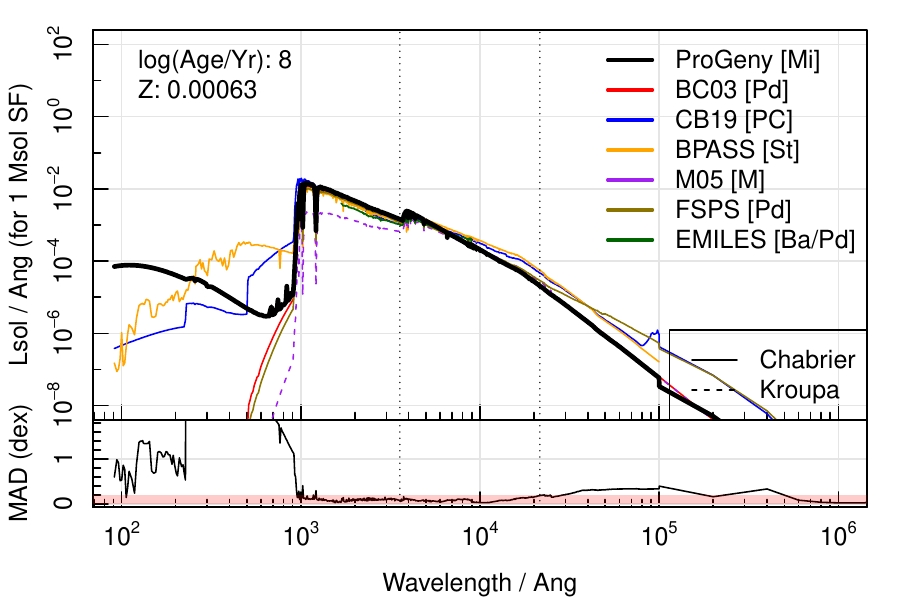}
\\
\includegraphics[width=8.5cm]{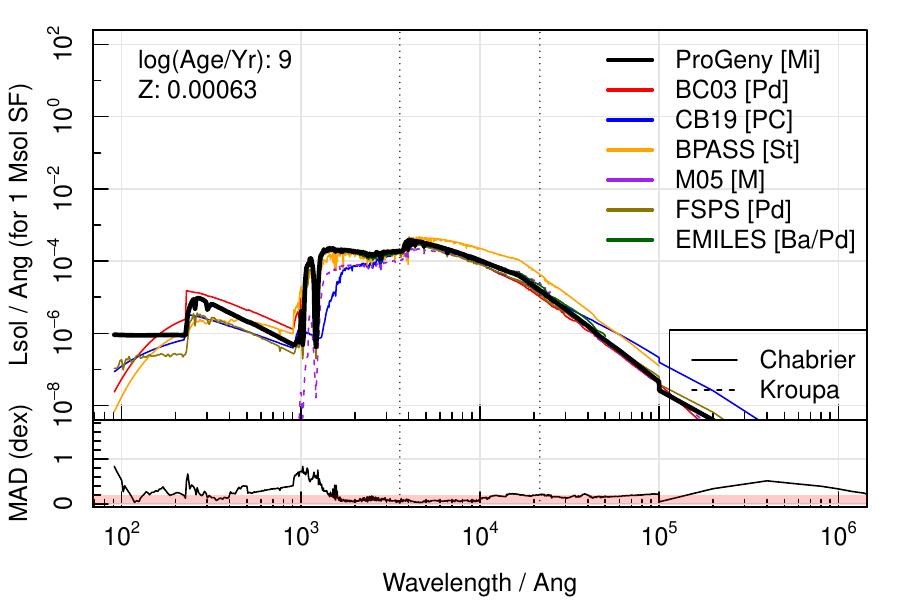}
\includegraphics[width=8.5cm]{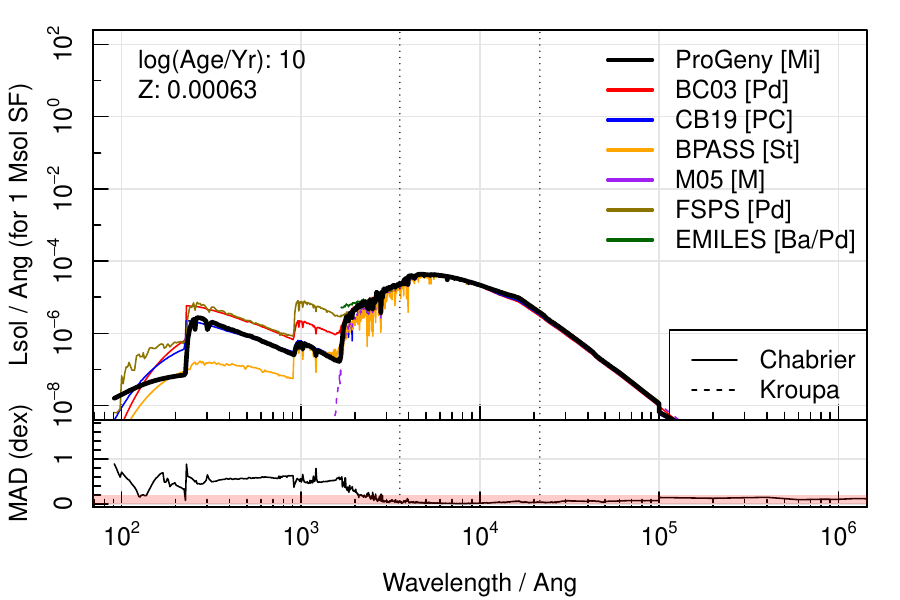}
\\
\caption{Comparison of \Zsol{} -1.5 dex metallicity spectra at different stellar population ages for different SSPs. See Figure \ref{fig:comp_spec_Z0} for additional context.}
\end{center}
\end{figure*}

\section{Glossary}
\label{sec:glossary}

For easy reference, common acronyms (including surveys and telescopes) and software are expanded and described in Tables \ref{tab:acronym} and \ref{tab:software}.

\begin{table*}
\begin{center}
\caption{Reference for common astronomy acronyms, surveys and telescopes used throughout this paper.}
\begin{tabular}{|l|l|l|l|}

Acronym & Expanded & Explanation & Reference (if relevant) \\

\hline

4MOST & 4m Multi-Object Spectroscopic Telescope & 2400 multi object spectrograph upgrade to VISTA & \citet{2019Msngr.175....3D} \\
AGB & Asymptotic Giant Branch & Phase of stellar evolution where the photosphere rapidly grows in size & NA \\
CHeB & Core Helium Burning & Phase of stellar evolution after RGB where hot helium burning occurs & NA \\
COSMOS & Cosmic Evolution Survey & Multi-facility survey targeting a $\sim$2 square degree field & \citet{2007ApJS..172..196K} \\
DEVILS & Deep Extragalactic VIsible Legacy Survey & Major spectroscopic survey on the Anglo Australian Telescope & \citet{2018MNRAS.480..768D} \\
FIR & Far Infrared & Wavelength range covering approximately 25--600 micron & NA \\
eAGB & early Asymptotic Giant Branch & Phase of stellar evolution just before AGB & NA \\
GALEX & Galaxy Evolutionary Explorer & UV space telescope & \citet{2007ApJS..173..682M} \\
GAMA & Galaxy And Mass Assembly & Major spectroscopic survey on the Anglo Australian Telescope & \citet{2011MNRAS.413..971D} \\
HB & Horizontal Branch & Phase of stellar evolution after RGB where hot helium burning occurs & NA \\
IMF & Initial Mass Function & Initial distribution of stellar mass in a burst of star formation & NA \\
logG & log$_{10}$(g) & logarithmic surface gravity (g) in CGS units ($\text{cm}/\text{s}^2$) & NA \\
logZ & log$_{10}$(Z/\Zsol) & logarithmic metallicity relative to solar (i.e. logZ = 0 is solar) & NA \\
NIR & Near Infrared & Wavelength range covering approximately 800--2500 nm & NA \\
MIR & Mid Infrared & Wavelength range covering approximately 2.5--25 micron & NA \\
RGB & Red Giant Branch & Phase of stellar evolution where the photosphere grows and cools & NA \\
SED & Spectral Energy Distribution & Electromagnetic radiation of an astronomical source & \citet{2020MNRAS.495..905R} \\
SPL & Stellar Population Library & A family of related SSPs (i.e.\ created with similar methodology) & NA \\
SSP & Simple/Single Stellar Population & Spectra as a function of stellar population age and metallicity (usually) & NA \\
TP-AGB & Thermally Pulsating Asymptotic Giant Branch & Phase of stellar evolution where the photosphere rapidly pulsates & NA \\
UV & Ultra Violet & Wavelength range covering approximately 100--400 nm & NA \\
VISTA & \makecell[l]{Visible and Infrared Survey \\ Telescope for Astronomy} & 4m NIR telescope located at Paranal Observatory (Chile) & NA \\
WAVES & Wide Area VISTA Extragalactic Survey & Major spectroscopic survey on 4MOST / VISTA & \citet{2019Msngr.175...46D} \\
WR & Wolf-Rayet star & Phase of stellar evolution in massive stars producing high temperatures & NA \\
ZAM & Zero Age Main sequence & The onset of main sequence star formation in a burst & NA \\

\end{tabular}
\end{center}
\label{tab:acronym}
\end{table*}%

\begin{table*}
\begin{center}
\caption{Quick reference for common astronomy software used throughout this paper.}
\begin{tabular}{|l|l|l|l|}

Name & About & Reference & Available \\

\hline

BaSTI & Bag of Stellar Tracks and Isochrones & \citet{2018ApJ...856..125H} & \url{basti-iac.oa-abruzzo.inaf.it/isocs.html} \\
BaSeL & BaseL Stellar Library & \citet{2002AA...381..524W} & \url{www.astro.mat.uc.pt/BaSeL/} \\
BPASS & Binary Population and Spectral Synthesis & \citet{2018MNRAS.479...75S} & \url{https://bpass.auckland.ac.nz} \\
COLIBRI & A total mystery...? Sorry & \citet{2017ApJ...835...77M} & \url{stev.oapd.inaf.it/cmd} \\
\FITS & Flexible Image Transport System file format standard & \citet{1999ASPC..172..487P} & NA  \\
FSPS & Flexible Stellar Population Synthesis & \citet{2018ApJ...854..139C} & \url{github.com/cconroy20/fsps} \\
PARSEC & PAdova and tRieste Stellar Evolutionary Code & \citet{2012MNRAS.427..127B} & \url{stev.oapd.inaf.it/cmd} \\
\prospect & \R{} package for global SED modelling & \citet{2020MNRAS.495..905R} & \url{github.com/asgr/ProSpect} \\
\R & High-level programming language focussed on data analysis & \citet{citeR} & \url{www.r-project.org} \\
TAMP & Tübingen NLTE Model-Atmosphere Package & \citet{2003ASPC..288...31W} & \url{svo2.cab.inta-csic.es/theory/newov2/}\\

\end{tabular}
\end{center}
\label{tab:software}
\end{table*}%

%%%%%%%%%%%%%%%%%%%%%%%%%%%%%%%%%%%%%%%%%%%%%%%%%%

% Don't change these lines
\bsp	% typesetting comment
\label{lastpage}
\end{document}